\documentclass[12pt]{article}
\pdfoutput=1
\usepackage{jheppub}

\usepackage{subfigure}
\usepackage{amsmath,amssymb,amsthm,amsfonts,array,bbm,mathdots}
\usepackage{graphicx}
\usepackage{color}
\usepackage[utf8]{inputenc}
\usepackage[american]{babel}
\usepackage[mathscr]{euscript}
\usepackage{MnSymbol}

\usepackage{tikz}
\usetikzlibrary{automata}
\usetikzlibrary{arrows}
\usetikzlibrary{calc}
\usetikzlibrary{decorations.markings}
\usetikzlibrary{decorations.pathreplacing}
\usetikzlibrary{intersections}
\usetikzlibrary{positioning}
\usetikzlibrary{topaths}
\usetikzlibrary{shapes.geometric}
\usetikzlibrary{shapes.misc}
\tikzset{->-/.style = {
		decoration = {markings, mark = at position #1 with {\arrow{>}}},
		postaction = {decorate}}}
\tikzset{color-group/.style = {
		shape = circle,
		minimum size = 2.5ex,
		inner sep = .5ex,
		draw}}
\tikzset{flavor-group/.style = {
		shape = rectangle,
		minimum size = 2.5ex,
		inner sep = .5ex,
		draw}}
\tikzset{cf-group/.style = {
		shape = rounded rectangle,
		rounded rectangle right arc = none,
		draw}}
\tikzset{fc-group/.style = {
		shape = rounded rectangle,
		rounded rectangle left arc = none,
		draw}}
\tikzset{cross/.style={minimum width=1pt, path picture={
			\draw[black, very thick]
			(path picture bounding box.south east)
			-- (path picture bounding box.north west)
			(path picture bounding box.south west)
			-- (path picture bounding box.north east);
}}}
\newcommand{\lp}{\left(}
\newcommand{\rp}{\right)}
\newcommand{\ti}{\widetilde}
\newcommand{\tr}{\textrm{Tr} \,}

\newcommand{\wat}{\widehat}

\newcommand{\p}{\partial}

\newcommand{\sh}{\text{sh}}

\newcommand{\w}{\text{w}}

\def\CP1{\bC\bP^1}

\numberwithin{equation}{section}

\newcommand{\be}{\begin{equation}} \newcommand{\ee}{\end{equation}}
\newcommand{\bea}{\begin{equation} \begin{aligned}} \newcommand{\eea}{\end{aligned} \end{equation}}

\newcommand{\vev}[1]{{\langle {#1} \rangle}}
\newcommand{\dd}{\text{d}}

\newcommand{\cC}{\mathcal{C}}

\newcommand{\cH}{\mathcal{H}}

\newcommand{\cL}{\mathcal{L}}

\newcommand{\cN}{\mathcal{N}}
\newcommand{\cO}{\mathcal{O}}

\newcommand{\cR}{\mathcal{R}}

\newcommand{\cW}{\mathcal{W}}

\newcommand{\cY}{\mathcal{Y}}

\newcommand{\bC}{\mathbb{C}}

\newcommand{\bN}{\mathbb{N}}

\newcommand{\bP}{\mathbb{P}}

\newcommand{\bR}{\mathbb{R}}
\newcommand{\bZ}{\mathbb{Z}}
\newcommand{\fg}{\mathfrak{g}}
\newcommand{\ft}{\mathfrak{t}}

\newcommand{\Gr}{\mathrm{Gr}}

\newcommand{\rk}{\mathrm{rk}}
\newcommand{\eps}{\mathrm{\epsilon}}

\def\repa{\raise4pt\hbox{$\square$}\mkern-14mu\raise-4pt\hbox{$\square$}}
\def\repab{\overline{\raise4pt\hbox{$\square$}\mkern-14mu\raise-4pt\hbox{$\square$}\mkern-1mu}}

\DeclareMathOperator{\sgn}{sgn}

\DeclareMathOperator*{\Res}{Res}

\numberwithin{equation}{section}       % equation numbers in each section

\begin{document}
	
	\begin{titlepage}

		\begin{center}

			\vskip .5in %.3in 
			\noindent

			\begin{center}
				\fontsize{18pt}{18pt}\selectfont\bfseries{Quantized Coulomb Branches, Monopole Bubbling
					and Wall-Crossing Phenomena in 3d $\mathcal{N}=4$ Theories}
			\end{center}		
			
			\bigskip\medskip
			
			Benjamin Assel$^1$, Stefano Cremonesi$^2$ and Matthew Renwick$^2$\\
			
			\bigskip\medskip
			{\small 
				$^1$ Theory Department, CERN, CH-1211, Geneva 23, Switzerland \\
				$^2$ Department of Mathematical Sciences, Durham University, Durham DH1 3LE, UK
			}
			
			\vskip .5cm %.3cm
			{\small \tt benjamin.assel@gmail.com, stefano.cremonesi@durham.ac.uk, matthew.renwick@durham.ac.uk}
			\vskip .9cm %.6cm
			{\bf Abstract }
			\vskip .1in
		\end{center}
		
\noindent
To study the quantized Coulomb branch of 3d $\mathcal{N}=4$ unitary SQCD theories, we propose a new method to compute correlators of monopole  and Casimir operators that are inserted in the $\bR\times\bR^2_\epsilon$ Omega background. This method combines results from supersymmetric localization with inputs from the brane realisation of the correlators in type IIB string theory. The main challenge is the computation of the partition functions of certain Super-Matrix-Models (SMMs), which appear in the contribution of monopole bubbling sectors and are realised as the theory living on the D1 strings in the brane construction. We find that the non-commutativity arising in the monopole operator insertions is related to a wall-crossing phenomenon in the FI parameter space of the SMM. 
We illustrate our method in various examples and
we provide explicit results for arbitrary correlators of non-bubbling bare monopole operators.
We also discuss the realisation of the non-commutative product as a Moyal (star) product and use it to successfully test our results.
		
		\vfill

	\end{titlepage}

	\setcounter{page}{1}

	\noindent\hrulefill
	
	\setcounter{tocdepth}{2}
	\tableofcontents
	
	\noindent\hrulefill
	
	\bigskip
	
	%%%%%%%%%%%%%%%%%%%%%%%%%%%%%%%%%%%%%%%%%%
	%%%%%%%%%%%%%%%%%%%%%%%%%%%%%%%%%%%%%%%%%%
	
	%%%%%%%%%%%%%%%%%%%%%%%%%%%%%%%%%%%%%%%%%%
	%%%%%%%%%%%%%%%%%%%%%%%%%%%%%%%%%%%%%%%%%%
	
\section{Introduction}\label{sec:Intro}
	
Coulomb branches of three-dimensional (3d) supersymmetric gauge theories are subvarieties of the moduli space of supersymmetric vacua, in which vector multiplet scalars acquire VEVs. It is not an easy task to characterise their geometry. For instance, the metric on the Coulomb branch of non-abelian theories receives non-perturbative quantum corrections, which are notably hard to compute. Nevertheless, significant progress has been made in recent years for 3d theories with $\cN=4$ supersymmetry, see \cite{Cremonesi:2013lqa, Cremonesi2014b,CremonesiFerlitoHananyEtAl2014,Cremonesi:2014uva, Bullimore:2015lsa, Bullimore:2016hdc,  Assel:2017jgo, Dedushenko:2017avn, Assel:2018exy,  Hanany:2018xth, Dimofte:2018abu, Dedushenko:2018icp} for a physics perspective and \cite{Nakajima:2015txa,Braverman:2016wma, Braverman:2016pwk,Braverman2017,Nakajima:2017bdt,Braverman2018} and references thereof for a rigourous mathematical construction. We will take the physics perspective in this paper. The majority of these results have utilised the description of the Coulomb branch as a complex algebraic variety, hence bypassing difficulties related to the metric. The key players in this description are chiral operators, including standard Casimir invariant operators, and also importantly monopole operators \cite{Borokhov:2002cg}. The VEVs of these chiral operators parametrise the Coulomb branch and the algebraic relations that they satisfy give the chiral ring relations. Although there is always an infinite number of monopole operators, it is believed (but not proved) that the Coulomb branch is finitely generated, namely that every Coulomb branch admits a finite basis of generators, which are constrained by some algebraic relations. The goal then becomes to isolate such a basis and to extract the relations.

	One can go a step further. The coordinate ring $\bC[\cC]$ of the Coulomb branch $\cC$ has the structure of a Poisson algebra and admits a deformation quantization $\bC_\epsilon[\cC]$ \cite{Bullimore:2015lsa, Beem:2016cbd, Braverman:2016wma}, with parameter $\epsilon$, which is an associative, non-commutative algebra.%
	\footnote{In the mathematics literature the quantization parameter is denoted $\hbar$, not to be confused with the Planck constant of the three-dimensional quantum field theory.}  This is referred to as the ``quantized Coulomb branch". It is a richer object compared to the simple Coulomb branch, and thus it is desirable to study. In particular, obtaining the deformed Coulomb branch relations is a useful task. 
	Several methods have been proposed to study the quantized Coulomb branch and we will briefly review some of these proposals in section \ref{sec:CBandLoc}.

	In this paper we study the quantized Coulomb branch of 3d $\cN=4$ SQCD theories by leveraging inputs from supersymmetric localization and brane constructions \cite{Hanany:1996ie}. Our setup allows us in principle to compute any correlator of monopole and Casimir operators on $\bR\times\bR^2_{\epsilon}$, where $\bR^2_{\epsilon}$ is the Omega background with deformation parameter $\epsilon$, and the operators are inserted along the line at the origin of $\bR^2_{\epsilon}$ as in \cite{Bullimore:2015lsa}. These correlators are topological in the sense that they depend only on the ordering of the insertions along this line, but not on their actual positions. We find (following \cite{Bullimore:2015lsa}) that they are expressed as rational functions of abelian coordinate VEVs, which are the VEVs of Cartan scalars and dual photons in the vector multiplet.%
	\footnote{As we will explain in the core of the paper, the correct abelian variables are actually abelian monopole VEVs  and complex scalar VEVs.} The knowledge of these correlators can be used to extract a monopole basis and reconstruct the quantized Coulomb branch relations. In practice however, we will only be able to give explicit expressions for correlators of certain monopole operators of low magnetic charges. As we will explain, more general correlators are obtained by computing certain matrix models, whose precise contour of integration remains to be studied. For SQCD theories it turns out that these low charge monopoles contain a basis of generators and thus we are still able to describe the quantized Coulomb branch. We will devote this paper only to the study and computation of such correlators.
	
	First, we exploit the power of supersymmetric localization, recycling the computations of 't Hooft loops in 4d from \cite{Ito:2011ea} and performing the dimensional reduction to 3d, to extract an expression for a monopole operator VEV and then a correlator of monopole operators, written in terms of a sum over monopole bubbling sectors. These are the sectors of the path integral where the magnetic charge of the defect is screened by the magnetic charge of a smooth 't Hooft-Polyakov monopole \cite{Kapustin2007}. A given bubbling sector contribution contains a product of several factors. All of these factors are well understood, apart from a rational function of the abelian VEVs, $Z_{\rm SMM}$, which is the partition function of an $\cN = (0,4)$ Super-Matrix-Model, a 0d supersymmetric gauge theory.%
	\footnote{The same factor appears in the computation of supersymmetric 't Hooft loops in 4d $\cN=2$ theories, except that it is a partition function of a SQM theory, rather than SMM theory. The computation of such a factor can be quite subtle, as has been demonstrated recently in \cite{Chang2018,Brennan2018b,Brennan2018,Brennan2019a} (see also \cite{Assel:2019iae}).}
	Here lies the difficulty in the computation. First, it is not obvious to determine what these SMM are, and it is certainly not obvious how to evaluate them. 
	
	To solve these issues we rely on a realisation of monopole insertions in a type IIB brane setup (see \cite{Assel:2017hck} for a first study of this setup in abelian theories). We are able to map each bubbling (and non-bubbling) contribution in the monopole VEV expansion, or in the correlator expansion, to a corresponding brane setup. We then read off the SMM as the theory living on the D1 strings. These are gauged quiver SMM with (bi)fundamental hypermultiplets and Fermi multiplets. Once the SMM is known, it remains to determine a contour of integration for the eigenvalues of the matrix model $Z_{\rm SMM}$. For SMM that appear in generic correlators, this is a difficult problem that we do not address in this paper. It is related to the fact that these SMM have vanishing Fayet-Iliopoulos (FI) parameters. However, for correlators of non-bubbling monopoles, the contour is given by the Jeffrey-Kirwan (JK) residue prescription \cite{Jeffrey1995} and we are able to give the final evaluation of such correlators. 
	
	We apply our method in several examples and provide some general results in the $U(N)$ SQCD theory. Since we study the quantized Coulomb branch, we find non-commuting monopole operators and we compute their commutators. Interestingly, the non-commutativity is tied to a wall-crossing phenomenon in the gauged SMM, which is similar to an observation in \cite{Hayashi:2019rpw} for gauged SQM related to non-commutative 't Hooft loop operators. The choice in the ordering of the monopole insertions along the line in $\bR\times\bR^2_\epsilon$ is directly related to a choice in the signs of the FI parameters in the gauged SMM.
	When we reverse the order of two insertions, we cross a codimension one hyperplane -- a wall -- in the FI parameter space and the JK contour changes, instructing us to pick contributions at different poles, and the partition function of the SMM may change. In simple cases, the commutators are simply related to the residues of poles at infinity in the matrix models. We provide several examples of this wall-crossing phenomenon. 
	Beyond pedagogical examples, we give explicit results for arbitrary correlators of ``minimal" bare monopole operators (non-bubbling bare monopoles), whose magnetic charges are highest weights of minuscule representations. We prove that all positively charged (or all negatively charged) monopole operators commute among themselves. On the other hand, we show that positively and negatively charged monopole operators generically do not commute.
	
	Another important feature is the observation that the non-commutative product of monopole operators can be effectively computed as a Moyal (star) product. This can be inferred from the localization results in 4d, and imported to 3d. A similar property was also central in the bootstrap approach in \cite{Dedushenko:2018icp}. The explicit results that we obtain using the brane construction are all in agreement with this Moyal product representation. 
	
	This paper is organised as follows. In section \ref{sec:CBandLoc} we review some results in the literature on Coulomb branches and we extract the 3d localization formulae from the reduction of the 4d ones. We also introduce and discuss the non-commutative product and the Moyal product formulae. In section \ref{sec:Branes_Bubbing_WC} we propose a type IIB brane setup for realising bare monopole operators in $U(N)$ SQCD theories. We relate brane setups to monopole bubbling contributions in the localization formula and explain how to read the SMM and compute $Z_{\rm SMM}$ (at non-zero FI parameters). We then illustrate the method by providing several examples of computations of two and three-point functions in $U(2)$ SQCD. In section \ref{sec:U(N)SQCD} 	we use our brane construction to provide a closed formula for correlation functions of non-bubbling bare monopole operators in $U(N)$ SQCD. We also study various examples of correlators containing bare monopole operators of minimal positive and negative charge, investigating the wall-crossing phenomenon and making connection with the Moyal product. Along the way we outline a correspondence between the data of our brane construction and the geometry of the affine Grassmannian, which plays a key role in the mathematical definition of the Coulomb branch, although we do not rely on this correspondence for any of our computations. 
	In section \ref{sec:CasimirAndDressedMonop} we extend the discussion to include Casimir operators and dressed monopoles. We propose a brane realisation for those operators and compute some simple correlators. Finally, in section \ref{sec:Discussion} we conclude by discussing the remaining issues and possible future work. In particular, the analysis carried out in this paper using brane constructions has the potential to be extended to gauge theories with $\mathcal{N}<4$ supersymmetry, where a mathematical description  along the lines of \cite{Nakajima:2015txa,Braverman:2016wma} has not been developed. We collect several details of the computations of SMM partition functions in various appendices.

	%%%%%%%%%%%%%%%%%%%%%%
	
	\bigskip
	
	{\bf Note}: During the completion of this paper we became aware of an independent related work by T. Okuda and Y. Yoshida \cite{Okuda2019}. We are grateful to them for agreeing to a coordinated submission.

	%%%%%%%%%%%%%%%%%%%%%%%
	
	\section{Quantized Coulomb branches and Localization formulae}
	\label{sec:CBandLoc}

	\subsection{The Coulomb branch in 3d $\cN=4$ theories}
	
	In this paper we consider 3d $\cN=4$ gauge theories. The Lagrangian theory is fixed by the choice of a gauge group $G$, which comes with an $\cN=4$ vector multiplet, and a pseudo-real representation $\cR_{\rm p-r}$ of $G$, under which matter multiplets transform. We will consider only matter hypermultiplets that come in pairs of chiral multiplets transforming in complex conjugate representations, namely $\cR_{\rm p-r} = \cR \oplus \cR^\ast$.
	
	The space of vacua is a union of intersecting branches. Generically there are two distinguished branches: the Higgs branch $\cH$, on which only the $SU(2)_C$ R-symmetry does not act, and the Coulomb branch $\cC$, on which only the $SU(2)_H$ R-symmetry does not act. The remaining mixed branches are products of subspaces of the Higgs and Coulomb branches, on which the full R-symmetry $SU(2)_H\times SU(2)_C$ acts non-trivially.
	
	The Higgs branch is parameterised by VEVs of scalar fields in hypermultiplets, subject to the D- and F-term constraints. It is elegantly described as a hyperk\"ahler quotient $\bR^{4 \, \dim \cR} //// G$ \cite{HitchinKarlhedeLindstromEtAl1987} and is protected against quantum corrections \cite{Argyres:1996eh}.
	
	In this paper we focus instead on the Coulomb branch, which is parameterised by the VEVs of scalar fields in vector multiplets, which take values in the Lie algebra $\mathfrak{g}$ of the gauge group $G$, and of dual photons. 
	The potentials in the action impose that only scalars valued in a Cartan subalgebra $\ft\subset \fg$ can take VEV. The Coulomb branch moduli are thus captured by complex scalars $\varphi_a$, real scalars $\sigma_a$ and dual photons $\gamma_a$, which arise from the dualization of Cartan gauge fields, with $a=1,\ldots,\rk (G)$. Dual photons are compact scalars, which we normalise to have periodicity $2\pi$. These combine with the non-compact real scalars $\sigma_a$ to form the single-valued chiral VEVs
	\be\label{abelian_monopole_class}
	e^{\chi_a} = e^{\frac{2\pi}{g^2}\sigma_a + i \gamma_a} \,,
	\ee
	where $g^2$ is the Yang-Mills coupling (of the relevant gauge group factor).
	
	This leads to the ``classical" Coulomb branch, which is parameterised by the Cartan VEVs $\varphi_a$ and $e^{\chi_a}$ modulo residual gauge transformation in the Weyl group $\cW$:
	\be
	\cC_{\rm classical} = (\bC \times \bC^\ast)^{\rk(G)} / \cW \,.
	\ee
	The classical Coulomb branch receives quantum corrections, which, roughly speaking, encode the fact that the dualization that defines the dual photons $\gamma_a$ is valid at generic points on the Coulomb branch but fails at loci where matters fields or W-bosons become massless. Along these subspaces of $\cC$, the radii of the circles parameterised by some dual photons shrink to zero or diverge, and therefore the $e^{\chi_a}$ are not good coordinates on the full Coulomb branch. Instead, there should be combinations of $\varphi_a$ and $e^{\chi_a}$ that are well-defined everywhere and parametrize $\cC$. In general, $\cC$ will be described as a complex algebraic variety, with the ``good coordinates" as generators and relations between them.  
	
	\medskip
	
	\noindent{\bf Deformation quantization}
	
	Before discussing approaches describing the ``quantized" Coulomb branch, it is important to mention that, as a hyperk\"ahler manifold, $\cC$ has a holomorphic symplectic structure which defines a Poisson bracket. Thus, the coordinate ring of holomorphic functions on $\cC$, which is physically realised in terms of the VEVs of Coulomb operators, also has the structure of a Poisson algebra. It admits a natural quantization, where the VEVs of Coulomb operators are replaced by operators and the Poisson bracket is replaced by a Lie bracket, or commutator, with quantization parameter $\epsilon$. Equivalently, the coordinate ring $\bC[\cC]$ of $\cC$ admits a deformation quantization which gives it the structure of an associative, non-commutative algebra, with deformation parameter $\epsilon$.%
	\footnote{See \cite{Beem:2016cbd}  and references therein for more background on deformation quantization.} % in the context of 3d $\cN=4$ theories}
	We will denote the non-commutative product between VEVs by a star $\star$, and the deformed coordinate algebra by $\bC_\epsilon[\cC]$.

	\medskip
	
	\noindent{\bf Abelianization approach}
	
	In the physics literature indirect methods have been proposed to compute the quantum corrected Coulomb branch.%
	\footnote{Here we are referring to the standard quantum corrections in the gauge theory that are weighted by Planck's constant $\hbar$, and not by the deformation quantization parameter $\epsilon$, which instead physically controls an Omega deformation.} In \cite{Bullimore:2015lsa} it was proposed that one should work in the complement of the loci where W-bosons and hypermultiplets respectively become massless  and use abelian variables $u_v$, with integer vectors $v = (v^a)_{a=1,\ldots,N}$, replacing all the combinations $ e^{v.\chi} := e^{\sum_a v^a\chi_a}$, and consider first the ``abelianized" Coulomb branch $\cC_{\rm abelian}$ parameterised by the $u_v$, subject to a set of quantum relations. 
	The $u_v$ are understood as the VEVs of {\it abelian monopole operators} and are labelled by magnetic charges $v \in \Lambda_{\rm cochar}$, the cocharacter lattice of $G$.
	The precise conjectured relations are\footnote{These relations can be seen as quantum corrections of the classical relations $u_{v_1} u_{v_2} = u_{v_1+v_2}$ for the classical abelian monopole operators $u_v = e^{v^a\chi_a}$.} 
	\be
	u_{v_1} u_{v_2} = u_{v_1+v_2} 
	\frac{\prod\limits_k\prod\limits_{w^{(k)}\in \cR_k} ( w^{(k)}.\varphi - m_k )^{h(w^{(k)},v_1,v_2)}}
	{\prod\limits_{\alpha \in G} ( \alpha.\varphi )^{h(\alpha,v_1,v_2)}} \,,
	\label{abelrel}
	\ee
	where $\alpha \in G$ denotes the non-zero roots $\alpha$ of the gauge algebra and $w^{(k)} \in \cR_k$ denotes the weights $w^{(k)}$ of the representation $\cR_k$. The relations also depend on complex masses $m_k$ of the hypermultiplets in the theory. Finally we have 
	\be
	h(\sigma, v_1, v_2) := \frac 12 \big( |\sigma.v_1| + |\sigma.v_2| - |\sigma.(v_1+v_2)| \big)  \ \in \bZ_{\ge 0} \,.
	\ee
	
	The exact Coulomb branch $\cC$ should then be described as the quotient of $\cC_{\rm abelian}$ by the Weyl group, $\cC_{\rm abelian} / \cW$, extended to the loci which support massless W-bosons. The final description contains the generators $\Phi_n$ and $V_{B,p(\varphi)}$, the VEVs of Casimir invariant  operators (Weyl invariants of the $\varphi_a$) and the VEVs of non-abelian monopoles operators (Weyl invariants of the $u_v$ and $\varphi_a$) labelled by magnetic charges $B \in \Lambda_{\rm cochar}/\cW$ (dominant cocharacters), respectively, and a dressing polynomial $p(\varphi)$, which is an invariant of $\cW_B$, the stabiliser of $B$ in $\cW$. These generators are subject to complex algebraic relations which follow from the abelian relations \eqref{abelrel}. In all the cases that we know of, most of the $V_B$ are generated and the final description of $\cC$ is in terms of finitely many generators and relations.%
	\footnote{It is expected that Coulomb branches of 3d $\cN=4$ theories are affine varieties and therefore that their coordinate rings are finitely generated, but to the best of our knowledge this has not been proven.}
	
	The same construction generalises to the quantized Coulomb branch, where the abelian relations become abelian operator relations.

	\medskip
	
	While this description has passed many consistency tests, it was still observed, for instance in \cite{Assel:2018exy}, that some ingredients in the construction are missing. In particular, one should allow certain rational functions of the $\varphi_a$ (instead of only polynomial functions) in the construction of the monopole generators $V_B$. This is tied to ambiguities of the extension of $\cC_{\rm abelian} / \cW$ to the loci where massless W-bosons arise.
	
	A refined construction was put forward in \cite{Dimofte:2018abu}, based on mathematical works, where one includes a few simple rational functions of the $\varphi_a$ in the form of BGG-Demazure operators before performing the Weyl quotient. Precisely, the space constructed there is a subspace of $\bC_\epsilon[\cC]$ and it is believed to be exactly $\bC_\epsilon[\cC]$, at least in quiver theories (it was shown in linear quiver theories). We refer to the paper for details and references on this approach.
	
	It would be interesting to justify this construction by direct computations. One drawback of this approach is that the physics of monopole bubbling is not manifest. 
	
	\medskip
	
	\noindent{\bf Localization-bootstrap approach}
	
	Another construction was developed in the papers \cite{Dedushenko:2018icp, Dedushenko:2017avn}.%
	\footnote{See \cite{Dedushenko:2016jxl} for earlier work on localization results and the Higgs branch.}
	The sphere partition function with BPS monopole singularities inserted at the two poles of $S^3$ was computed using supersymmetric localization. An essential ingredient in the construction is a certain overlap $\langle \sigma, B | \Psi_b \rangle$, computing the hemisphere partition function with a BPS monopole defect of magnetic charge $b$ inserted at the pole of the hemisphere and BPS boundary conditions at the $S^2$ boundary, imposing (in particular) a magnetic flux $B$ and a constant vector multiplet scalar $\sigma$. The sphere partition function is then obtained by a gluing formula, which includes a sum over Weyl transformations and a sum over monopole bubbling sectors, which we will explain in a moment.
	The results are completely known for abelian gauge theories, which have no bubbling sectors, but it is still unknown how to compute directly the %factors dressing 
	bubbling contributions in non-abelian theories. The authors of \cite{Dedushenko:2018icp} bypass this issue by bootstrapping the form of these terms. 
	
	First, they consider the algebra of (dressed) monopole and Casimir operators acting on the hemisphere partition function, introducing the corresponding operator insertion at the pole. They then observe that the bubbling terms can be absorbed as abelian bubbling operators into the monopole operators. The exact form of these bubbling operators is fixed, up to ambiguities, by requiring operator products, acting on the hemisphere partition function, to be polynomials of a chosen basis of operators. The ambiguities can be understood as operator mixings and fixing such ambiguities is synonymous with choosing a basis of (dressed) monopole operators. 
	
	In this construction the algebra of monopole operators is associative but not commutative, with the inverse radius of the three-sphere $\epsilon = r^{-1}$ playing the role of the non-commutative parameter. The algebra is thus a quantization of $\cC$ and the undeformed Coulomb branch chiral ring is obtained by taking the limit $\epsilon \to 0$. 
	In this limit the operator product becomes commutative and taking the VEV of the operator relations yields the description of $\cC$ as an algebraic variety. The mixing ambiguities discussed above disappear in this limit, since they arise from operators multiplied by positive powers of $r^{-1}$. 
	
	%\medskip
	
	This construction is very general. It can be applied to any Lagrangian gauge theory and it reproduces the abelian relations \eqref{abelrel}.
	We refer to \cite{Dedushenko:2018icp} for details and examples.
	
	\noindent The main drawback of this approach is that the computations involved are case-by-case and become rapidly involved as the rank of the gauge group increases. Moreover, monopole operators in this approach are considered up to operator mixings, whereas one might want to select a distinguished basis of monopole operators, which are those defined by BPS monopole singuarities in the path integral. Such a preferred basis cannot be found from this method. It would also be desirable to find a direct method for computing monopole bubbling factors, compared to a bootstrap approach. Finally, one might prefer to compute correlators of monopoles on $\bR\times\bR^2_\epsilon$ rather than on $S^3$, studying directly $\bC_\epsilon[\cC]$.
	
	\medskip
	
	\noindent{\bf  A first principle approach}
	
	In this paper we propose a new approach to the quantized Coulomb branch $\bC_\epsilon[\cC]$, using results from localization computations on $\bR \times \bR^2_\epsilon$ with monopole insertions along the $\bR$ line, which we parametrize by a Euclidean time coordinate $x^0$. The $\bR^2_\epsilon$ directions stand for the Omega background  \cite{Nekrasov:2010ka} with parameter $\epsilon$. Mathematically, this amounts to working equivariantly with respect to rotations in the $\bR^2_\epsilon$ plane, with equivariant parameter $\epsilon$.
	The insertion of a BPS monopole of charge $B \in \Lambda_{\rm cochar}/\cW$ is defined by requiring in the path integral the singular BPS profile
	\be
	F = -\frac{B}{2} \star \dd\lp \frac 1 r \rp \,, \quad \sigma = \frac{B}{2r} \,, 
	\label{MonoProfile}
	\ee
	in the vicinity of the insertion point (with $r$ the radial distance). Due to the Weyl average, Weyl equivalent profiles ($B \to \w(B)$, for $\w \in \cW$) are also allowed and summed over. Strictly speaking \eqref{MonoProfile} is valid at $\epsilon=0$. When $\epsilon \neq 0$ the Omega background localizes the flux to the line at the origin of the Omega plane, with a discontinuity at the insertion point.

	The background with Coulomb operators $\cO_i$ inserted at the origin of $\bR^2_\epsilon$ and at arbitrary but different positions $x^0_i$ preserves two supercharges.%
	\footnote{The 3d $\cN=4$ theory preserves eight Poincar\'e supercharges. The insertion of half-BPS Coulomb operators breaks supersymmetry by a half, and the Omega background on $\bR^2_\epsilon$ by another half. The resulting two supercharges generate a 0d $\cN=2$ supersymmetry algebra.}
	The correlators $\vev{\cO_1(x^0_1) \cO_2(x^0_2) \cdots}$ computed in this way are topological, in the sense that they do not depend on the actual positions $x^0_i$, {\it except (possibly) for their ordering} along $\bR$. 
	\smallskip
	
	For a single insertion, the VEV is simply independent of the position and we define
	\be
	\cO := \vev{\cO(x^0)} \,.
	\ee
	The VEVs $\cO$ of Coulomb operators define elements of the ring $\bC_\epsilon[\cC]$.
	\smallskip
	
	For two local operator insertions, the correlators define an associative but non-commutative product, with
	\be
	\cO_1 \star \cO_2 := \vev{\cO_1(x^0 + \delta) \cO_2(x^0 - \delta)} \quad \,, \ \delta > 0 \,.
	\label{NonComProd}
	\ee
	The right hand side is independent of $\delta$, as long as $\delta >0$, and the limit $\delta \to 0^+$ yields the VEV of a local chiral operator. It therefore belongs to $\bC_\epsilon[\cC]$. Thus the star product is an associative non-commutative product on $\bC_\epsilon[\cC]$. 
	The Omega background parameter $\epsilon$ is interpreted as the deformation parameter, as anticipated in the notation.
	In the limit $\epsilon \to 0$, we obtain Coulomb operator insertions on flat $\bR^3$, which are independent of the positions in $\bR^3$. There is no ordering anymore and the star product becomes the usual commutative product between holomorphic functions on $\cC$.
	
	From the localization results, one can determine the quantized Coulomb branch (namely $\bC_\epsilon[\cC]$), since localization provides exact computations for correlators, such as $\cO_1 \star \cO_2$, and allows us to determine the (deformed) Coulomb branch relations.  In addition, one may want to characterise the commutators $[\cO_1, \cO_2]_\star$, for any $\cO_1, \cO_2$. 
	Our first objective will be to compute correlators of monopole operators in SQCD theories, using localization and some insights from brane constructions.
	As we will see, the results from localization on $\bR \times \bR^2_\epsilon$ will be closely related to the abelianization approach that appeared in previous works.

	\subsection{Localization formulae}
	
	The localization of Coulomb operators in the $\bR \times \bR^2_\epsilon$ background has not been performed to our knowledge. This is a long computation that would deserve a paper in itself. Fortunately a closely related computation has been performed.  In \cite{Ito:2011ea}  the VEV of 't Hooft lines of 4d $\cN=2$ theories on $\bR \times \bR^2_\epsilon \times S^1$,%
	\footnote{What we denote schematically as $\bR^2_\epsilon \times S^1$ is more precisely an $S^1$ fibration over $\bR^2$, where the periodicity conditions on the circle are twisted by a rotation in the plane.}  wrapping $S^1$ and inserted at points on $\bR$, were computed using supersymmetric localization. Upon reduction along $S^1$, 't Hooft loops become local 't Hooft monopole operator insertions. Thus, the 3d localization results can be inferred from the 4d results by taking an appropriate limit, which truncates the Kaluza-Klein towers on $S^1$ to their zero modes.
	\medskip
	
	\noindent{\bf The 4d localization result}
	
	The result of the computation in \cite{Ito:2011ea} can be recast as follows (in the spirit of \cite{Dedushenko:2018icp}). The VEV of a 't Hooft loop $L_B$ of magnetic charge $B \in \Lambda_{\rm cochar}/\cW$ takes the form\footnote{For a vector $x = (x_1, \ldots, x_n)$, we define $x^\w = (x_{\w(1)}, \ldots, x_{\w(n)})$ for $\w \in \cW$, $n=\,\rk(G)$.}
	\bea
	\vev{L_{B}} & = \frac{1}{|\cW_B|}\sum_{\w \in \cW} e^{B^\w. b} Z_{\rm 1-loop}(\epsilon, a,m; B^\w)  \cr
	&+  \sum_{|v| < |B|} \frac{1}{|\cW_v|} \sum_{\w \in \cW}  e^{v^\w.b} \, Z_{\rm 1-loop}(\epsilon,a,m;v^\w) \, Z_{\rm bub}(\epsilon,a,m;B^\w,v^\w) \,,
	\eea
	where the first sum is over the Weyl group and carries the contributions of the non-bubbling abelian magnetic sectors $B^\w$. The second sum is over monopole bubbling sectors labelled by $v$, which are the dominant weights (other than $B$) appearing in the representation of highest weight $B$ of the dual gauge group $G^\vee$ (this is symbolised by the notation $|v| < |B|$). This implies $B-v \in \Lambda_{\rm cr}$, the coroot lattice of $G$.  
	The electric chemical potentials $a_a$, with $a=1,\ldots,\, \rk(G)$, are the VEVs of the real parts of the eigenvalues of the adjoint complex scalars in the 4d $\cN=2$ vector multiplet, complexified by the holonomies of their photon superpartners. Similarly, the magnetic chemical potentials $b_a$ are the VEVs of the imaginary parts of the eigenvalues of the adjoint complex scalars, complexified by the holonomies of the dual photons (which are gauge bosons in four dimensions). The masses $m_k$ are analogues of $a_a$ for flavour symmetries. Finally, $\cW_B$ is the stabiliser of $B$ in $\cW$ and $|\cW_B|$ is its order.
	
	The one-loop contribution is given by a product of contributions from the vector multiplet and hypermultiplets, 
	\bea
	Z^{\rm vec}_{\rm 1-loop}(v) &= \prod_{\alpha \in G \atop \alpha >0} \prod_{j=0}^{|\alpha.v|-1} \sh \Big[ \pm(\alpha.a) + \big(|\alpha.v|-2j \big)\epsilon \Big]^{-1/2} \cr
	Z^{\rm hyp}_{\rm 1-loop}(v) &= \prod_{w \in \cR } \prod_{j=0}^{|w.v|-1} \sh\Big[ w.a-m + \big(|w.v|-1-2j \big)\epsilon \Big]^{1/2} \,,
	\eea
	with $\sh(x) := 2\sinh(\frac x2)$ and $f(x\pm y) := f(x+y)f(x-y)$.
	
	The factor $Z_{\rm bub}(B,v)$, on the other hand, arises physically from the modes living on the defect when a screening smooth monopole collapses onto the 't Hooft loop. It is computed as the Witten index of a specific  $\cN = (0,2)$ deformation of an ADHM $\cN = (0,4)$ supersymmetric quantum mechanics (SQM). The $\cN=(0,2)$ deformation, which we may call $\cN=(0,2)^*$ borrowing a common terminology in related contexts, is constructed by selecting an $\cN=(0,2)$ subalgebra of the $\cN=(0,4)$ supersymmetry algebra and turning on a constant background proportional to $\epsilon$ for the Cartan generator of the R-symmetry of the $\cN=(0,4)$ supersymmetry algebra, which commutes with the selected $\cN=(0,2)$ subalgebra. The computation of this Witten index can be rather subtle \cite{Brennan2019a, Assel:2019iae}.
	\medskip
	
	\noindent{\bf From 4d to 3d}
	
	In the above formulae, the $S^1$ radius $R$ has been set to one. We can re-introduce it by rescaling the dimensionful parameters by $R$ to build the dimensionless quantities: $a_a \to R a_a, m_k \to R m_k, \epsilon \to R\epsilon$. The 3d limit is then obtained by taking $R \to 0$, keeping $a_a$, $m_k$, $\epsilon$ and $b_a$ fixed, and renormalising the leading order term by an appropriate power of $R$ that is fixed by dimensional analysis to obtain a finite result. After a change of complex structure,%
	\footnote{The purely three-dimensional $\cN=4$ theory (with no Kaluza-Klein states) has three equivalent complex structures in a triplet of the $SU(2)_C$  R-symmetry that acts on its hyperk\"ahler Coulomb branch, so this change of complex structure is inconsequential.}
	the complex scalars $a_a$ and $b_a$ can be identified with the 3d complex scalars $\varphi_a$ and $\chi_a = \frac{1}{g^2}\sigma_a + i \gamma_a$, respectively. 
	
	In this limit the form of the result as a sum over monopole bubbling configurations is preserved and the one-loop determinants simplify from trigonometric to rational functions, which can alternatively be obtained by truncating the Kaluza-Klein towers on the circle to their zero-modes. 
	We obtain the following 3d localization formulae. The VEV of a bare monopole operator $V_B$ of magnetic charge $\Lambda_{\rm cochar}/\cW$ is given by
	\bea
	V_B & = \frac{1}{|\cW_B|}\sum_{\w \in \cW} e^{B^\w.\chi} Z_{\rm 1-loop}(\epsilon, \varphi,m; B^\w)  \cr
	&+  \sum_{|v| < |B|} \frac{1}{|\cW_v|} \sum_{\w \in \cW}  e^{v^\w. \chi} \, Z_{\rm 1-loop}(\epsilon,\varphi,m;v^\w) \, Z_{\rm bub}(\epsilon,\varphi,m;B^\w,v^\w) \,,
	\label{locresult}
	\eea
	and the one-loop contributions are 
	\bea
	Z^{\rm vec}_{\rm 1-loop}(\varphi;v) &= \prod_{\alpha \in G \atop \alpha >0} \prod_{j=0}^{|\alpha.v|-1} \Big[ \pm(\alpha.\varphi) + \big(|\alpha.v|-2j \big)\epsilon \Big]^{-1/2} \cr
	Z^{\rm hyp}_{\rm 1-loop}(\varphi;v) &= \prod_{w \in \cR } \prod_{j=0}^{|w.v|-1} \Big[ w.\varphi-m + \big(|w.v|-1-2j \big)\epsilon \Big]^{1/2} \,.
	\eea
	The factors $Z_{\rm bub}(\varphi;B,v)$ now refer to the 3d bubbling factors, each of which is computed by the partition function of an $\cN=2$ deformation of an $\cN=(0,4)$ SMM (Super-Matrix-Model) ADHM-like theory. This theory can be read off from brane constructions for theories with classical gauge groups. We will discuss it further in section \ref{sec:Branes_Bubbing_WC}.
	
	\smallskip
	
	We notice that the one-loop factors are square roots and as such have sign ambiguities. It is not clear to us how the localization computation should be modified to lift these sign ambiguities. We will make a proposal for how to resolve this issue shortly.

	\medskip
	
	\noindent{\bf Dressed monopoles}
	
	As already mentioned, a monopole insertion with magnetic charge $B$ can be dressed with the extra insertion of a polynomial $p(\varphi)$ that is invariant under $\cW_B$, the stabiliser of $B$ in the Weyl group. This means that in the path integral, the sector with monopole singularity $B^\w$, for $\w \in \cW/\cW_B$, carries the insertion of $p(\varphi^{\w^{-1}})$ at the same point in space. This defines the dressed monopole $V_{B,p(\varphi)}$. For a given $B$, there are always only a finite number of independent dressed monopoles $V_{B,p(\varphi)}$ \cite{Cremonesi:2013lqa}.
	
	The localization argument in the presence of the polynomial insertion $p(\varphi)$ goes through without modification and the result is simply the same as before, with only two differences: the non-bubbling terms carry the factors $p(\varphi)$ and the SMM computing the monopole bubbling terms may be modified. This last point is visible from the brane picture that we will develop in the next section.%
	\footnote{A more uniform way to look at the localization expression is to understand the dressing terms $p(\varphi)$ also as SMM factors, as will stem from the brane discussion.}
	
	The localization result thus takes the form
	\bea
	V_{B,p(\varphi)} & = \frac{1}{|\cW_B|}\sum_{\w \in \cW} p(\varphi^{\w^{-1}}) e^{B^\w.\chi} Z_{\rm 1-loop}(\epsilon, \varphi,m; B^\w)  \cr
	&+  \sum_{|v| < |B|}  \frac{1}{|\cW_v|} \sum_{\w \in \cW}  e^{v^\w. \chi} \, Z_{\rm 1-loop}(\epsilon,\varphi,m;v^\w) \, Z_{\rm bub}(\epsilon,\varphi,m;B^\w,v^\w, p^{\w^{-1}}) \,.
	\eea

	\medskip
	
	\noindent{\bf Abelian variables}
	
	In the localization result \eqref{locresult}, one immediately identifies the abelian monopole VEVs discussed in \cite{Bullimore:2015lsa},
	\be
	u_v = e^{v.\chi} Z_{\rm 1-loop}(\varphi;v) = e^{v.\chi} \left( \frac{\prod\limits_{w\in \cR} \prod\limits_{j=0}^{|w.v|-1}[ w.\varphi - m + (|w.v| - 1 - 2j)\epsilon ]}
	{\prod\limits_{\alpha \in G} \prod\limits_{j=0}^{|\alpha.v|-1}[ \alpha.\varphi  + (|\alpha.v| - 2j)\epsilon ]} \right)^{1/2} \,.
	\label{abelvar}
	\ee
	In terms of the abelian monopole variables $u_v$, the localization formula for bare monopole VEVs takes the simpler form
	\bea
	V_B &= \frac{1}{|\cW_B|}\sum_{\w \in \cW} u_{B^\w}  +  \sum_{|v| < |B|}  \frac{1}{|\cW_v|} \sum_{\w \in \cW} u_{v^\w} \, Z_{\rm bub}(\epsilon,\varphi,m;B^\w,v^\w) \,,
	\label{locresult2}
	\eea
	reminiscent of the formulae in \cite{Dedushenko:2018icp}. The formula for dressed monopoles simplifies similarly.
	
	Now we notice that the sign ambiguities, due to the square roots in the formulae, have been absorbed into the definition of the abelian variables $u_v$, which we regard as single valued (lifting sign ambiguities).

	\medskip
	
	\noindent{\bf Comment on the action of PT}
	
	An important consistency check of our results will be related to the action of the spacetime symmetry PT on monopole operators, where we define PT to act as a reflection on Euclidean time $x^0 \to -x^0$ and one coordinate $x^1 \to - x^1$, which is nothing but a rotation by $\pi$ in the $01$ plane. The reflection leaves invariant a monopole operator sitting at the origin (the reflection P or T alone would instead reverse the magnetic flux). On $\bR\times\bR^2_\epsilon$ the P symmetry can be implemented as the reversal of the Omega background parameter $\epsilon \to -\epsilon$, while the $T$ action simply reverses the locations of insertion points on the $\bR$ axis.
	
	For a single monopole operator sitting at the origin, this symmetry implies that the exact VEVs obey the property
	\be
	V_{B}(\epsilon)  = V_{B}(-\epsilon) \,.
	\label{PTrel1}
	\ee
	For the insertion of two monopoles, that are brought to the origin, the $T$ action reverses their ordering, therefore we have the relation
	\be
	V_{B_1} \star V_{B_2} (\epsilon)   = V_{B_2} \star V_{B_1} (-\epsilon) \,.
	\label{PTrel2}
	\ee
	Other identities in the same vein hold for higher point functions. Similar observations were made in \cite{Beem:2016cbd}. 
	
	\medskip
	
	Examining the various terms in the localization formula \eqref{locresult}, we observe that the one-loop contributions are invariant under $\epsilon \to -\epsilon$. The relation \eqref{PTrel1} then implies that each bubbling term has the symmetry
	\be
	Z_{\rm bub}(\epsilon;B,v) = Z_{\rm bub}(-\epsilon;B,v) \,.
	\label{PTrel3}
	\ee
	In addition, we observe that the abelian monopole variables $u_v$ are also invariant under $\epsilon\to - \epsilon$, extending the property \eqref{PTrel1} to abelian operators VEVs.
	
	\bigskip

	\noindent{\bf Formula for star products/correlators}
	
	There is a variation of the localization formula \eqref{locresult2} that computes correlators of ``minimal" monopoles, compared to the VEV of a single monopole. 
	Instead of considering the VEV of a single monopole,  one can consider a correlation function%
	\footnote{In such expressions where we do not write the insertion positions $x^0_i$ explicitly, it is assumed that the monopoles are already time ordered.} $\vev{\prod_i (V_{h_i})^{n_i}} = \vev{(V_{h_1})^{n_1} (V_{h_2})^{n_2} \cdots }$, with $n_i \in \bZ_{>0}$, where $h_i$ are generators in the various chambers of the magnetic lattice $\Lambda_{\rm cochar}$. We will refer to the $h_i$ as ``minimal" magnetic charges.  One important property of minimal monopoles is that they do not have bubbling sectors, so that we already know their exact VEVs.
	
	The correlator $\vev{\prod_i (V_{h_i})^{n_i}}$ defines a local chiral operator (since the insertion points can be collapsed at a single point) and it can be expanded in a linear combination of monopole VEVs. This linear combination must involve a monopole operator of magnetic charge $B = \sum_i n_i h_i$, which can be a dressed monopole, plus monopoles of lower magnetic charges that arise from the bubbling sectors. Such an expansion is formally similar to \eqref{locresult2}.
It takes the generic form
	\bea
	\vev{\prod_i (V_{h_i})^{n_i}} \Big|_\tau &= \sum_{|v| \le |B|}  \frac{1}{|\cW_v|} \sum_{\w \in \cW} u_{v^\w} \, Z^{(\tau)}_{\rm SMM}(\epsilon,\varphi,m;B^\w,v^\w) \,,  \quad \tau \in S_N \,,
	\label{locresult3}
	\eea
	where the sum over $v$ comprises all the weights that appear in the representation of highest weight $B$, namely the bubbling and non-bubbling sectors. Each abelian monopole sector is weighted with a Super-Matrix-Model factor $Z^{(\tau)}_{\rm SMM}$, which can be a factor 1, or a dressing term in a non-bubbling sector, or a non-trivial SMM factor in a bubbling sector. 
	Again, the SMM computing the various factors can be found by using the brane realisation of the monopole correlator, when available.
	
	This formula involves a permutation $\tau \in S_N$, where we define $N = \sum_i n_i$, which describes the ordering of the minimal monopoles $V_{h_i}$ in the correlator. Correspondingly, the right hand side should depend on $\tau$.
	In the next section we will argue, from the brane realisation of monopole correlators, that the dependence of the ordering $\tau$
	is directly connected to the chamber in which the FI parameter vector of the SMM belongs. 
	We will provide the explicit relation between the ordering $\tau$ and the FI signs in examples.

	As we will see, the SMM are readily evaluated when the FI parameters are away from certains walls (hyperplanes in FI space), allowing us to compute correlators of minimal monopoles with \eqref{locresult3} efficiently. We will thus relate the non-commutativity in monopole correlators to a wall-crossing phenomenon in the matrix models. 
	
	On the other hand, the evaluation of matrix models on the FI space walls, which is necessary to evaluate VEVs of higher charge monopoles with \eqref{locresult2}, is a more challenging task and we will postpone it to a future work.

	\subsection{Star product as a Moyal product and abelian relations}
	\label{sec:Star product as a Moyal product and abelian relations}
	
	Another nice input from the 4d analysis in \cite{Ito:2011ea} is the explicit definition of the non-commutative star product as a Moyal product, 
	\be\label{Moyal}
	(f\star g)(\varphi,\chi) := e^{\epsilon \sum_a (\p_{\chi'_a}\p_{\varphi_a} - \p_{\varphi'_a}\p_{\chi_a} )} f(\varphi,\chi) g(\varphi',\chi') \Big|_{\varphi'=\varphi \atop \chi'=\chi} \,.
	\ee
	The star product between two VEVs can be computed via the following formula. With the VEV of a monopole operator $V$ given by the expansion
	\be
	V = \sum_{v} e^{v.\chi} Z_V (\varphi;v) := \sum_{v} Z_{V,\text{tot}} (\varphi,\chi;v) \,,
	\ee
	we have 
	\be
	\begin{split}
		\vev{V_1 V_2} := V_1\star V_2 &= \sum_{v_1} \sum_{v_2}  Z_{V_1,\text{tot}} \big( \varphi + \epsilon v_2 ,\chi ;v_1 \big) Z_{V_2,\text{tot}} \big( \varphi - \epsilon v_1 , \chi ;v_2 \big) \\
		&= \sum_{v_1} \sum_{v_2}  e^{(v_1+v_2).\chi} Z_{V_1} \big( \varphi + \epsilon v_2 ;v_1 \big) Z_{V_2} \big( \varphi - \epsilon v_1 ;v_2 \big) \,.
	\end{split}
	\label{starprod}
	\ee
	This formula will prove very useful in our analysis.
	
	Computing the star product of two abelian monopole variables we find the general abelian relations
	\be
	\begin{split}
		\label{star_mon}
		u_{v_1} \star u_{v_2} &= u_{v_1+v_2} 
		\frac{\prod\limits_{w\in \cR} \prod\limits_{j_w=-\frac{h^-(w,v_1,v_2)-1}{2}}^{\frac{h^-(w,v_1,v_2)-1}{2}} \big[ w.\varphi - m - \mathrm{sgn}(w.v_{12}) h^+(w,v_1,v_2)\epsilon + 2j_w\epsilon \big]}
		{\prod\limits_{\alpha \in G} \prod\limits_{j_\alpha=-\frac{h^-(\alpha,v_1,v_2)-1}{2}}^{\frac{h^-(\alpha,v_1,v_2)-1}{2}}  \big[ \alpha.\varphi  - \mathrm{sgn}(\alpha.v_{12})h^+(\alpha,v_1,v_2)\epsilon + (2j_\alpha + 1)\epsilon \big]} \,,
	\end{split}
	\ee
	with 
	\be \label{hpm}
	h^\pm(\sigma, v_1, v_2) := \frac 12 \big( |\sigma.v_1| + |\sigma.v_2| \pm |\sigma.(v_1+v_2)| \big) \,,
	\ee
	and $v_{12} = v_1 - v_2$. 
	More generally 
	\be
	\label{star_rel}
	(u_{v_1} f_1(\varphi)) \star (u_{v_2}f_2(\varphi)) = (u_{v_1} \star u_{v_2}) f_1(\varphi+ v_2 \epsilon) f_2(\varphi- v_1 \epsilon)~,
	\ee
	and in particular
	\be
	\label{q_abelian_relations}
	u_v \star f(\varphi) = u_v \cdot f(\varphi- v\epsilon )~, \qquad f(\varphi) \star u_v = u_v \cdot f(\varphi+v\epsilon)~.
	\ee
	These are the ``quantized" abelian relations. They nicely reduce to the abelian relations \eqref{abelrel} conjectured in \cite{Bullimore:2015lsa} in the commutative limit $\epsilon \to 0$. 
	
	\medskip
	
	Since the star product is a product on $\bC_\epsilon[\cC]$, it can be used to generate monopole operators with higher magnetic charge from products of monopole operators with lower magnetic charges, or dressed monopoles from products of bare monopoles and Casimir operators. However, to identify the precise operators that appear in a product, one first needs  to know the explicit expression \eqref{locresult2} for monopole operators in terms of abelian variables.

	\subsection{Some applications in $U(N)$ SQCD}
	\label{sec:UNsimpleProd}
	
	To close this section we would like to put this machinery to use for the SQCD theory with gauge group $U(N)$ and $N_f$ flavours of hypermultiplets in the fundamental representation, which is the main focus in this paper.
	
	The cocharacter lattice allows for non-abelian magnetic charges $B \in \bZ^N/S_N$. The bare monopoles of smallest charge are $V_{(1,0^{N-1})}$ and $V_{(0^{N-1},-1)}$. The coroot lattice is generated by the simple coroots $\alpha^\vee_n=(0^{n-1},1,-1,0^{N-n-1})$ for $n=1,\dots,N-1$, and the above monopoles have no bubbling sector. Their VEVs are simply
	\bea\label{V+-}
	V_{(\pm 1, 0^{N-1})} &= \sum_{a=1}^N u_{\pm e_a} \,, 
	\eea
	where $e_a  := (0^{a-1}, 1, 0^{N-a})$. Explicitly, the abelian monopole operators of minimal charge are 
	\be
	u_{\pm e_a} =  e^{\pm\chi_a} \left( \frac{\prod\limits_{k=1}^{N_f}  (\varphi_a -m_k)}{\prod\limits_{b \neq a}  (\pm \varphi_{ab} + \epsilon) } \right)^{1/2}\,.
	\ee 
	These obey the star product relations
	\bea
	\label{starproducts}
	u_{e_a} \star u_{-e_a}  &= (-1)^{N-1} \frac{\prod\limits_{k=1}^{N_f} (\varphi_a - m_k - \epsilon)}{\prod\limits_{b\neq a} \varphi_{ab} (\varphi_{ab} - 2\epsilon)} \,, \quad a=1,\ldots, N\,, \cr
	u_{-e_a}  \star u_{e_a}  &= (-1)^{N-1} \frac{\prod\limits_{k=1}^{N_f} (\varphi_a - m_k + \epsilon)}{\prod\limits_{b\neq a} \varphi_{ab} (\varphi_{ab} + 2\epsilon)} \,, \quad a=1,\ldots, N\,, \cr
	u_{e_a} \star u_{-e_b} &= u_{-e_b} \star u_{e_a}  =   u_{e_a - e_b}  \,, \quad a\neq b \,, \cr
	u_{e_a} \star u_{e_b} &= -\frac{1}{\varphi_{ab} (\varphi_{ab} - 2\epsilon)} u_{e_a + e_b} \,, \cr
	u_{-e_a} \star u_{-e_b} &= -\frac{1}{\varphi_{ab} (\varphi_{ab} + 2\epsilon)} u_{-e_a - e_b} \,, \cr
	(u_{\pm e_a})^{\star n} &= u_{\pm ne_a}\,,  \quad n>0 \,,
	\eea
	where we have explicitly 
	\bea
	u_{e_a+e_b} &= e^{\chi_a + \chi_b} \left( \frac{\prod\limits_{k=1}^{N_f} (\varphi_a-m_k)(\varphi_b-m_k)}{\prod\limits_{c\neq a,b} (\pm \varphi_{ac} +\epsilon) (\pm \varphi_{bc} +\epsilon)} \right)^{1/2}\,, \\
	u_{e_a-e_b} &= e^{\chi_{ab}} \left( \frac{\prod\limits_{k=1}^{N_f} (\varphi_a-m_k)(\varphi_b-m_k)}{(\pm \varphi_{ab})( \pm \varphi_{ab} + 2\epsilon) \prod\limits_{c\neq a,b} (\pm \varphi_{ac} +\epsilon) (\pm \varphi_{bc} +\epsilon)} \right)^{1/2}\,, \\
	u_{ne_a} &=  e^{n \chi_a} \left( \frac{\prod\limits_{k=1}^{N_f}  \prod\limits_{j=0}^{|n|-1}[\varphi_a -m_k + (|n|-1 - 2j)\epsilon]}{\prod\limits_{b \neq a}  \prod\limits_{j=0}^{|n|-1}[\pm \varphi_{ab} + (|n| - 2j)\epsilon] } \right)^{1/2}\,, \qquad n \in \bZ\,.
	\eea
	From these relations we compute for instance 
	\bea
	\label{star_product_2_minimal}
	V_{(1, 0^{N-1})} \star V_{(1, 0^{N-1})}  &=  \sum_{a=1}^N u_{2e_a} - \sum_{a\neq b} \frac{1}{\varphi_{ab} (\varphi_{ab} - 2\epsilon)} u_{e_a+e_b}  \,, \cr
	V_{(1, 0^{N-1})} \star V_{(0^{N-1}, -1)} &=  \sum_{a \neq b} u_{e_a-e_b} + (-1)^{N-1} \sum_{a=1}^N  \frac{\prod\limits_{k=1}^{N_f} (\varphi_a - m_k - \epsilon)}{\prod\limits_{b\neq a} \varphi_{ab} (\varphi_{ab} - 2\epsilon)} \,, \cr
	V_{(0^{N-1}, -1)} \star V_{(1, 0^{N-1})} &=  \sum_{a \neq b} u_{e_a-e_b} + (-1)^{N-1} \sum_{a=1}^N  \frac{\prod\limits_{k=1}^{N_f} (\varphi_a - m_k + \epsilon)}{\prod\limits_{b\neq a} \varphi_{ab} (\varphi_{ab} + 2\epsilon)}  \,,
	\eea
	where the third product is obtained from the second by reversing the sign of $\epsilon$, in agreement with \eqref{PTrel2}.
	
	We would like to identify the right hand sides with combinations of monopole operators $V_{(2, 0^{N-1})}$, $V_{(1^2, 0^{N-2})}$, $V_{(1, 0^{N-2},-1)}$ and Casimir operators $P(\varphi)$. To do this we need to know the exact bubbling factors for these monopoles (except for $V_{(1^2, 0^{N-2})}$, which has no bubbling). To compute these bubbling factors we are going to propose to use a type IIB brane realisation of monopole operators. Unfortunately, we will not be able to extract explicit results for the bubbling factors of non-minimal monopoles. What we will be able to do instead is to reproduce the above formulae, by computing correlators of minimal monopole operators, including all bubbling factors.

%%%%%%%%%%%%%%%%%%%%%%%%

\section{Brane constructions and monopole bubbling}
\label{sec:Branes_Bubbing_WC}

In this section we explain how to realise monopole operators in brane systems and how to read off the SMM which encodes the monopole bubbling contributions. We focus here on $U(2)$ SQCD with $N_f$ flavours and generalise to $U(N)$ in section \ref{sec:U(N)SQCD}.

\subsection{Brane realisation of monopoles and localization formula}

The brane realisation of monopole operators in abelian 3d $\cN=4$ gauge theories was introduced and studied in \cite{Assel:2017hck} (see also \cite{Assel:2017eun} for the analogue in $\cN=3$ Chern-Simons-Matter theories). Here we extend the construction of \cite{Assel:2017hck} to non-abelian SQCD theories.

\medskip

To realise the 3d theory we consider the usual Hanany-Witten brane setup with D3, D5 and NS5' branes oriented as the three first entries in Table \ref{tab:orientations}. $n$ D3 branes stretched between two NS5' branes realise a 3d $U(n)$ gauge group. $m$ D5 branes intersecting the $n$ D3 segments realise $m$ fundamental hypermultiplets for this gauge group.

\begin{table}[h]
	\begin{center}
		\begin{tabular}{|c||ccc|c|ccc|ccc|}
			\hline
			& 0 & 1 & 2 & 3 & 4 & 5 & 6 & 7 & 8 & 9 \\ \hline
			D3  & X & X & X & X &   &   &   &   &   &   \\
			D5  & X & X & X &   & X & X & X &   &   &   \\
			NS5' & X & X & X &   &   &   &   & X & X & X \\  \hline
			D1 &  &  &  & X  &   &   &   & X &  &  \\ 
			D3' &  &  &  &   & X  & X  & X  & X &  &  \\ 
			NS5 & & & & X & X & X & X &  & X & X \\ \hline
		\end{tabular}
		\caption{\footnotesize Brane array for 3d $\cN=4$ theories and half-BPS local operators.}
		\label{tab:orientations}
	\end{center}
\end{table}

The D1 string, D3' and NS5 branes that appear in Table \ref{tab:orientations} realise the insertion of Coulomb branch operators in the theory. 

An important property is that the triple (NS5, D3, D1) is a Hanany-Witten triple, which means that there is a D1 creation/annihilation effect as an NS5 and a D3 cross each other. It also means that there is an s-rule: at most one D1 string can be stretched between an NS5 and a D3 \cite{Hanany:1996ie}.

\medskip

We can define a linking number $\ell$ for a D3 brane as follows:
\be
\ell = \frac 12 [n(NS5_{L}) - n(NS5_{R})] + n(D1_R) - n(D1_L) \,,
\label{linkingD3}
\ee
where $n(NS5_L)$, or $n(NS5_R)$, is the number of NS5 on the left, or on the right, of the D3 brane along the $x^7$ direction. Similarly, $n(D1_L)$, or $n(D1_R)$, is the number of D1 strings ending on the D3 brane from the left, or from the right, along $x^7$.
For the $U(2)$ theory we have only two linking numbers $(\ell_1, \ell_2)$ for the two D3 branes. 

Similarly we can define a linking number $h$ for an NS5 brane,
\be
h = \frac 12 [\wat n(D3_R) - \wat n(D3_L)] + \wat n(D1_L) - \wat n(D1_R) \,,
\label{linkingNS5}
\ee
with similar definitions for the $\wat n$ numbers. These definitions are such that $\sum_b h_b = \sum_a \ell_a$.

We denote the two partitions of linking numbers by $\rho = p[(h_b)]$ and $\sigma =p[(\ell_a)]$, with $p$ representing the operation of ordering in non-increasing fashion. In principle, $h_b$ and $\ell_a$ can be half-integers, but we will only consider situations where they are integers. 
In this section we will only consider situations where the $h_b$ are already ordered, so that $\rho=(h_b)$. On the other hand, the $\ell_a$ can be unordered and we define $v=(\ell_a)$. 

In a given brane configuration, the D3 branes are taken separated, indicating that we consider a Coulomb branch configuration where the gauge group is broken to a maximal torus. With this point of view, we associate a given brane setup to the realisation of an {\bf abelian monopole operator}. 
In the above notation, the brane configuration realises the insertion of an abelian monopole of charge $v = (\ell_a)$.

\medskip 

In the following we will call NS5$_+$, or NS5$_{-}$, an NS5 brane placed to the right, or to the left, of all the D3 branes and D5 branes in the $x^7$ direction. An ``NS5 pair" will refer to a pair (NS5$_+$, NS5$_-$) of NS5 branes. 
We also label the D3 branes as D3$_{a}$, $a=1, \dots, N$. 

\medskip

To realise a configuration with partitions $(\rho, \sigma)$, we add NS5 pairs to the brane setup and we stretch D1 strings between the D3s and the NS5$_\pm$. For $\ell_a > 0$, we let $\ell_a$ D1s end on the right of the $a$-th D3, for $\ell_a < 0$, we let $-\ell_a$ D1s end on the left of the $a$-th D3. The other ends of those D1 strings terminate on the NS5$_\pm$ branes in agreement with their linking numbers $h_b$. We add as many NS5 pairs as needed for the construction and we order the NS5$_\pm$ such that linking numbers $h_b$ decrease towards $x^7 \to +\infty$. 

\bigskip

\noindent{\bf Brane configuration - abelian VEV dictionary}

\medskip

A given brane configuration is described by two vectors $(\rho, \sigma)$ -- up to ordering of the $\ell_a$ -- collecting the linking numbers of the NS5 branes and the D3 branes.%
\footnote{Strictly speaking the positions of the D5 branes relative to the NS5 branes is also important, because they form a Hanany-Witten triple (NS5, D5, D3') with the D3' brane. Here we always take the D5s to lie between the NS5$_-$ and NS5$_+$ branes.} 
In the following we will use a more convenient characterization of a brane configuration in terms of two pairs of partitions $(\rho^+, \sigma^+)$ and $(\rho^-,\sigma^-)$.

The partitions $\rho^+ = (h^+_i)$ and $\rho^-=(h^-_i)$ collect {\it different} linking numbers of the NS5$_+$ branes and NS5$_-$ branes respectively, defined by $h^+ = \wat n (D3_R) + \wat n(D1_L) - \wat n(D1_R)$ for an NS5$_+$ and $h^- = \wat n (D3_L) + \wat n(D1_R) - \wat n(D1_L)$ for an NS5$_-$.%
\footnote{These linking numbers $h^\pm$ are related but not equal to the linking numbers $h$ defined in (\ref{linkingNS5}). The relation is $h_a=\pm(h^\pm_a-N/2)$ for NS5$_\pm$ branes, respectively.} NS5$_\pm$ branes with vanishing $h^\pm$ linking numbers are spectator branes (decoupled from the other branes) and we do not include them in $\rho^\pm$.
We will always consider partitions $\rho^\pm$ ordered in non-increasing fashion, namely $h^\pm_i \ge h^\pm_{i+1}$.

The partitions $\sigma^\pm$ are defined more simply by the schematic split $\sigma = (\sigma^+, \vec 0, -\sigma^-)$, where  $\sigma^+ = p(\ell^+_a)$ collects the positive D3 linking numbers $\ell_a^+ = \ell_a > 0$, as defined above, and $\sigma^-= p(\ell^-_a)$ collects the negative D3 linking numbers  $\ell_a^- = - \ell_a > 0$, ordered non-increasingly. 

These definitions are such that the four partitions $\rho^\pm, \sigma^\pm$ contain only strictly positive integers, ordered non-increasingly. 
The partitions are simply read from the pattern of D1 strings stretched between the D3 and NS5$_+$ branes for $(\rho^+, \sigma^+)$, and of D1 strings stretched between the D3 and NS5$_-$ branes for $(\rho^-, \sigma^-)$.

\medskip

We then relate a given brane configuration to a specific abelian monopole VEV. Let's assume for simplicity that $\sigma = v$, namely the linking numbers $\ell_a$ are ordered. Then we have the relation
\be
\text{Brane setup } (\rho,\sigma)  \quad \longleftrightarrow \quad u_{\sigma} Z_{\rho,\sigma}(\varphi)  \,,
\label{BraneVEV_dict}
\ee
where  $u_{\sigma}$ is the abelian monopole VEV of charge $v=\sigma=(\sigma^+, \vec 0, -\sigma^-)$ and  $Z_{\rho,\sigma}$ is the matrix model that describes the low-energy theory living on the D1 strings.
When $\sigma \neq v$, the right hand side becomes $u_v Z_{\rho,\sigma}(p(\varphi))$, where $p$ is the ordering operation $\sigma = p(v)$. 

To read off the SMM living on the D1s, one has to move the D3s along $x^7$, taking into account Hanany-Witten string creation/annihilation effects, until no D1 ends on any D3s anymore. This leads to configurations with D1 strings stretched between NS5 branes, supporting unitary gauge nodes of an $\cN=(0,4)$ SMM (or rather $\cN=2^\star$ when $\epsilon \neq 0$). D3 branes are the source of hypermultiplets in the SMM, while D5 branes are the source of Fermi mulitplets. If the configuration reached has no D1 strings left, we simply have $Z_{\rho,\sigma}=1$.

\medskip

\begin{figure}[t]
	\centering
	\vspace{-.5cm}
	\includegraphics[scale=0.75]{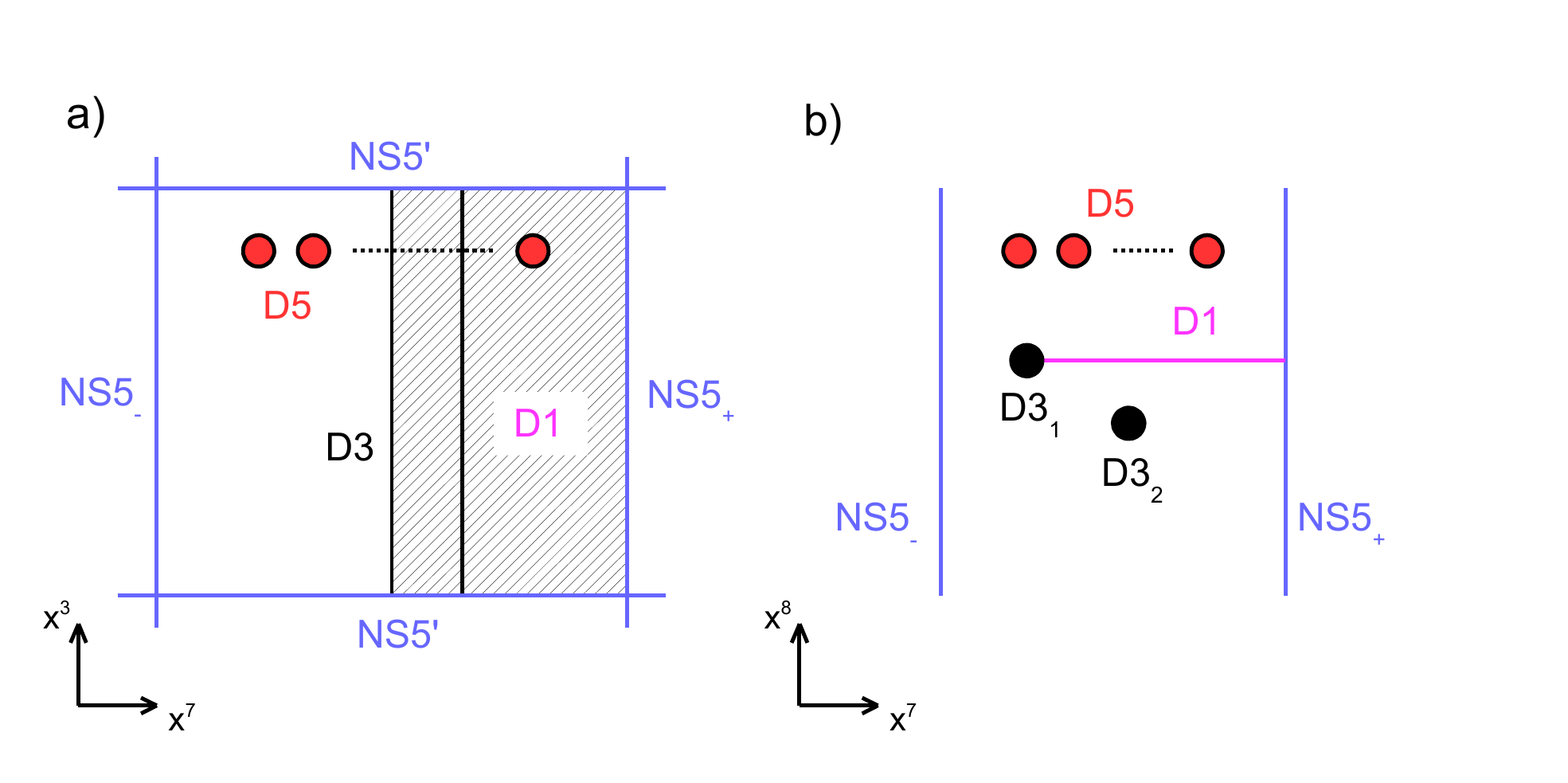}
	\vspace{-.5cm}
	\caption{\footnotesize{Brane setup realising the insertion of the abelian monopole $u_{e_1}$. We depicted the setup in the $x^{37}$ plane and in the $x^{78}$ plane (in this case the NS5' branes span the whole plane).}}
	\label{Branes1}
\end{figure}

Let us give a simple example.
In Figure \ref{Branes1}-a we illustrate the brane setup for the $U(2)$ SQCD theory with $N_f$ flavours, realised with D3, NS5' and D5 branes. In addition, there is an NS5 pair and a stretched D1 string realising an abelian monopole operator. Since the string ends on the first D3, i. e. D3$_1$, the abelian magnetic charge is $v=(1,0)$ and the setup realises the abelian monopole $u_{e_1}$. The partitions for this setup are $\rho = \sigma = (1,0)$, or $\rho_+= \sigma_+ = (1)$ and $\rho_- = \sigma_- = ()$ (empty vector). Figure \ref{Branes1}-b represents the same setup but in the $x^{78}$ plane, which is more convenient (we will always draw configurations in this plane henceforth). There, the NS5' branes are not visible as they fill the whole plane.

There is no dressing term $Z_{\rho,\sigma}$ here since, after moving the D3 brane to the right (in Figure \ref{Branes1}-b) the D1 string is annihilated, so $Z_{\rho,\sigma}=1$.

\bigskip

\noindent{\bf Non-abelian VEV}

The evaluation of the VEV of a monopole operator in the non-abelian theory is then obtained by summing over the contributions of abelian monopole sectors, as in the localization formula \eqref{locresult2}. 

\medskip

To obtain the VEV of a non-abelian monopole $V_B$, one has to sum over all the brane setups with fixed $\rho^\pm = \rho^\pm_B$. This means that we sum over patterns of the D1 strings attached to the NS5$_\pm$ in the specific arrangement given by the partition $\rho_B$, and ending on the D3 branes in any possible way, compatible with the $s$-rule. In the above notation, we sum over all possible $v$.
The partitions $\rho^\pm_B$ are defined as $\rho^\pm_B = \wat B^\pm$, where $\wat B^\pm$ are the transpose of $B^\pm$ seen as Young tableaux, and $B^\pm$ are the partitions obtained by the split $B = (B^+, \vec 0, -B^-)$, which is shorthand for  $B_a=B^+_a - B^-_{N+1-a}$.

This definition of $\rho_B$ is such that the abelian configurations contributing with trivial SMM ($Z_{\rho_B,\sigma}=1$), i.e. the non-bubbling contributions, are those with $v=B$ and permutations, as expected. We will observe it in examples.

\medskip

The (non-abelian) monopoles of smallest magnetic charge in the $U(2)$ SQCD theory are $V_{(1,0)}$ and $V_{(0,-1)}$. 

\begin{figure}[t]
	\centering
	%\vspace{-.5cm}
	\includegraphics[scale=0.75]{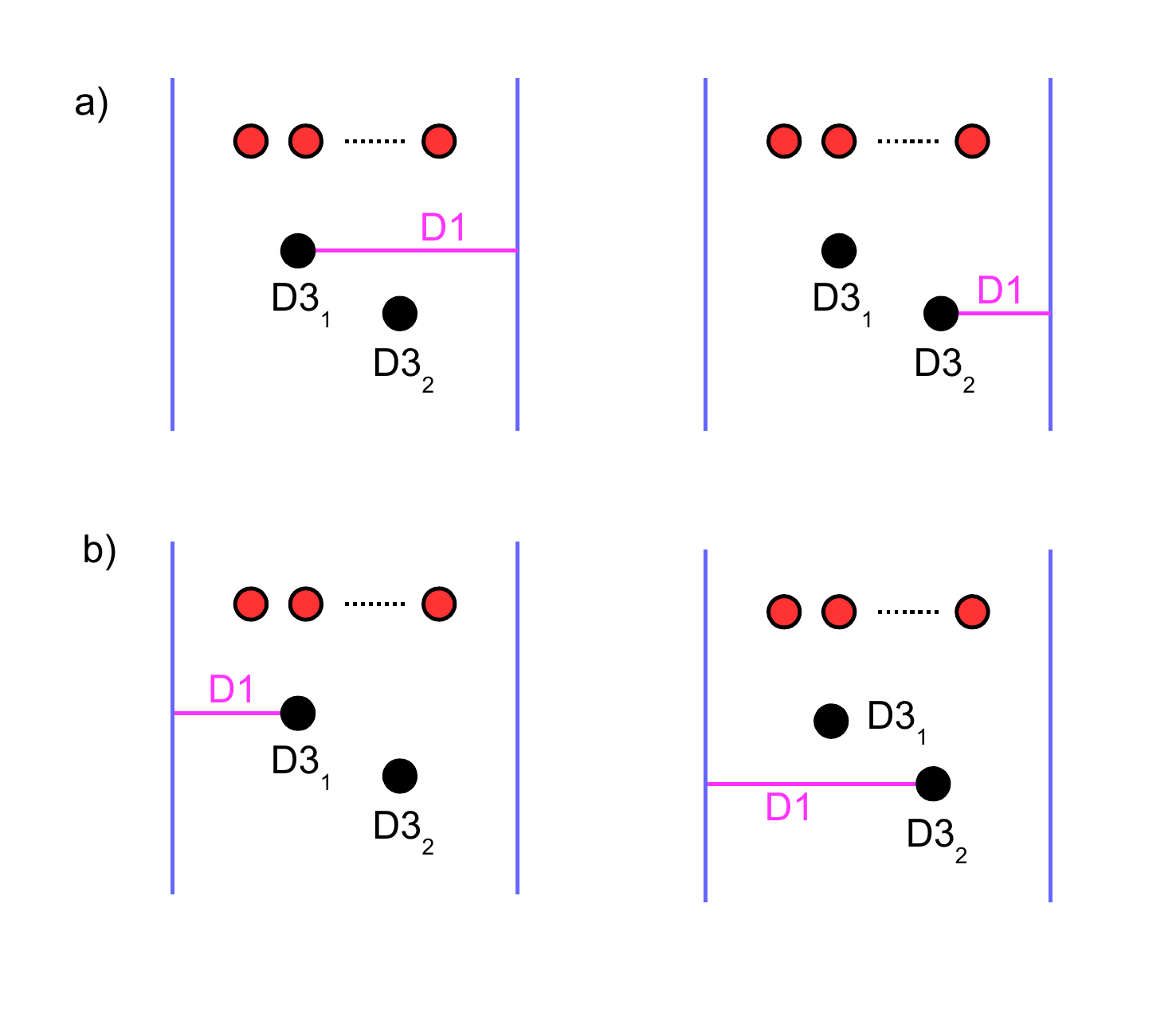}
	\vspace{-1cm}
	\caption{\footnotesize{Brane setups for the two abelian contributions to a) $V_{(1,0)} = u_{e_1} + u_{e_2}$ and b) $V_{(0,-1)} = u_{-e_1} + u_{-e_2}$, depicted in the $x^{78}$ plane.}}
	\label{Branes1_0}
\end{figure}

For $V_{(1,0)}$ we have $\rho^+ = (1)$ and $\rho^- =()$ (empty). Thus, there is one NS5 pair with a D1 string emanating from the NS5$_+$. This D1 string can end on either of the two D3s, leading to two possible abelian configurations, with $v = (1,0) = e_1$ or $v = (0,1) = e_2$. Both configurations correspond to $\sigma^+ = (1)$ and $\sigma^- =()$ (because the two $v$'s are identical after reordering). These two configurations are depicted in Figure \ref{Branes1_0}-a. According to the previous discussion they are associated to the abelian VEVs $u_{e_1}$ and $u_{e_2}$ with trivial dressing factor $Z=1$. Consequently, $V_{(1,0)}$ is then given by the sum
\be
V_{(1,0)} = u_{e_1} + u_{e_2} \,,
\ee
reproducing the known formula.

Similarly, for $V_{(0,-1)}$ we have $\rho^- =(1)$ and $\rho^+=()$. There is a single NS5$_-$ brane with one D1 attached. This D1 can end on either of the D3s (see Figure \ref{Branes1_0}-b), leading to two abelian configurations, for $u_{-e_1}$ and $u_{-e_2}$. We obtain
\be
V_{(0,-1)} = u_{-e_1} + u_{-e_2} \,.
\ee

\begin{figure}[t]
	\centering
	\includegraphics[scale=0.7]{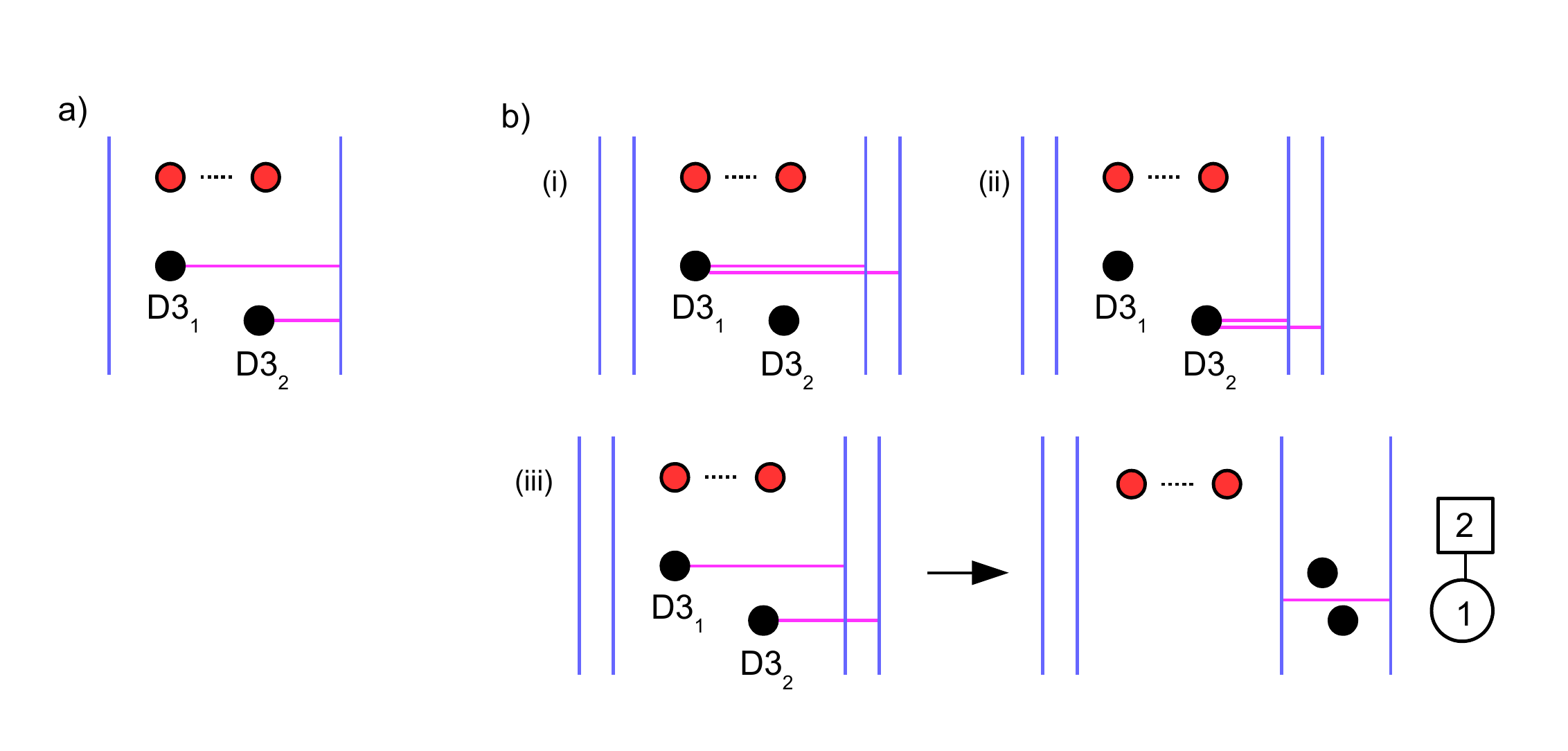}
	\vspace{-1cm}
	\caption{\footnotesize{Setup realising abelian contributions:  a) A single contribution for $V_{(1,1)} = u_{e_1+e_2}$; b) Three contributions for $V_{(20)} = u_{2e_1} + u_{2e_2} + u_{e_1+e_2} Z_1$. The bubbling factor $Z_1$ is computed as the SMM described by the quiver in Figure (iii).}}
	\label{Branes1_1}
\end{figure}

Let us now consider the monopole $V_{(1,1)}$. We have $\rho^+ = (2)$, $\rho^-=()$, with two D1 strings ending on the NS5$_+$ of an NS5 pair. Because of the $s$-rule the two D1s must end on different D3s, leading to a single configuration with abelian charge $v=(1,1) = e_1 + e_2$. This is shown in Figure \ref{Branes1_1}-a. If we move the D3s to the right, both D1s disappear, leaving a configuration without D1s, so that the dressing is again trivial $Z =1$. We obtain
\be
V_{(1,1)} = u_{e_1 + e_2} \,.
\ee

\medskip

A more elaborate example is the computation of $V_{(2,0)}$. We have $\rho^+ = (1,1)$, $\rho^-=()$. There are two NS5$_+$, each with a D1 ending on it. The two D1s can end on the two D3s in three ways, as shown in Figure \ref{Branes1_1}-b. 
Both D1s can end on D3$_1$, or on D3$_2$, leading to the undressed abelian VEVs $u_{2e_1}$ and $u_{2e_2}$, respectively.
The third possibility is to have one D1 ending on each D3, realising $u_{e_1+e_2}$. In this third case we observe (see Figure \ref{Branes1_1}-b) that after moving the D3s to the right, there is a D1 string remaining, stretched between the two NS5$_+$. This configuration is associated to the $\cN=(0,4)$ SMM with a $U(1)$ gauge node and two hypermulitplets (from the D3-D1 modes) with masses $\varphi_1$ and $\varphi_2$ (the distances between D1 and D3s along $x^{8+i9}$). Therefore, there is a non-trivial dressing by a factor $Z_{\rm SMM} := Z_1$ equal to the matrix model of this SMM. We obtain
\be
V_{(2,0)} = u_{2e_1} + u_{2e_2} + u_{e_1+e_2} Z_1 \,.
\label{V20}
\ee

\medskip

\begin{figure}[t]
	\centering
	\includegraphics[scale=0.8]{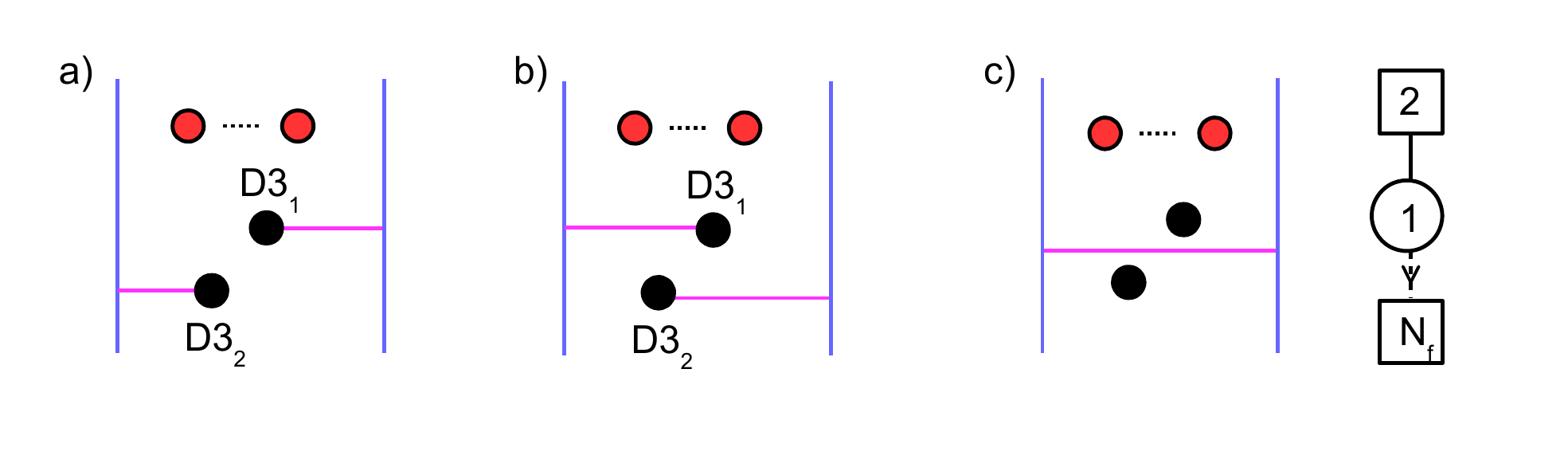}
	\vspace{-1cm}
	\caption{\footnotesize{The three configurations contributing to $V_{(1,-1)}$. a) $u_{e_1-e_2}$, b) $u_{-e_1+e_2}$, c) $Z_2$.}}
	\label{Branes1_2}
\end{figure}

Finally, let us look at $V_{(1,-1)}$, which has $\rho^+ = \rho^-=(1)$. It is realised with a single NS5 pair, with one D1 string attached to the NS5$_+$ and another one attached to the NS5$_-$. The D1s can either end on different D3s, leading to abelian VEVs $u_{e_1-e_2}$ or $u_{-e_1+e_2}$, or alternatively they can reconnect, leaving the D3s unconnected. In that case we have a single configuration with a D1 string stretched all the way from the NS5$_-$ to the NS5$_+$. There is no monopole charge, but only the dressing factor $Z_{\rm SMM} := Z_2$ given by the $\cN=(0,4)$ SMM living on the D1s. This is a $U(1)$ theory with two hypermultiplets, with mass $\varphi_{a=1,2}$ from the D1-D3 modes and $N_f$ Fermi multiplets with mass $m_{k=1,\cdots, N_f}$ from the D1-D5 modes. This is illustrated in Figure \ref{Branes1_2}.
We obtain
\be
V_{(1,-1)} = u_{e_1-e_2} + u_{-e_1+e_2} + Z_2 \,.
\label{V1-1}
\ee
The hypermultiplet masses $\varphi_a$ are identified with the Coulomb parameters of the 3d theory and the Fermi masses $m_k$ are identified with the 3d hypermultiplet masses. This is read from the brane picture. It indicates that there are couplings between 0d and 3d fields at the location of the defect, which identify the 0d flavour symmetries with 3d gauge or flavour symmetries.
These results reproduce the structure of the localization formula \eqref{locresult2}, supporting our proposal for the brane realisation of monopoles. In principle, we have all the ingredients to write down the exact VEVs of monopole operators in the $U(2)$ SQCD theory. 

\medskip

To complete the computation one would like to evaluate the matrix models $Z_{\rm SMM}$ that appear in the bubbling terms. Unfortunately, a complication arises. As we will explain shortly, these SMM can be readily evaluated from the Jeffrey-Kirwan recipe {\it when the FI parameters are non-vanishing}. The FI parameters of the SMM are read from the brane configurations as the separations between the NS5 branes in the $x^0$ direction. In the realisation of a given monopole $V_B$, the NS5 branes sit at the same position in $x^0$ and the FI parameters of the SMM for the bubbling terms are all vanishing. This is the precise situation when we do not know how to evaluate the matrix models from standard recipes. We therefore cannot at the moment provide more explicit formulae, for the VEVs $V_B$, than those in terms of the SMM factor $Z_{\rm SMM}$. 

On the other hand, we know how to evaluate the SMM at non-zero FI, which means when NS5 branes are separated along $x^0$. These situations correspond to the insertion of multiple ``minimal" monopoles, here $V_{(\pm 1,0)}$ and $V_{\pm(1,1)}$, at different positions on $x^0$. Each NS5 pair is responsible for the insertion of one monopole. This means that we can evaluate explicitly correlators of these monopoles. We will focus on these computations in the rest of the paper.

\medskip

We will now pause to explain how to compute SMM at non-zero FI and then we will explicitly compute some monopole correlators in the $U(2)$ SQCD theory.

\subsection{Bubbling SMM}

The Super-Matrix-Model (SMM) multiplets and Lagrangian in the $\epsilon\to 0$ limit can be viewed as the dimensional reduction of 2d $\cN=(0,4)$ multiplets and Lagrangian (see \cite{Tong:2014yna, Putrov:2015jpa} for reviews) down to zero dimensions. The Omega deformation further breaks (the dimensional reduction of) $\cN=(0,4)$ supersymmetry down to  (the dimensional reduction of) $\cN=(0,2)$ supersymmetry as will be discussed below.%
\footnote{The dimensional reduction of 2d $\cN=(0,2)$ supersymmetry to 0d $\cN=2$ was described in \cite{Franco:2016tcm}. See also \cite{Closset:2017xsc} for $\cN=1$ SMM details. }

We will follow the conventions of \cite{Tong:2014yna} for 2d $\cN=(0,2)$ and $\cN=(0,4)$ Minkowski supersymmetry. The R-symmetry automorphism of the 2d $\cN=(0,4)$ SUSY algebra is $SU(2)_1 \times SU(2)_2$, under which the right-moving supercharges $Q_{\alpha\dot\beta,+}$ transform as $(2,2)$. Here $\alpha=\pm$ and $\dot\beta=\dot\pm$ are doublet indices for $SU(2)_1$ and $SU(2)_2$ respectively, whereas the Lorentz spin $+$ indicates that the supercharges are right-moving. We then select an $\cN=(0,2)$ subalgebra of the $\cN=(0,4)$ superalgebra, with $U(1)_R$ R-symmetry generated by $R=J_{3}^{(1)}+J_{3}^{(2)}$, the diagonal combination of the Cartan generators of $SU(2)_1$ and $SU(2)_2$. The $\cN=(0,2)$ supercharges are $Q_+ \equiv Q_{- \dot -,+}$ and $\bar Q_+ \equiv Q_{+ \dot +,+}$, which carry charges $-1$ and $+1$ under the $U(1)_R$ R-symmetry and are invariant under a flavour symmetry generated by $F=2(J_{3}^{(1)}-J_{3}^{(2)})$. The Omega deformation with parameter $\eps$ breaks SUSY to $\cN=(0,2)$ by turning on a background for the $U(1)_F$ flavour symmetry: $\cN = (0,2)$ multiplets with flavour charge $F$ acquire a mass $F\epsilon$. 

We then Wick rotate to Euclidean signature, after which we use a tilde to denote what was Hermitian conjugation in Minkowski space, and finally we dimensionally reduce to zero dimensions. The resulting 0d theories gain an extra $U(1)_r$ R-symmetry arising from the 2d Lorentz group, which we normalise to have integer charges. Thus, the R-symmetry of a 0d $\cN=(0,4)$ theory is $SU(2)_1 \times SU(2)_2 \times U(1)_r$, under which the supercharges $Q_{\alpha\dot\beta}$ transform as $(2,2)_1$.%
\footnote{We denote by $\cN=(0,4)$ the supersymmetry of the 0d theory, even though there is no notion of chirality in 0d. This refers to the R-symmetry and multiplet content of the theory obtained by dimensional reduction from 2d $\cN=(0,4)$, which differs from, for instance, the reduction from 2d $\cN=(2,2)$.} 
 In the brane construction we identify $SU(2)_1 \sim SO(3)_{\rm 012}$, $SU(2)_2 \sim SO(3)_{\rm 456}$ and $U(1)_r \sim SO(2)_{89}$. The Omega background breaks 0d $\cN=(0,4)$ supersymmetry to an $\cN=2$ subalgebra, which is the dimensional reduction of 2d $\cN=(0,2)$ supersymmetry, with a $U(1)_R\times U(1)_r$ R-symmetry under which the supercharges $Q=Q_{-\dot -}$, $\ti Q=Q_{+\dot +}$ have charges $(R,r)=(-1,1)$ and $(R,r)=(1,1)$. In addition, there is a $U(1)_F$ flavour symmetry under which the supercharges are neutral. Following common conventions, we denote  the supersymmetry of the deformed model in zero dimensions by $\cN=2^\star$.
 
Next, we describe the content of the relevant $\cN=2$ and $\cN=(0,4)$ supermultiplets in zero dimensions. 
The $\cN = 2$ multiplets are: 
\begin{itemize}
	\item {\it Gauge multiplet} (in Wess-Zumino (WZ) gauge): a complex scalar $\tilde\sigma$, fermions $\zeta$ and  $\ti\zeta$ and an auxiliary field $D$ in $\rm{Adj}(G)$ representation. A second (SUSY singlet) complex scalar $\sigma$ is needed to write an action.%
\footnote{$\zeta$ and $\tilde\zeta$ are \emph{not} Hermitian conjugates. The $u_0-u_1$ component of the 2d Minkowski gauge field $u_\mu$ gives rise to $-\tilde\sigma$ in 0d. The $u_0+u_1$ component is a SUSY singlet in WZ gauge and becomes $\sigma$ in 0d.}
	\item {\it Chiral multiplet}: a complex scalar $\phi$ and a complex fermion $\psi$. 
	\item {\it Fermi multiplet}: a complex fermion $\lambda$ and a complex auxiliary field $G$.
\end{itemize}
The $\cN=(0,4)$ multiplets that arise from string excitations in our brane setups are:
\begin{itemize}
	\item {\it Gauge multiplet}: an $\cN=2$ gauge superfield $U$ with charges $(R,F)=(0,0)$ and an $\cN=2$ Fermi superfield $\Theta$ with charges $(R,F)=(0,-2)$ in $Adj(G)$ representation. It arises from D1-D1 string modes, with the D1 stretched between NS5 branes in both the $x^3$ and $x^7$ directions.
	\item {\it Hypermultiplet}: two $\cN=2$ chiral superfield $\Phi$, $\ti\Phi$ with charges $(R,F)=(\frac{1}{2},1)$ in conjugate representations of $G$.  A fundamental hypermultiplet arises from D1-D3 string modes; a bifundamental hypermultiplet arises from D1-D1' string modes across an NS5 brane. 
	\item {\it Fermi multiplet}: a \emph{single} $\cN=2$ Fermi superfield $\Lambda$ with charges $(R,F)=(0,0)$ \cite{Tong:2014yna}. This is sometimes called an $\cN=(0,4)$ half-Fermi multiplet \cite{Lee:2019skh}.  It arises from D1-D5 string modes, or from strings stretched between two D1s across an NS5 brane, or from D3-D3' intersections.
\end{itemize}

The components of the $\cN=(0,4)$ multiplets transform as follows under the $SU(2)_1\times SU(2)_2\times U(1)_r$ R-symmetry:%
\footnote{See \cite{Witten:1993yc} for details on the 2d parent theories and \cite{Witten:1994tz} for a thorough discussion of Fermi multiplets.}

\begin{center}
	\begin{tabular}{|c|c|c|}
		\hline
		$\cN=(0,4)$ Multiplet & Components & $SU(2)_1 \times SU(2)_2 \times U(1)_r$ \cr
		\hline
		Gauge & $(\tilde \sigma,\lambda_{\alpha\dot\beta},  D_{(\alpha\alpha')};\sigma)$ & $(1,1)_{-2}$, $(2,2)_{-1}$,  $(3,1)_0$; $(1,1)_2$ \cr
		\hline
		Hyper & $(q_\alpha,\psi_{\dot\beta})$ & $(2,1)_0$, $(1,2)_{1}$ \cr
		\hline
		Fermi  & $(\lambda, G)$ & $(1,1)_{-1}$, $(1,1)_0$ \\
		\hline
	\end{tabular}
\end{center}

In terms of the $\cN=2$ subalgebra, its $U(1)_R\times U(1)_r$ R-symmetry and the $U(1)_F$ flavour symmetry, the charges are:
\begin{center}
	\begin{tabular}{|c|c|c|}
		\hline
		$\cN=2$ Multiplet & Components & $U(1)_R\times U(1)_r\times U(1)_F$ \cr
		\hline
		Gauge $U$ & $(\tilde \sigma,\zeta, \tilde\zeta, D;\sigma)$ & $(0,-2,0)$, $(1,-1,0)$,  $(-1,-1,0)$, $(0,0,0)$; $(0,2,0)$ \cr
		Fermi $\Theta$ & $(\vartheta,g)$ & $(0,-1,-2)$, $(-1,0,-2)$ \cr
		\hline
		Chiral $\Phi$ & $(\phi,\psi)$ & $(\frac{1}{2},0,1)$, $(-\frac{1}{2},1,1)$ \cr
		Chiral $\ti\Phi$ & $(\tilde\phi,\tilde\psi)$ & $(\frac{1}{2},0,1)$, $(-\frac{1}{2},1,1)$ \cr
		\hline
		Fermi $\Lambda$  & $(\lambda, G)$ & $(0,-1,0)$, $(-1,0,0)$ \\
		\hline
	\end{tabular}
\end{center}

\medskip

The action for the $\cN=(0,4)$ SMM is easily obtained from \cite{Tong:2014yna} upon Wick rotation and dimensional reduction, so we omit the details. An important deformation term in our analysis is the FI term
\be\label{FI}
S_{\rm FI} = \xi \tr(D) \,,
\ee
with real parameter $\xi$ and $D$ is the auxiliary field in the $\cN=2$ gauge multiplet. This FI term breaks the $SU(2)_1$ R-symmetry to $U(1)\sim SO(2)_{12}$.

\bigskip

\noindent{\bf Contributions to the matrix model}

The resulting supersymmetric matrix model can be localized using the $\cN=2$ actions that descend from 2d kinetic terms. This type of computation is by now standard and even easier in our zero dimensional setup, so we will be schematic.%
\footnote{See \cite{Benini2014,Benini:2013xpa,Hori:2014tda,Closset2016a} for related localization computations in higher dimensions.} Localizing the gauge multiplet first leads to a BPS locus parameterised by commuting $\sigma$ and $\tilde\sigma$ (all other fields vanish). They can therefore be diagonalized simultaneously by a complexified gauge transformation, reducing to an integral over a Cartan subalgebra of the gauge group, modulo the action of the Weyl group. Supersymmetry ensures that the dependence on $\tilde\sigma$ drops out, so the partition function is computed by a holomorphic contour integral in  $z_a$, the eigenvalues of $\sigma$.

The determination of the contour of integration for the resulting integral is subtle and can be worked out by carefully analysing the matrix model when integrating out the auxiliary field $D$ (see for instance \cite{Hori:2014tda} in the context of SQM). 
The outcome of such a careful analysis is the Jeffrey-Kirwan (JK) prescription \cite{Jeffrey1995}. We refer to \cite{Benini:2013xpa} for a review of the JK prescription. In the following we identify the  unphysical JK parameter with the physical FI parameter $\xi$ in (\ref{FI}), which ensures that the integration cycle has no contributions at infinity. The multi-dimensional poles that are encircled by the JK integration cycle are then in one-to-one correspondence with the Higgs vacua of the theory.
When the vector of FI parameters $\xi$ is generic (that is, it lies is in the interior of a chamber in FI space, where the Higgs vacua are isolated), the prescription is simple: the choice of contour only depends on the chamber that $\xi$ belongs to. When $\xi$ lies on a wall separating two different chambers, the contour integral is more subtle to determine and we will not attempt such computations in this paper.

At generic non-zero FI parameters, the matrix model takes the form 
\bea
Z_{\rm SMM} (\xi)= \frac{1}{|\cW|}\oint_{\mathrm{JK}(\xi)} \prod_a \frac{dz_a}{2\pi i} \, Z_{\rm gauge} Z_{\rm hyper} Z_{\rm F} \,.
\eea
The integral is over $z_a$ taking values in a complexified Cartan subalgebra of the gauge group $G$ and $|\cW|$ denotes the order of the Weyl group of $G$. The precise contour $\mathrm{JK}(\xi)$ is given by the JK prescription with JK parameter identified with the FI parameter $\xi$.

The integrand factors come from integrating out the fields of the corresponding $\cN=(0,4)$ multiplets near the BPS locus. Alternatively, one may simply borrow the results of SQM matrix models and take the 1d $\to$ 0d limit, replacing trigonometric functions by rational functions (of the complex masses).
For $G =U(k)$, we find
\bea\label{Z_1-loop}
&\text{Gauge}:   & Z_{\rm gauge} &= (2\epsilon)^k \prod_{a \neq b} z_{ab} (z_{ab}+2\epsilon) \,, \cr
&\text{Fundamental hyper}:    & Z_{\rm hyper} &= \prod_{a=1}^k \frac{1}{\pm (z_a - m) + \epsilon} \,, \cr
&\text{Bifundamental hyper}:   &  Z_{\rm b-hyper} &= \prod_{a,b} \frac{1}{\pm (z_a - \hat z_b) + \epsilon} \,, \cr
&\text{Fundamental Fermi}:    & Z_{\rm F} &= \prod_{a=1}^k  (z_a - m) \,,
\eea
where $z_{ab} := z_a - z_b$, $(\pm x + y) := (x+y)(-x+y)$ and $m$ is the complex mass of the multiplet (we will not need masses for bifundamental hypermultiplets).%
\footnote{There are sign ambiguities in the one-loop determinants, which are counterparts of sign ambiguities in the definition of abelian monopoles. We picked a convention compatible with the star product structure. The signs in the one-loop determinants above follow from the action in \cite{Tong:2014yna} using the convention $\int d\tilde\eta d\eta \,\eta \tilde\eta=1$ for integrals of Grassann variables $\eta$, $\tilde\eta$ if the Fermi multiplets $\Lambda$ are actually \emph{anti}-fundamentals. We will nevertheless call them fundamentals in the following.}

\bigskip

\noindent{\bf Computing $Z_1$ and $Z_2$ at $\xi \neq 0$}

\medskip

As an example we compute the SMM $Z_1$ and $Z_2$ that appeared as bubbling factors in \eqref{V20} and \eqref{V1-1}, but at non-zero FI parameters. As explained, these are not the actual bubbling factors of $V_{(2,0)}$ and $V_{(1,-1)}$, which would be the SMM at zero FI parameter, instead they are factors in monopole correlators, as we explain in the next subsection.

The SMM for $Z_1$ is the $U(1)$ theory with two fundamental hypermultiplets of respective masses $\varphi_1$ and $\varphi_2$. The matrix model is thus
\be
Z_1(\xi) = \oint_{\mathrm{JK}(\xi)} \frac{dz}{2\pi i} \frac{2\epsilon}{\prod\limits_{a=1,2}[\pm (z - \varphi_a) + \epsilon]} \,.
\ee
The JK prescription for an FI parameter $\xi >0$ is to pick the residues at $z= \varphi_a - \epsilon$, for $a=1,2$, while at $\xi <0$ we pick the residues at $z = \varphi_a + \epsilon$. After simplification, this leads to the same result in both cases:
\be
Z_1(\xi) = \frac{2}{\pm \varphi_{12} + 2\epsilon} \,,   \quad \xi \neq 0  \,.
\label{Z1}
\ee
Although it may be tempting to conjecture that this is also the result at $\xi=0$, we will not do so since we have no evidence for it. In fact, we suspect that the result at $\xi=0$ is different.%
\footnote{We thank D. Dorigoni for instructive discussions on this point.}

For $Z_2$ the SMM has $N_f$ extra Fermi multiplets with masses $m_k$,
\be
Z_2(\xi) = \oint_{\mathrm{JK}(\xi)} \frac{dz}{2\pi i} \frac{2\epsilon \prod\limits_{k=1}^{N_f}(z-m_k)}{\prod\limits_{a=1,2}[\pm (z - \varphi_a) + \epsilon]} \,.
\ee
The JK prescription is the same as before. We now obtain two different results, depending on the sign of $\xi$:
\bea
Z_2(\xi) = \left\lbrace
\begin{array}{cc}
	-\frac{\prod\limits_{k=1}^{N_f}(\varphi_1 - m_k - \epsilon)}{\varphi_{12} (\varphi_{12} - 2\epsilon)}  + (\varphi_1 \leftrightarrow \varphi_2) \,, &  \quad \xi >0 \vspace{1mm} \cr
	-\frac{\prod\limits_{k=1}^{N_f}(\varphi_1 - m_k + \epsilon)}{\varphi_{12} (\varphi_{12} + 2\epsilon)}  + (\varphi_1 \leftrightarrow \varphi_2) \,,  & \quad  \xi <0  \ .
\end{array}
\right.
\label{Z2}
\eea
We will see examples of non-abelian SMMs in later sections.

\subsection{Monopole correlators and Wall-Crossing}

We will now focus on correlators of the ``minimal" monopoles $V_{(1,0)}, V_{(0,-1)}, V_{(1,1)},$ and $V_{(-1,-1)}$ in the $U(2)$ SQCD theory, which are the generators of the chambers in the magnetic weight lattice and are the non-bubbling monopoles in the theory. Their VEVs, expressed in terms of abelian monopoles, are
\bea
V_{(1,0)} &= u_{e_1} + u_{e_2} \,, \quad
V_{(0,-1)} = u_{-e_1} + u_{-e_2} \,, \cr
V_{(1,1)} & = u_{e_1+e_2} \,, \qquad  V_{(-1,-1)} = u_{-e_1-e_2} \,.
\label{BasicMonopVEVs}
\eea
We can evaluate any correlator of these monopoles inserted along the $x^0$ line, at the origin of the Omega background $\bR^2_{\epsilon}$, using the tools developed in the previous sections.

The method starts by considering the brane realisation of the monopole correlator. The correlator is then given by a sum of contributions associated to each allowed pattern of D1 strings in the brane configuration. To each pattern of D1s corresponds a given abelian monopole VEV $u_v$ and a dressing factor $Z_{\rm SMM}$ equal to the matrix model (or partition function) 
of the SMM theory living on the D1 strings. The resulting contribution to the correlator is $u_v Z_{\rm SMM}$. We have already showed that the simple expressions \eqref{BasicMonopVEVs}, which are one-point correlators, can be derived from this perspective.

This leads to the same structure for the VEVs of correlators as that of arbitrary monopole operators \eqref{locresult3} derived from localization, except that the dressing factors $Z_{\rm SMM}$ cannot easily be interpreted as ``bubbling" contributions.
Importantly, the matrix models $Z_{\rm SMM}$ that appear in correlators of minimal monopoles will have non-vanishing FI parameters, which will allow us to evaluate them.

\medskip

Let us start with two-point correlators. We will look at $\vev{V_{(1,0)}^2}$, $\vev{V_{(1,1)}^2}$, $\vev{V_{(1,0)}V_{(1,1)}}$, $\vev{V_{(1,0)}V_{(0,-1)}}$, $\vev{V_{(1,1)}V_{(0,-1)}}$, and $\vev{V_{(1,1)}V_{(-1,-1)}}$, with both orderings when relevant. We do not write the insertion points $x^0_i$ of the operators, assuming they are time-ordered.

\medskip

\begin{figure}[t]
	\centering
	\includegraphics[scale=0.65]{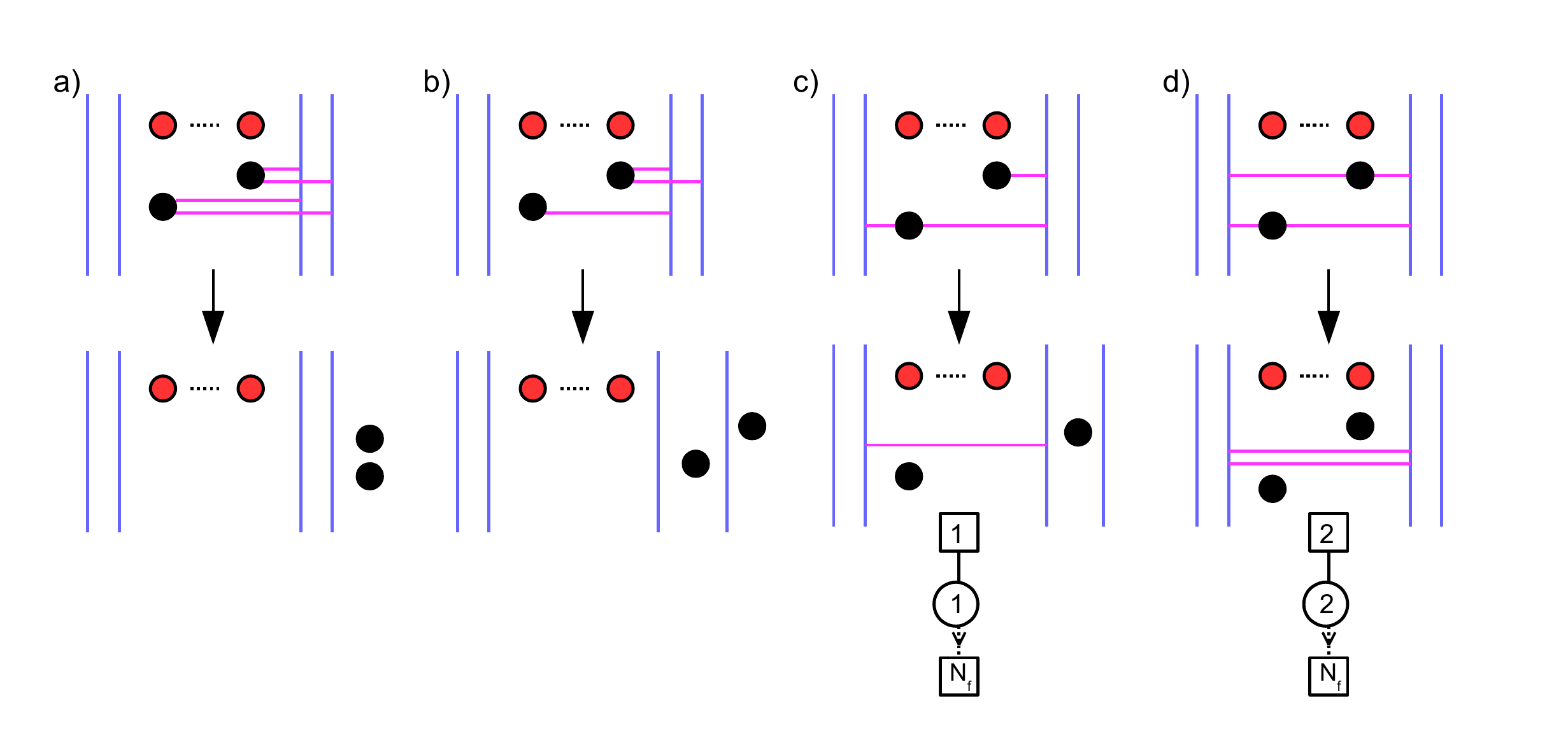}
	\vspace{-1cm}
	\caption{\footnotesize{Brane patterns for two-point correlators in the $U(2)$ SQCD theory and associated SMM. a) $\vev{V_{(1,1)}^2}$, b) $\vev{V_{(1,0)}V_{(1,1)}}$, c) $\vev{V_{(1,1)}V_{(0,-1)}}$, d) $\vev{V_{(1,1)}V_{(-1,-1)}}$.}}
	\label{Branes_U2_2p}
\end{figure}

\noindent\underline{$\vev{V_{(1,0)}^2}$}: 
\smallskip

The brane realisation is the same as for $V_{(2,0)}$. There are two NS5 pairs, with one D1 emanating from each NS5$_+$ (and the NS5$_-$s are spectators). The difference with the $V_{(2,0)}$ configuration is that the NS5$_+$ sit at different positions along $x^0$. Here the ordering along $x^0$ should not matter since we are inserting the same monopole $V_{(1,0)}$ twice. The two D1s can end on the D3s in three different patterns, leading to three contributions to $\vev{V_{(1,0)}^2}$. This is all the same as for $V_{(2,0)}$ (see Figure \ref{Branes1_1}-b and \eqref{V20}). The only difference is the matrix model factor $Z_1(\xi)$ should be evaluated at $\xi>0$ or $\xi <0$, as in \eqref{Z1} (instead of $\xi =0$). Consequently, we obtain 
\bea
\vev{V_{(1,0)}^2} &= u_{2e_1} + u_{2e_2} + u_{e_1+e_2} Z_1(\xi >0) \cr
&=u_{2e_1} + u_{2e_2} +\frac{2}{\pm \varphi_{12} + 2\epsilon} u_{e_1+e_2} \,.
\eea

\noindent\underline{$\vev{V_{(1,1)}^2}$}: 
\smallskip  

The brane configuration has two NS5 pairs (one for each $V_{(1,1)}$ insertion) with two D1s emanating from each NS5$_+$. Because of the s-rule, there is no choice in the pattern of D1s ending on D3s: each D3 has two D1s ending on it (see Figure \ref{Branes_U2_2p}-a). The abelian monopole is $u_{2e_1 + 2 e_2}$ and the dressing factor is trivial (since all the D1s disappear after moving the D3s to the right of the NS5$_+$s). Consequently, we obtain
\bea
\vev{V_{(1,1)}^2} &= u_{2e_1+2e_2} \,.
\eea

\noindent\underline{$\vev{V_{(1,0)}V_{(1,1)}}$}: 
\smallskip

The brane configuration has two NS5 pairs with two D1s attached to the left-most NS5$_+$ and one D1 attached to the right-most NS5$_+$ (the ordering of NS5$_+$ along $x^7$ is non-increasing in linking numbers, as discussed). There are two possible D1 patterns: two D1s end on D3$_1$ and one on D3$_2$, or vice versa (see Figure \ref{Branes_U2_2p}-b). They correspond to the abelian monopoles $u_{2e_1+e_2}$ and $u_{e_1+2e_2}$. After moving the D3 branes, we see that the dressing factors are trivial, since there are no D1s left. We thus obtain
\bea
\vev{V_{(1,0)}V_{(1,1)}} &= u_{2e_1+e_2} + u_{e_1+ 2e_2}  \,.
\eea
Reversing the order of the insertions does not change the argument, so we conclude that the two operators commute
\be
\vev{V_{(1,1)}V_{(1,0)}} = \vev{V_{(1,0)}V_{(1,1)}} \,.
\ee

\noindent\underline{$\vev{V_{(1,0)}V_{(0,-1)}}$}: 
\smallskip

The brane configuration necessitates only a single NS5 pair,%
\footnote{To be consistent we should always introduce one pair of NS5 brane per minimal monopole inserted, with the two NS5 of a pair sitting at the same $x^0$ position. Among each pair, one NS5 is a spectator. Here this results in having a spectator NS5$_+$ and a spectator NS5$_-$, which we simply suppress. But the remaining NS5$_+$ and NS5$_-$ sit at different positions in $x^0$.} with one D1 emanating from the NS5$_+$ and one emanating from the NS5$_-$. This is the same as for the monopole $V_{(1,-1)}$, except that the NS5$_+$ and NS5$_-$ sit at different positions along $x^0$ (each NS5 inserts one monopole operator). As in the case of $V_{(1,-1)}$ and \eqref{V1-1}, there are three patterns of D1s (see Figure \ref{Branes1_2}) and the correlator is given by the right hand side of \eqref{V1-1}. The difference is that the FI parameter $\xi$ is positive:
\bea
\vev{V_{(1,0)}V_{(0,-1)}} &= u_{e_1-e_2} + u_{-e_1+e_2} + Z_2(\xi >0) \cr
&= u_{e_1-e_2} + u_{-e_1+e_2} -\frac{\prod\limits_{k=1}^{N_f}(\varphi_1 - m_k - \epsilon)}{\varphi_{12} (\varphi_{12} - 2\epsilon)}  + (\varphi_1 \leftrightarrow \varphi_2) \,.
\label{V1V-1}
\eea
We used our evaluation of $Z_2(\xi>0)$ in \eqref{Z2}.
Exchanging the ordering of the insertions corresponds to having $\xi <0$ in the matrix model and we get instead
\bea
\vev{V_{(0,-1)}V_{(1,0)}} &= u_{e_1-e_2} + u_{-e_1+e_2} + Z_2(\xi <0) \cr
&= u_{e_1-e_2} + u_{-e_1+e_2} -\frac{\prod\limits_{k=1}^{N_f}(\varphi_1 - m_k + \epsilon)}{\varphi_{12} (\varphi_{12} + 2\epsilon)}  + (\varphi_1 \leftrightarrow \varphi_2) \,.
\label{V-1V1}
\eea
We observe that in this case the two operators do not manifestly commute. A closer inspection, with the help of Mathematica, reveals that the two expressions are actually equal for $N_f =0,1,2$, but start to differ for $N_f \ge 3$, with the difference being a polynomial in $\varphi_a$ and $\epsilon$, symmetric in $\varphi_a$. Thus, it is a combination of Casimir operator VEVs. This is a non-trivial result. For small $N_f$ we have 
\bea
\vev{[V_{(1,0)},V_{(0,-1)}]} &= 0 \,, \quad  N_f \in \{0,1,2\} \,, \cr
 &= 2\epsilon  \,, \quad  N_f =3 \,, \cr
  &= -2\epsilon (\sum_{k=1}^4 m_k - 2\varphi_1 - 2\varphi_2)   \,, \quad  N_f =4 \,, \cr
    &= \epsilon \Big[4\epsilon^2+\big(2\varphi_1^2+ 2\varphi_2^2-\sum_k m_k^2\big)+\big(2\varphi_1+2\varphi_2-\sum_k m_k\big)^2\Big]   \,, \quad  N_f =5 \,.
\eea
This is our first encounter with a wall-crossing phenomenon in SMM: as the FI parameter crosses the $\xi=0$ wall, the SMM changes with contributions coming from residues at different poles. This is the 0d analogue of having a change in the spectrum of BPS states in SQM. As a result, the two monopole operators do not commute.
\medskip

\noindent\underline{$\vev{V_{(1,1)}V_{(0,-1)}}$}: 
\smallskip

The brane configuration has two NS5 pairs with two D1s emanating from the left-most NS5$_+$ and one D1 emanating from the right-most NS5$_-$. There are two D1 patterns, each with two D1s reconnecting and one D1 ending on D3$_1$ or on D3$_2$ from the right (see Figure \ref{Branes_U2_2p}-c). We get
\be
\vev{V_{(1,1)}V_{(0,-1)}} = u_{e_1} Z_3(\varphi_2, \xi >0) + u_{e_2} Z_3(\varphi_1, \xi >0) \,,
\ee
where $Z_3$ is the SMM living on the D1 strings. It is a $U(1)$ theory with a hypermultiplet of mass $x=\varphi_2$ or $x=\varphi_1$, and $N_f$ fundamental Fermi multiplets of masses $m_k$. The matrix model is
\be
Z_3 (x,\xi) = \oint_{\mathrm{JK}(\xi)} \frac{dz}{2\pi i} \frac{2\epsilon \prod_{k=1}^{N_f} (z-m_k)}{\pm (z - x) + \epsilon} \,.
\ee
With $\xi > 0$, we pick the residue at $z = x - \epsilon$, leading to
\be
\label{Z3}
Z_3(x, \xi >0) = \prod\limits_{k=1}^{N_f} (x-m_k -\epsilon) \,,
\ee
and 
\be
\vev{V_{(1,1)}V_{(0,-1)}} = u_{e_1} \prod_{k=1}^{N_f} (\varphi_2-m_k -\epsilon) + u_{e_2} \prod_{k=1}^{N_f} (\varphi_1-m_k -\epsilon) \,.
\ee
Permuting the order of the operator insertions changes the sign of the FI parameter in the SMM ($\xi <0$) and we must pick the poles at $z = x  + \epsilon$ instead. This leads to the same result with $\epsilon$ reversed, as expected:
\be
\vev{V_{(0,-1)}V_{(1,1)}} = u_{e_1} \prod_{k=1}^{N_f} (\varphi_2-m_k +\epsilon) + u_{e_2} \prod_{k=1}^{N_f} (\varphi_1-m_k +\epsilon) \,.
\ee
As soon as  $N_f \geq 1$, we observe a wall-crossing phenomenon, meaning a non-trivial commutator $\vev{[V_{(1,1)},V_{(0,-1)}]} \neq 0$. This commutator is manifestly a polynomial in $\varphi_a,\epsilon$, symmetric in $\varphi_a$,  i.e. it is a combination of (VEVs of) Casimir operators.

\medskip

\noindent\underline{$\vev{V_{(1,1)}V_{(-1,-1)}}$}: 
\smallskip

The brane configuration has two NS5 pairs with two D1s emanating from the left-most NS5$_+$ and two D1s emanating from the right-most NS5$_-$. There is a single D1 pattern, with each D3 having a D1 ending on its left and another on its right (see Figure \ref{Branes_U2_2p}-d). This corresponds to the setup where the D1s fully reconnect and the abelian magnetic charge vanishes. We obtain
\be
\vev{V_{(1,1)}V_{(-1,-1)}} = Z_4(\xi >0)  \,,
\ee
where $Z_4$ is the SMM living on the D1 strings. It is a $U(2)$ theory with two hypermultiplets of masses $\varphi_1, \varphi_2$, and $N_f$ fundamental Fermi multiplets of masses $m_k$. The matrix model is
\be
Z_4 (x,\xi) = \oint_{\mathrm{JK}(\xi)} \frac{dz_1dz_2}{(2\pi i)^2} \frac 12 \frac{(2\epsilon)^2 (\pm z_{12})(\pm z_{12} + 2\epsilon) \prod\limits_{k=1}^{N_f}\prod\limits_{i} (z_i-m_k)}{\prod\limits_a\prod\limits_i (\pm (z_i - \varphi_a) + \epsilon)} \,.
\ee
In the evaluation of $Z_3$ at $\xi>0$, the only poles contributing are at $(z_1,z_2)=(\varphi_1-\epsilon,\varphi_2-\epsilon)$ and the permutation $z_1 \leftrightarrow z_2$. After simplification, we obtain
\be
Z_4 (x,\xi>0) = \prod\limits_{a=1,2}\prod\limits_{k=1}^{N_f} (\varphi_a-m_k -\epsilon) \,,
\ee
and  so
\be
\vev{V_{(1,1)}V_{(-1,-1)}} = \prod\limits_{a=1,2}\prod\limits_{k=1}^{N_f} (\varphi_a-m_k -\epsilon)  \,.
\ee
For the commuted correlator, we compute $Z_4(\xi<0)$ and find the same result with $\epsilon \to -\epsilon$ as expected,
\be
\vev{V_{(-1,-1)}V_{(1,1)}} = \prod\limits_{a=1,2}\prod\limits_{k=1}^{N_f} (\varphi_a-m_k +\epsilon)  \,.
\ee
Again, the two monopoles do not commute and we observe wall-crossing, as soon as $N_f\geq1$. The commutator is a well-defined Casimir operator.

\medskip

Finally, to extend a little the range of examples that are meant to illustrate the general procedure, we are going to compute two random instances of three-point correlators: $\vev{V_{(1,0)}^2V_{(1,1)}}$ and $\vev{V_{(1,0)}V_{(1,1)}V_{(0,-1)}}$. 

\medskip

\begin{figure}[t]
	%\centering
	\hspace*{-18pt}
	\includegraphics[scale=0.62]{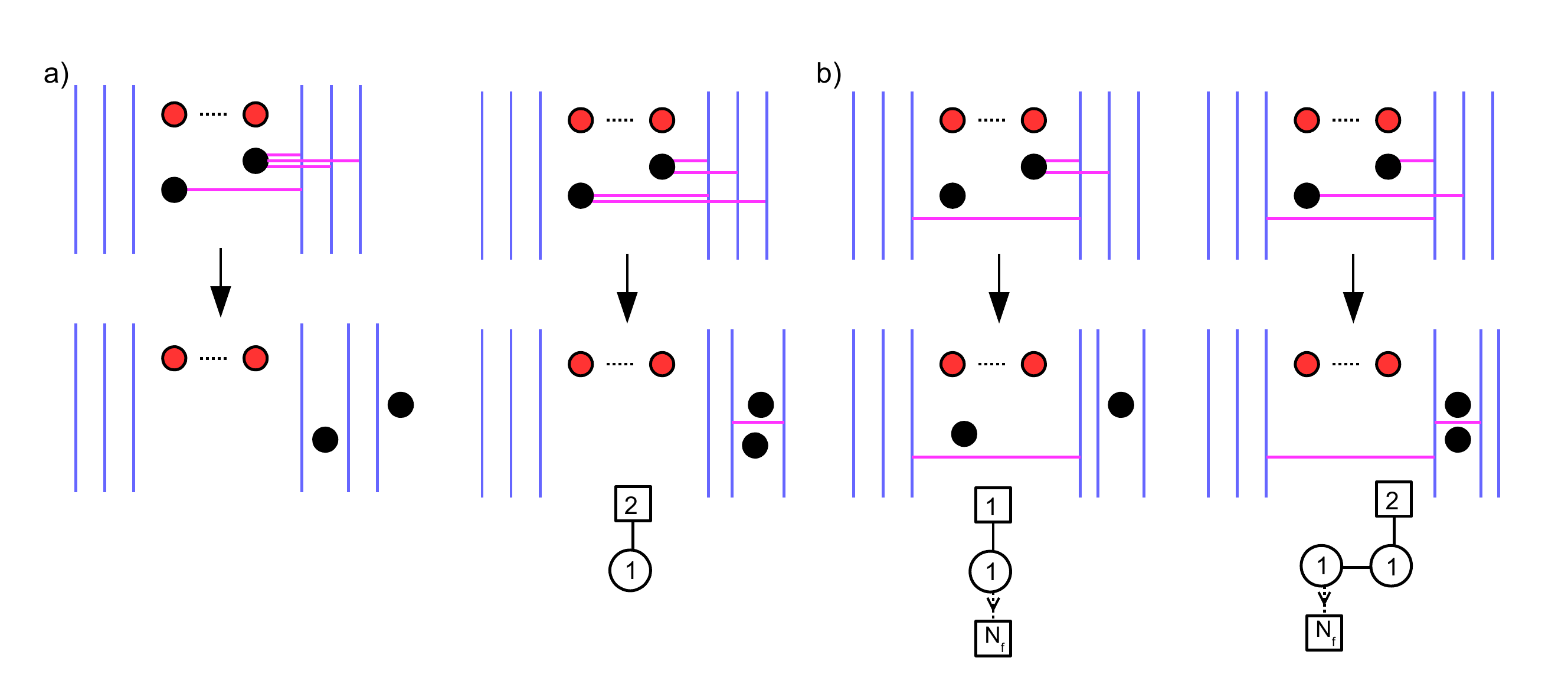}
	\vspace{-1cm}
	\caption{\footnotesize{Brane patterns for three-point correlators in the $U(2)$ SQCD theory and associated SMM. a) $\vev{V_{(1,0)}^2V_{(1,1)}}$, b) $\vev{V_{(1,0)}V_{(1,1)}V_{(0,-1)}}$.}}
	\label{Branes_U2_3p}
\end{figure}

\noindent\underline{$\vev{V_{(1,0)}^2V_{(1,1)}}$}: 
\smallskip

The brane configuration has three NS5 pairs, one for each minimal monopole inserted, with two D1s emanating from the left-most NS5$_+$ and one D1 emanating from the two other NS5$_+$. There are three D1 patterns, where the numbers of D1 strings ending on (D3$_1$,D3$_2$) are given by $(3,1), (2,2)$ and $(1,3)$ respectively (see Figure \ref{Branes_U2_3p}-a). We obtain
\bea
\vev{V_{(1,0)}^2V_{(1,1)}} &= u_{3e_1+e_2} + u_{e_1+3e_2} + u_{2e_1+2e_2} Z_1(\xi >0) \cr
&= u_{3e_1+e_2} + u_{e_1+3e_2} + u_{2e_1+2e_2}\frac{2}{\pm \varphi_{12} + 2\epsilon}  \,,
\eea
where $Z_1$ is the SMM with gauge group $U(1)$ theory and two hypermultiplets of mass $\varphi_1,\varphi_2$, that we encountered before.

Changing the order of the insertions of the operators does not affect the final result. This is because, as found earlier, $V_{(1,0)}$ and $V_{(1,1)}$ commute as operators.

\medskip

\noindent\underline{$\vev{V_{(1,0)}V_{(1,1)}V_{(0,-1)}}$}: 
\smallskip

The brane realisation has three NS5 pairs, with two D1s emanating from the left-most NS5$_+$ and one D1 emanating from the middle NS5$_+$ and right-most NS5$_-$. The remaining NS5 branes are spectators. In the language of partitions this is $\rho_+ = (2,1)$, $\rho_- = (1)$. There are three patterns of D1s, with a D1 reconnection always across the D3s and the remaining D1s ending on the D3s, with either two D1s ending on a single D3 ($\sigma^+=(2)$), or one D1 ending on each D3 ($\sigma^+=(1,1)$) -- see Figure \ref{Branes_U2_3p}-b. 
The correlator is thus
\bea
\vev{V_{(1,0)}V_{(1,1)}V_{(0,-1)}} &= u_{2e_1} Z_3(\varphi_2,\xi >0) + u_{2e_2} Z_3(\varphi_1,\xi >0) + u_{e_1+e_2} Z_5(\xi_{1,2} >0) \,,
\eea
where $Z_3$ is the SMM that already appeared in the two-point correlator $\vev{V_{(1,1)}V_{(0,-1)}}$ and $Z_5$ is the SMM described by the right-most quiver in Figure \ref{Branes_U2_3p}-b. The quiver SMM $Z_5$ is computed by
\be
Z_5(\xi) = \oint_{\mathrm{JK}(\xi)} \frac{dz}{2\pi i} \frac{d\hat z}{2\pi i} \frac{(2\epsilon)^2 \prod\limits_{k=1}^{N_f} (z - m_k)}{[\pm(z-\hat z) + \epsilon] \prod\limits_{a=1,2} [\pm  (\hat z-\varphi_a) + \epsilon]} \,.
\ee
There are two FI parameters  $\xi := (\xi_1,\xi_2)$, one for each gauge node. In the case when they are both positive, we pick residues at $z=\hat z - \epsilon$ and $\hat z = \varphi_a - \epsilon$, for $a=1,2$. This gives
\be
Z_5(\xi_{1,2} >0) = -\frac{\prod\limits_{k=1}^{N_f} (\varphi_1 - m_k - 2\epsilon)}{\varphi_{12}(\varphi_{12} - 2\epsilon)} + (\varphi_1 \leftrightarrow \varphi_2) \,.
\ee
{\it In fine}, we have
\bea
\vev{V_{(1,0)}V_{(1,1)}V_{(0,-1)}} &= u_{2e_1} P(\varphi_2 -\epsilon) + u_{2e_2} P(\varphi_1 -\epsilon) - u_{e_1+e_2} \Big[\frac{P(\varphi_1 - 2\epsilon)}{\varphi_{12}(\varphi_{12} - 2\epsilon)} + (\varphi_1 \leftrightarrow \varphi_2) \Big] \,,
\eea
with 
\be\label{P_poly}
P(x) := \prod\limits_{k=1}^{N_f} (x-m_k)~.
\ee

\begin{figure}[t]
	\center{\includegraphics[scale=0.5]{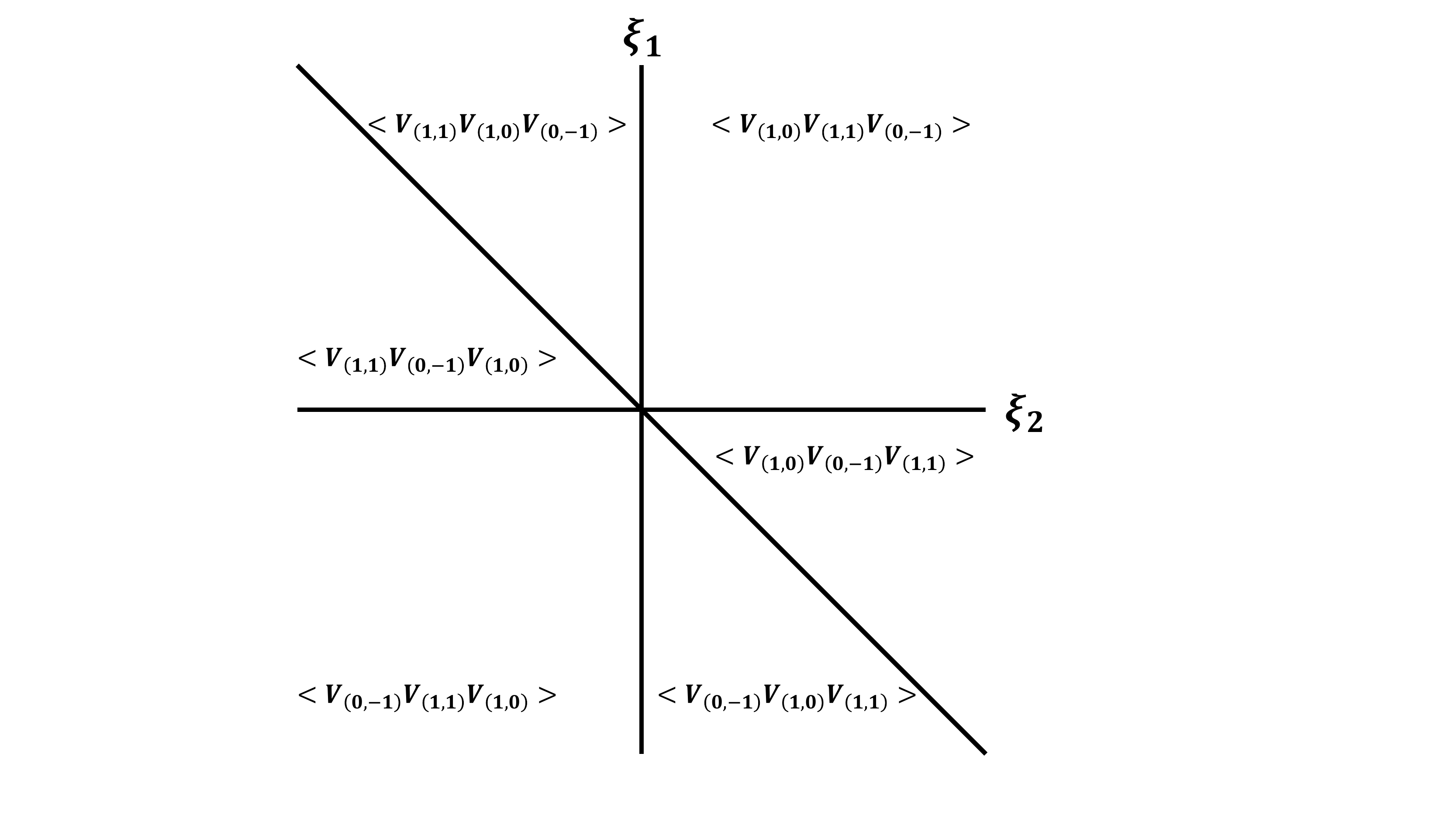}}
	\vspace{-0.8cm}
	\caption{\footnotesize{This figure illustrates the six distinct chambers in FI space. Each chamber corresponds to a specific ordering of the monopole operators $V_{(1,0)}, V_{(1,1)}$ and $V_{(0,-1)}$. } }
	\label{chambers_and_orderings}
\end{figure}

The other operator orderings in the correlator are given by the same computation but with FI parameters in different chambers. There are six possible orderings of these monopole operators, which we illustrate in Figure \ref{chambers_and_orderings}. Here $\xi_1$ is both the single FI parameter of $Z_3$ and the left node FI parameter of $Z_5$, while $\xi_2$ is only the FI parameter of the right node in $Z_5$. Exchanging $V_{(1,0)}$ and $V_{(1,1)}$ corresponds to crossing the $\xi_1 = 0$ axis. In addition, exchanging $V_{(1,1)}$ and $V_{(0,-1)}$ corresponds to crossing the $\xi_2 = 0$ axis.  Finally, exchanging $V_{(1,0)}$ and $V_{(0,-1)}$ corresponds to crossing the $\xi_1 + \xi_2 =0$ line.

This is all worked out from the brane setup. Each SMM node corresponds to D1 strings stretched between two NS5s. These two NS5s are associated to two monopole operator insertions and their positions along $x^0$ determine the FI parameter of the node $\xi = x^0_{\rm right} - x^0_{\rm left}$. By looking at the ordering in the operator insertions in the correlator, we can determine the chamber in FI space.

We leave the computation of the SMM is these various cases as an exercise to the enthusiastic reader. In the end, there is wall-crossing when $V_{(0,-1)}$ is commuted with either $V_{(1,0)}$ (wall at $\xi_1+\xi_2=0$) or $V_{(1,1)}$ (wall at $\xi_2=0$).

\bigskip

\noindent{\bf Star product}

To conclude this section, we want to show that the above results are compatible with the Moyal product structure on the quantized Coulomb branch, which we derived from the dimensional reduction of 4d to 3d. 

\medskip

In section \ref{sec:Star product as a Moyal product and abelian relations} we described the Moyal, or star, product structure that should exist on the Coulomb branch. The explicit formula \eqref{starprod} allows us to compute the star product between any two monopole operators, and, by iteration, the star product of any number of operators. The star product is supposed to compute correlators (see \eqref{NonComProd}), therefore we can check whether it agrees with the computations presented above, based on the brane contructions.

Remember that in order to use the formula \eqref{starprod}, one needs to first express the abelian monopoles $u_v$ in terms of the abelian variables $\chi_a,\varphi_a$, with \eqref{abelvar}.

In section \ref{sec:UNsimpleProd} we already computed some star products in the $U(N)$ SQCD theory, which we can evaluate for $N=2$:
\bea
V_{(1,0)}\star V_{(1,0)} &= u_{2e_1} + u_{2e_2} + u_{e_1+e_2} \frac{2}{\pm\varphi_{12} + 2\epsilon}  \ = \ \vev{V_{(1,0)}^2}   \,,\cr
V_{(1,0)}\star V_{(0,-1)}  &= u_{e_1-e_2} + u_{-e_1+e_2} - \Big[ \frac{P(\varphi_1 - \epsilon)}{\varphi_{12} (\varphi_{12} - 2\epsilon)}  + (\varphi_1 \leftrightarrow \varphi_2) \Big] \ = \ \vev{V_{(1,0)}V_{(0,-1)}}  \,,\cr
V_{(0,-1)}\star V_{(1,0)}  &= u_{e_1-e_2} + u_{-e_1+e_2} - \Big[\frac{P(\varphi_1 + \epsilon)}{\varphi_{12} (\varphi_{12} + 2\epsilon)}  + (\varphi_1 \leftrightarrow \varphi_2)\Big]  \ = \ \vev{V_{(0,-1)}V_{(1,0)}} \,.
\eea
We find agreement in all cases.

Let us show some other examples. 
\bea
V_{(1,0)} \star V_{(1,1)} &= \Big[e^{\chi_1} \left( \frac{P(\varphi_1)}{ \pm \varphi_{12} + \epsilon } \right)^{1/2}  + (..._1 \leftrightarrow ..._2)\Big] \star \Big[ e^{\chi_1+\chi_2} (P(\varphi_1)P(\varphi_2))^{1/2} \Big] \cr
&= e^{2\chi_1+\chi_2} \left( \frac{P(\varphi_1 + \epsilon)P(\varphi_1-\epsilon)P(\varphi_2)}{ \pm \varphi_{12} + \epsilon } \right)^{1/2} +  (..._1 \leftrightarrow ..._2) \cr
&= u_{2e_1+e_2} + u_{e_1+2e_2} \,.
\eea
\bea
& V_{(1,0)} \star V_{(1,1)} \star V_{(0,-1)} = \Big[e^{2\chi_1+\chi_2} \left( \frac{P(\varphi_1 + \epsilon)P(\varphi_1-\epsilon)P(\varphi_2)}{ \pm \varphi_{12} + \epsilon } \right)^{1/2} +  (..._1 \leftrightarrow ..._2)\Big]  \cr
& \hspace{4cm}  \star \Big[ e^{-\chi_1} \left( \frac{P(\varphi_1)}{ \pm \varphi_{12} + \epsilon } \right)^{1/2}  + (..._1 \leftrightarrow ..._2) \Big] \cr
& \qquad = e^{\chi_1+\chi_2} \left( \frac{P(\varphi_1 )P(\varphi_1-2\epsilon)^2 P(\varphi_2)}{\varphi_{12}^2 (\varphi_{12} - 2\epsilon)^2 } \right)^{1/2} 
   \cr
&\qquad  + e^{2\chi_1} \left( \frac{P(\varphi_1 + \epsilon)P(\varphi_1-\epsilon)P(\varphi_2-\epsilon)^2}{(\pm\varphi_{12} + 2\epsilon) (\pm\varphi_{12})} \right)^{1/2}+ (..._1 \leftrightarrow ..._2) \cr
&\qquad= -u_{e_1+e_2} \Big[\frac{P(\varphi_1-2\epsilon)}{\varphi_{12} (\varphi_{12} - 2\epsilon)} + (\varphi_1 \leftrightarrow \varphi_2)\Big] + u_{2e_1} P(\varphi_2-\epsilon) +u_{2e_2} P(\varphi_1-\epsilon)\,.
\eea
This is in perfect agreement once again. It is easy to check that the other correlators computed through the star product formula all agree with the computations based on branes. This is a strong consistency check of both our brane method and the star product formula.

\medskip

Now that we have a clear procedure for computing monopole correlators in SQCD theories, we derive in the next section general formulae for arbitrary correlators of minimal monopoles in the $U(N)$ SQCD theory, from brane constructions and SMM computations.

%%%%%%%%%%%%%%%%%%%%%%%

\section{$U(N)$ SQCD theories}
\label{sec:U(N)SQCD}

In this section we use the brane construction from section \ref{sec:Branes_Bubbing_WC} to write down the topological correlator of bare monopole operators for $U(N)$ SQCD in terms of abelian monopole operators and the bubbling terms associated to gauged SMMs. We also discuss how branes encode the relevant aspects of the geometry of the affine Grassmannian. We then present the general result for correlators containing bare monopole operators of positive and negative charges. Finally, we study some examples of such correlators focusing on their wall-crossing behaviour.\footnote{See \cite{Kodera:2016faj} for a detailed mathematical analysis of the Coulomb branch of the related 3d $\mathcal{N}=4$ SQCD theory with a massive adjoint hypermultiplet.}

\subsection{SMMs for a general bare monopole correlator}\label{sec:SMM_general_U(N)}

In this section we use the type IIB brane construction to derive the SMMs whose partition functions encode the coefficients (valued in the field of rational functions of $\varphi$) in the expansion of the VEV of a product of non-abelian bare monopole operators in terms of abelian monopole operator VEVs. We will consider the insertion at separated points of bare monopole operators. These generate the chambers in the magnetic charge lattice, which are domains of linearity of the R-charge formula for monopole operators. The location of the operators along the line transverse to the 2d Omega background is specified by the FI parameters of the SMM. The partition function of the SMM is a piecewise constant function of the FI parameters, which is constant in the chambers of FI space and jumps at codimension-one walls where a 0d Coulomb branch opens up, corresponding to two operators crossing each other along the line. Generic monopole operators can in principle be obtained by colliding multiple monopole generators in a given chamber. The associated bubbling factors are on-the-wall partition functions of the relevant SMMs, which we cannot compute using the Jeffrey-Kirwan prescription (but see \cite{Brennan2019a} for the analogous computation in one dimension higher).

\medskip

\noindent{\bf The bare monopole operators}

Before we discuss the brane construction, let us briefly recall that the R-charge of a bare monopole operator is a piecewise linear function of the magnetic charge $B=(B_a)$,
whose domains of linearity define subchambers of the positive Weyl chamber in the cocharacter lattice.%
\footnote{We are interested in gauge invariant monopole operators, so we restrict to the positive Weyl chamber in the cocharacter lattice (the R-charge formula is gauge invariant). The discussion can be easily extended to abelian monopole operators with charge in the full cocharacter lattice.}
For $U(N)$ SQCD with $N_f$ fundamental hypermultiplets, the R-charge 
\be\label{R-charge_formula}
R[B]=\frac{N_f}{2}\sum_{a=1}^N |B_a| - \sum_{1\le a<b\le N} |B_a-B_b| 
\ee 
is linear in  $N+1$ subchambers of the positive Weyl chamber 
\be\label{charge_chambers}
C_k=\{(B_1,\dots,B_N)\in \bZ^N~|~ B_1 \ge B_2 \ge \dots \ge B_k \ge 0 \ge B_{k+1}\ge \dots \ge B_N \}
\ee
labelled by $k=0,1,\dots,N$. The $k$-th subchamber is generated by the $N$ lattice vectors $(1^a,0^{N-a})$ and $(0^{N-b},(-1)^b)$ with $a=1,\dots,k$ and $b=1,\dots,N-k$, respectively.%
\footnote{We use $x^a$ to denote $x,\dots,x$ (repeated $a$ times).} These are the magnetic charges of the $N$ bare monopole generators for chamber $C_k$ referred to above. If we consider instead the union of all subchambers, we need a total of $2N$ bare monopole generators, $N$ with positive charges and $N$ with negative charges: 
\be\label{monop_generators}
\begin{split}
	U^+_a&:=V_{(1^a,0^{N-a})}\qquad \qquad (a=1,\dots,N)~,\\  U^-_b&:=V_{(0^{N-b},(-1)^{b})} \qquad~ ~~\,(b=1,\dots,N)~.
\end{split}
\ee
We remark that these are ``generators'' only within the sector of bare monopole operators that we are discussing so far. When dressed monopole operators and Casimir invariants are included, only the operators $U^\pm_1$ among (\ref{monop_generators}) are actual generators \cite{Cremonesi:2013lqa,Bullimore:2015lsa}.

In this section we will be general and consider the insertion of bare monopole operators (\ref{monop_generators}) along the $x^0$ line transverse to the Omega deformed plane, rather than restricting to a specific subchamber (\ref{charge_chambers}). We will therefore consider VEVs of operators which are words in the letters (\ref{monop_generators}), made of $n^+_a$ letters $U^{+}_{a}$ and $n^-_b$ letters $U^{-}_{b}$:
\be\label{VEV_prod}
\left\langle T\left(\prod_{a=1}^N (U^+_a)^{n_a^+} \prod_{b=1}^N (U^-_b)^{n_b^-} \right)           \right\rangle ~,
\ee 
where $T$ denotes the time ordering of the operators along $x^0$ (and we have suppressed the insertion points of the operators along $x^0$ to ease the notation). 

We would like to express (\ref{VEV_prod}) as a linear combination of abelian variables $u_v$. The coefficients in this linear combination are specific rational functions of $\varphi_a$, $m_k$ and $\epsilon$, which are given by the partition function of SMMs that we will identify using a brane construction. 

\medskip

\noindent{\bf The brane construction}

$U(N)$ SQCD with fundamental hypermultiplets is realised by a brane construction with $N$ D3 branes and $N_f$ D5 branes suspended between two NS5' branes, see Table \ref{tab:orientations} \cite{Hanany:1996ie}. To insert a gauge invariant bare monopole operator \eqref{monop_generators} in the three-dimensional SQCD theory, we introduce an NS5 pair in this brane construction. An NS5 pair with linking numbers $(h^-,h^+)=(0,a)$ inserts a gauge invariant monopole operator $U^+_{a}=V_{(1^a,0^{N-a})}$, whereas an NS5 pair with linking numbers $(h^-,h^+)=(b,0)$ inserts a gauge invariant monopole operator $U^-_{b}=V_{(0^{N-b},(-1)^b)}$.%
\footnote{One of the two NS5 branes in each pair is a spectator. We introduce it to ensure that $h^\pm$ are integers, but it plays no role in the following.} 
In our convention, we will always keep the NS5$_+$ (respectively NS5$_-$) branes  to the right (resp. left) of all the D3 and D5 branes along the $x^7$ direction. Therefore, an NS5$_+$ (resp. NS5$_-$) brane  with linking number $h^+=a$ (resp. $h^-=a$) has $a$ D1 strings attached to its left (resp. right). In a departure from the previous section, we will \emph{not} necessarily order the NS5$_\pm$ with non-increasing linking numbers as we move towards $x^7\to\pm\infty$. We collect the NS5$_\pm$ linking numbers in two integer vectors $\rho^\pm=(\rho^\pm_i)=(h^\pm_i)$ representing \emph{unordered} partitions. 

We can then rewrite the topological correlator \eqref{VEV_prod} by making the insertion points manifest:
\be\label{VEV_prod_2}
\left\langle T\left(~\prod_i U^-_{\rho^-_i}(x^0_{-i})\prod_i U^+_{\rho^+_i}(x^0_{+i}) \right) \right\rangle ~.
\ee
We identify the insertion points $x^0_{\pm i}$ of the monopole operators with the positions  of the $i$-th NS5$_\pm$ branes in the $x^0$ direction. (Note that in this convention there is no $x^0_0$ insertion point and that the spectator NS5 branes trivially insert the identity operator.)

The D1 strings emanating from the NS5 branes are not allowed to extend to $x^7=\pm\infty$, so they must either end on other NS5 branes or on D3 branes, in a way consist with the s-rule. We sum over all  configurations with fixed linking numbers for the NS5 branes. D1 strings stretched between NS5 and D3 branes realise abelian monopole operators $u_v$ in the effective abelian 3d gauge theory on the Coulomb branch. The vector of linking numbers $(\ell_a)$ of the D3 branes, which encodes how the D1 strings end on the $N$ D3 branes, determines the magnetic charge $v=(\ell_a)$ of the abelian monopole operator $u_v$. On the one hand, it turns out to be  convenient to separate the $N$ D3 branes into three sets with positive, vanishing and negative linking numbers, respectively. We collect the positive D3 brane linking numbers $\ell_a^+=\ell_a >0$ in an \emph{unordered} partition $v^+=(\ell^+_a)$ and a corresponding \emph{ordered} partition $\sigma^+=(\sigma^+_i)=p(v^+)$, and similarly the absolute values of the negative  D3 brane linking numbers $\ell^-_a=-\ell_a>0$ in an \emph{unordered} partition $v^-=(\ell^-_a)$ and a corresponding \emph{ordered} partition $\sigma^-=(\sigma^-_i)=p(v^-)$. By construction, the sum of the lengths of $\sigma^+$ and $\sigma^-$  must be less than or equal to $N$. 

We remark that the partitions $\sigma^+$ and $\rho^+$ (and similarly $\sigma^-$ and $\rho^-$) do not necessarily have the same magnitude (the magnitude of a partition $\mu$ is $|\mu|:=\sum\limits_k \mu_k$). Rather $|\sigma^\pm|\le |\rho^\pm|$, as will become clearer below. There is however a constraint 
\be\label{constraint_sigmarho+-}
|\rho^-|-|\sigma^-|=|\rho^+|-|\sigma^+| \ ,
\ee
which arises from requiring that all D1 strings end on NS5 or D3 branes on both sides. 

\medskip

Finally, we introduce an alternative notation for the brane configuration that we will use in the upcoming computations. In this notation, we drop all spectator NS5 branes and label the brane configuration by the linking numbers of all D3 branes, the linking numbers of all (active) NS5 branes, and an extra integer $L$ that specifies the NS5 brane interval in which the D5 branes lie. We choose not to move the D5 branes across the NS5 branes to avoid creating D3' branes, so the D5 branes still separate the NS5$_-$ from the NS5$_+$ branes. Therefore, the $N_f$ D5 branes lie in the interval between the $L$-th and  $(L+1)$-th NS5 along $x^7$, where
\be\label{r}
L=\ell(\rho^-)
\ee 
is the length (\emph{i.e.} the number of non-zero entries) of $\rho^-$, namely the number of active NS5$_-$ branes. Analogously, $R=\ell(\rho^+)$ is the number of active NS5$_+$ branes.

For future convenience, we redefine the linking numbers asymmetrically as follows:
\be
\begin{split}
	&\mathrm{D3}: \qquad ~~ \ell':=n(NS5_L)+n(D1_R)-n(D1_L) =\ell +L\ ,\\ 
	&\mathrm{NS5}: \qquad h':=\wat n(D3_R)+\wat n(D1_L)-\wat n(D1_R) =h+N/2 ~.
\end{split}
\ee
These linking numbers can be read off by moving all the D3 branes across all the NS5$_-$ branes, so that they lie to the left of all the NS5 branes, and counting the net number of D1 strings ending on the D3 branes from the right and on the NS5 branes from the left. The D3 brane linking numbers are $v'=(\ell'_a)=v+L \equiv(\ell_a+L)$. The corresponding ordered partition is
\be\label{sigma}
\sigma'=p(v')=(\sigma^+ + L, L^{N-\ell(\sigma^+)-\ell(\sigma^-)},-\sigma^-+L)~,
\ee
where we use the shorthand notation  $-\mu:=(-\mu_\ell,\dots,-\mu_1)$ and $k+\mu:=(k+\mu_1,\dots,k+\mu_\ell)$ for a partition $\mu=(\mu_1,\dots,\mu_\ell)$ and an integer $k$.

Similarly, we label NS5 branes with increasing $x^7$ by an integer $I=1,\dots, L+R$ (so $x^7_{I+1}>x^7_I$) and we collect their linking numbers in an \emph{unordered} partition%
\footnote{Note that NS$5_+$ branes with $h^+=a$ and NS5$_-$ branes with $h^-=N-a$ have the same linking numbers $h'=a$ with D3 branes. They are however distinguished by their linking numbers with D5 branes (for the (NS5, D5, D3') Hanany-Witten triple), which differ by $N_f$.}
\be\label{rho}
\rho'=(\rho'_I)= (N-\rho^-,\rho^+)~.
\ee
Using the constraint (\ref{constraint_sigmarho+-}), we see that
$\sigma'$ and $\rho'$ are partitions of the same number $n$: 
\be\label{n}
|\sigma'|=NL+|\sigma^+|-|\sigma^-|=NL+|\rho^+|-|\rho^-|=|\rho'|\equiv n~.
\ee
From now on we will always use this second notation to describe the brane configurations and we will omit the primes from the notation.

\medskip

\noindent{\bf The gauged SMM}

The prefactor that multiplies the VEV of the abelian monopole operator $u_{v}$ in the expansion of the topological correlation function (\ref{VEV_prod_2}) is the partition function of the gauged $\cN=2^*$ SMM, which describes the low-energy physics on the worldvolume of the D1 strings. We denote the generic SMM by $T^{\sigma}_{\rho, L}[SU(n)]$ where $\sigma$ and $\rho$ are given by (\ref{sigma}) and (\ref{rho}), respectively (with primes omitted). When no Fermi fundamentals are present we use the standard notation $T^{\sigma}_{\rho}[SU(n)]$. 

To read off the gauge group and matter content of the gauged SMM, we move D3 branes along the $x^7$ direction, crossing NS5 branes until they no longer have any D1 strings attached, following the philosophy of \cite{Hanany:1996ie}. We then count the number of D1 strings (which contribute vector multiplets from D1-D1 strings), D3 branes (which contribute fundamental hypermultiplets from D3-D1 strings) and D5 branes (which contribute fundamental Fermi multiplets from D5-D1 strings) in each interval between two adjacent NS5 branes. The NS5 branes themselves contribute bifundamental hypermultiplets for adjacent gauge groups, from D1$_i-$D1$_{i+1}$ strings. 

The gauge and flavour nodes are labelled by a non-negative integer $I=1,\dots,L+R-1$, corresponding to the interval between the $I$-th and the $(I+1)$-th NS5 branes along $x^7$. The SMM quiver is then the same as for $T^{\sigma}_{\rho}[SU(n)]$ \cite{Gaiotto:2008ak}, reduced to zero dimensions and further decorated by $N_f$ extra fundamental Fermi multiplets attached to the $L$-th gauge node.%
\footnote{We relax the requirement that $\rho$ is ordered, which would make the undecorated quiver \emph{good}. The formulae for the ranks in the quiver apply nonetheless.} %\SC{(See also \cite{Webster:2020qnz} for the isomorphism of the Higgs branch of $T^{\sigma}_{\rho}[SU(n)]$ theories with slices to Schubert varieties in the affine Grassmannian of $PGL_n$.)}
The number of flavours of fundamental hypermultiplets $M_I$, the ranks $N_I$, and the FI parameters $\xi_I$ of the $I$-th gauge node ($I=1,\dots,L+R-1$) are given by 
\be\label{ranks_SMM_alt}
\begin{split}
	M_I &= \wat \sigma_I - \wat\sigma_{I+1} \ , \\[1ex] 
	N_I &= \sum_{K>I} \rho_K - \sum_{K>I} \wat\sigma_K  \ ,\\[1ex]
	\xi_I &= x_{I+1}- x_I \ ,
\end{split}
\ee
where $x_I \equiv x^0_I$ is the position of the $I$-th NS5 brane along $x^0$ and a hat denotes the dual (or transposed) partition. The FI parameters of the gauge nodes are related to the insertion points of the monopole operators along the line in the correlator (\ref{VEV_prod_2}). 

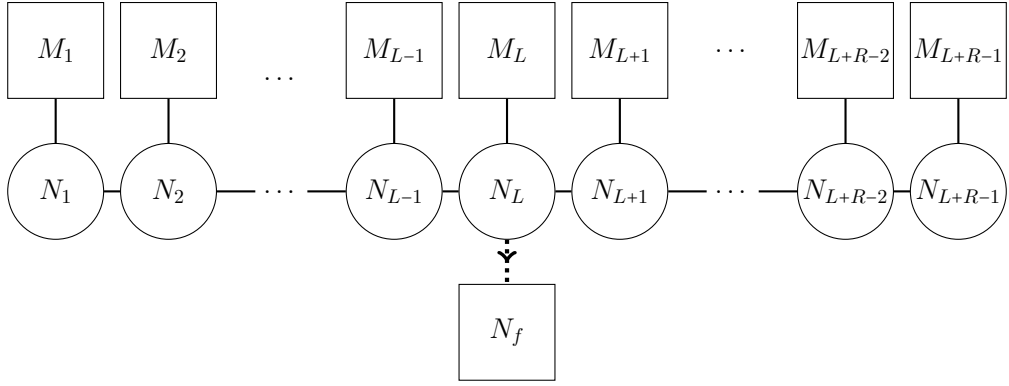
\begin{figure}[t]
	\begin{center}
		\begin{tikzpicture}[scale=0.75, transform shape]
		\node[circle,draw,fill=white,minimum size=1.7cm,inner sep=0pt] (a) at (0,-2) {\large $N_{1}$};
		\node[circle,draw,fill=white, minimum size=1.7cm,inner sep=0pt] (b) at (2,-2) {\large $N_{2}$};
		\node[draw=none,fill=none] (c) at (4,-2) {\large $\dots$};
		\node[circle,draw,fill=white, minimum size=1.7cm,inner sep=0pt] (d) at (6,-2) {\large $N_{L-1}$};
		\node[circle,draw,fill=white,minimum size=1.7cm,inner sep=0pt] (e) at (8,-2) {\large $N_{L}$};
		\node[circle,draw,fill=white,minimum size=1.7cm, inner sep=0pt] (f) at (10,-2) {\large $N_{L+1}$};
		\node[draw=none,fill=none] (g) at (12,-2) {\large $\dots$};
		\node[circle,draw,fill=white, minimum size=1.7cm,inner sep=0pt] (h) at (14,-2) {\large $N_{L+R-2}$};
		\node[circle,draw,fill=white,minimum size=1.7cm,inner sep=0pt] (i) at (16,-2) {\large $N_{L+R-1}$};
		\node[rectangle,draw,fill=white,minimum size=1.7cm,inner sep=0pt] (j) at (0,0.5) {\large $M_{1}$};
		\node[rectangle,draw,fill=white,minimum size=1.7cm,inner sep=0pt] (k) at (2,0.5) {\large $M_{2}$};
		\node[draw=none,fill=none] (l) at (4,0) {\large $\dots$};
		\node[rectangle,draw,fill=white,minimum size=1.7cm,inner sep=0pt] (m) at (6,0.5) {\large $M_{L-1}$};
		\node[rectangle,draw,fill=white, minimum size=1.7cm,inner sep=0pt] (n) at (8,0.5) {\large $M_{L}$};
		\node[rectangle,draw,fill=white,minimum size=1.7cm,inner sep=0pt] (o) at (10,0.5) {\large $M_{L+1}$};
		\node[draw=none,fill=none] (p) at (12,0.5) {\large $\dots$};
		\node[rectangle,draw,fill=white,minimum size=1.7cm,inner sep=0pt] (q) at (14,0.5) {\large $M_{L+R-2}$};
		\node[rectangle,draw,fill=white,minimum size=1.7cm,inner sep=0pt] (r) at (16,0.5) {\large $M_{L+R-1}$};
		\node[rectangle,draw,fill=white, minimum size=1.7cm,inner sep=0pt] (s) at +(8,-4.5) {\large $N_f$};
		\foreach \from/\to in {a/b,b/c,c/d,d/e,e/f,f/g,g/h,h/i,a/j,b/k,d/m,e/n,f/o,h/q,i/r}
		\draw [-, thick] (\from) -- (\to); 
		\draw [dotted,->-=.5, ultra thick] (e) -- (s);
		\end{tikzpicture}
	\end{center}
	\caption{\footnotesize{Quiver diagram for a generic gauged SMM from the brane construction. }}
	\label{a_generic_quiver}
\end{figure}

We encode the field content of the gauged SMM in a quiver,  as depicted in Figure \ref{a_generic_quiver}, or equivalently in the matrix notation by
\be\label{matrix_notation_2}
\begin{bmatrix}
	M_{1} & M_{2} & \dots & M_{L-1} & M_L & M_{L+1} & \dots & M_{L+R-2} & M_{L+R-1}\\
	N_{1} & N_{2} & \dots & N_{L-1} & \underline{N_L} & N_{L+1} & \dots & N_{L+R-2} & N_{L+R-1}
\end{bmatrix}~,
\ee
where the underline indicates the presence of $N_f$ extra fundamental Fermi multiplets.

To avoid violating the s-rule and breaking supersymmetry \cite{Hanany:1996ie}, the NS5 brane partition $\rho$ and the D3 brane partition $\sigma$ must satisfy the inequalities%
\footnote{Recall, $p(\rho)_K$ denotes the ordered partition associated to $\rho_K$.}
\be\label{ineq} 
\sum\limits_{K>I}p(\rho)_K\ge\sum\limits_{K>I}\wat\sigma_K \quad\forall ~I~ , \qquad\qquad \ell(\sigma)\le N~,
\ee
otherwise the SMM partition function vanishes.
The relevant D3 partitions $\sigma$, which appear for a given NS5 partition $\rho$, can be obtained by starting from the Young tableau associated to $\wat{p(\rho)}$ and consecutively moving boxes in $\sigma$ down to the next row or column.

\medskip

\noindent{\bf The partition function of the SMM $T^{\sigma}_{\rho, L}[SU(n)]$}

The partition function of $T^{\sigma}_{\rho, L}[SU(n)]$ is a meromorphic function of the complex masses (or equivariant parameters) for the global symmetries:
\begin{itemize}
	\item $m =\{m_\alpha\}_{\alpha=1}^{N_f}$ for the $U(N_f)$ flavour symmetry acting on the fundamental Fermi multiplets charged under the $L$-th gauge group;
	\item $\tilde\varphi_K=\{\tilde\varphi_{K,r}\}_{r=1}^{M_K}$ for the $U(M_K)$ flavour symmetry acting on the fundamental hypermultiplets of the $K$-th gauge group. We gather all the $\tilde\varphi_K$ in a vector  $\tilde\varphi=(\tilde\varphi_K)$. 
	\item $\epsilon$ for the $R$-symmetry of the $\cN=(0,4)$ superalgebra which commutes with the $\cN=2$ subalgebra preserved by the $\cN=2^*$ theory. 
\end{itemize}
With this notation and conventions, the partition function of  $T^{\sigma}_{\rho, L}[SU(n)]$ is computed by the integral
\be\label{Z_general}
\begin{split}
	Z_{T^{\sigma}_{\rho, L}[SU(n)]}(\varphi,m,\epsilon; & ~\xi) = \oint_{\mathrm{JK}(\xi)} \prod_K \left[\frac{d^{N_K}z_K}{(2\pi i)^{N_K} N_K!} (2\epsilon)^{N_K} \prod\limits_{I\neq J}^{N_K} z_{K,IJ}(z_{K,IJ}+2\epsilon) \right]\\ \\
	&\cdot \prod_K\frac{1}{\prod\limits_{I=1}^{N_K} \left[ \prod\limits_{J=1}^{N_{K+1}} (\pm (z_{K,I}-z_{K+1,J})+\epsilon)\prod\limits_{r=1}^{M_K}(\pm (z_{K,I}-\tilde\varphi_{K,r})+\epsilon))\right]}\\
	&\cdot \prod\limits_{I=1}^{N_L} \prod\limits_{\alpha=1}^{N_f}(z_{L,I}-m_\alpha)~,
\end{split}
\ee
where the products over gauge groups in the first and second line run from $K=1$ to $K=L+R-1$, and the $\tilde\varphi$ variables are related to the $\varphi$ variables as above. The three lines in the RHS of (\ref{Z_general}) account for the contributions of 0d vector multiplets, hypermultiplets and Fermi multiplets, respectively. The integral is over the Jeffrey-Kirwan cycle determined by the vector of FI parameters $\xi=(\xi_K)$. It is a piecewise constant function of $\xi$, which is constant in the interior of the chambers in FI space, but might jump at codimension-one walls separating different chambers, where the JK integral is ill-defined and a 0d Coulomb branch opens up.

\medskip

\noindent{\bf The abelian expansion of the topological correlator (\ref{VEV_prod_2})}

We are now ready to write down the expansion of the  topological correlator (\ref{VEV_prod_2}) in the abelian monopole operators $u_v$, where the abelian magnetic charge $v$ is encoded in the D3 brane linking numbers of the brane configuration. The brane construction described above shows that the 0d field theory that dresses the abelian variable $u_v$ in the expansion of the correlator (\ref{VEV_prod_2}) is the SMM  $T^{\sigma}_{\rho, L}[SU(n)]$  with $L=\ell(\rho^-)$, $\sigma=p(v)+L$, $\rho=(N-\rho^-,\rho^+)$, complex masses  $p(\varphi)$, $m$, $\epsilon$ and FI parameters $\xi$. We therefore deduce that
\be\label{general_expansion}
\left\langle T\bigg(\prod_{i=1}^{L}U^-_{\rho^-_i}(x^0_{-i})~\prod_{i=1}^{R}U^+_{\rho^+_i}(x^0_i)\bigg)\right\rangle
= \sum_{v}Z_{T^{p(v)+L}_{(N-\rho^-,\rho^+), L}[SU(n)]}(p(\varphi),m,\epsilon; \xi) \cdot u_v~.
\ee
The sum over the abelian magnetic charges $v$ is equivalent to the sum over ordered partitions $\sigma$ and their permutations as described above. Although the distinction between non-bubbling and bubbling sectors is not well-motivated when computing monopole correlators, we can still recognise the non-bubbling terms as those with trivial SMM ($Z=1$). They arise from the initial partitions $\wat\sigma = p(\rho)$. The bubbling contributions will arise from all the other partitions $\sigma$ (with non-vanishing contribution). The piecewise constant dependence of the SMM partition functions on the FI parameters in the RHS reflects the topological nature of the correlation function in the LHS.

\subsection{Affine Grassmannian, monopole operators and branes}

Before we explicitly calculate correlation functions of 't Hooft monopole operators in the next subsection, we now take a detour to outline how the brane construction that we have introduced to engineer these correlation functions is related to the geometry of the affine Grassmannian, which is exploited in the mathematical definition of Coulomb branches of 3d $\cN=4$ gauge theories of \cite{Braverman:2016wma,Braverman:2016pwk}. The relationship between the brane construction and the affine Grassmannian is not needed for any of the calculations that we perform in this paper, so readers who are more interested in the physics of the problem than in its mathematical interpretation may skip this subsection.

The affine Grassmannian is an infinite-dimensional complex algebraic variety which describes the moduli space of all Hecke modifications of $G$-bundles, which are implemented by the insertion of monopole operators for a gauge group $G$. 
We start by recalling some basic facts about the affine Grassmannian and monopole operators. We refer the reader to \cite{Kapustin2007} for a physicist-friendly introduction to the affine Grassmannian and a detailed explanation of the correspondence between 't Hooft operators and Hecke modifications. At the end of the subsection we will discuss the brane realisation of the main geometrical ingredients. 

The basic setup involves solutions with pointlike singularities of the Bogomol'nyi equations on the product of a Riemann surface $C$ with complex coordinate $z$ and a line (or an interval with suitable boundary conditions) with real coordinate $y$. In our case, the Riemann surface is the Omega-deformed complex plane $\bC_\eps$ with coordinate $z=x^1+ix^2$, and the transverse line is the Euclidean time $y\equiv x^0$ of the three-dimensional field theory. 
In the absence of singularities, the holomorphic type of the $G$-bundle on the Riemann surface $C_y=C\times \{y\}$ does not depend on $y$. If a monopole operator singularity is inserted at $(z_0,y_0)$, the holomorphic $G$-bundle on $C_y$ jumps at $y=y_0$ in a way that is trivial away from the insertion point $z_0$. For gauge group $G=U(N)$, we can describe the modification of the $G$-bundle in terms of its action on the fundamental representation $V$, which defines an $N$-dimensional vector bundle $E$. Given a local decomposition of $E$ into a sum of $N$ line bundles $\oplus_i \cL_i$, the insertion of a monopole operator of magnetic charge $B=(B_1,\dots,B_N)$ at $z_0$ twists the line bundles as
\be\label{Hecke}
\cL_i \to \cL_i \otimes \cO(z_0)^{B_i}~,
\ee
where $\cO(z_0)$ is a singular line bundle with unit curvature localized at $z=z_0$. This twist is called a Hecke modification of type $B$ at $z_0$.

The holomorphic perspective taken above is very natural in the context of the Omega deformation, where the physical configurations are holomorphic, monopole operators are forced to lie on the line $\{z=0\}\times \bR \subset \bC_\epsilon \times \bR$ transverse to the origin of the Omega deformation plane, and the magnetic flux emanating from monopole operators is confined to this line \cite{Ito:2011ea,Bullimore:2015lsa,Bullimore:2016hdc}. 

Since the Hecke modification is local, we can take $C=\bC$ and put the singularity at $z=0$. In terms of a basis of $N$ holomorphic sections $s_i(z)$ of  $\cL_i$ which are linearly independent at $z=0$, a general section of $E$ transforms locally as 
\be\label{Hecke_2}
s(z)=\sum_i g_i(z) s_i(z) \to s'(z)=\sum_i g_i(z) s_i(z) z^{-B_i}~,
\ee
under the Hecke modification, where $g_i(z)$ are  $N$ locally holomorphic functions. We can assume without loss of generality that $E$ is trivial before the modification, so that the section $s(z)$ is holomorphic. Then the Hecke modification introduces a polar part to $s'(z)$, where the magnetic charges $B_i$ control the change in the orders of the poles. The polar part of $s'(z)$ modulo holomorphic redefinitions of $\{g_i(z)\}$ parametrizes the moduli space $\cY(B)$ of Hecke modifications of type $B$. It is a finite-dimensional subvariety of complex dimension 
\be\label{Schubert cell}
\dim \cY(B) =\sum_{i<j}|B_i-B_j|~
\ee
of the infinite-dimensional affine Grassmannian $\Gr_N$, which is a union of strata $\cY(B)$ for all integer tuples $B_1\ge B_2 \ge \dots \ge B_N$.%
\footnote{To be more precise, the space of Hecke modification is unaffected if $E$ is tensored with  $\cO(z_0)^k$, which is invertible. This corresponds to an extra insertion of the 't Hooft operator $V_{(1^N)}^k$, which shifts $B_i\to B_i +k$ for all $i$. For the purpose of studying Hecke modifications we may set $B_N=0$ without loss of generality: this leads to the affine Grassmannian $\Gr_N$ of $PGL(N)$.} The strata of the affine Grassmannian, which are interpreted as spaces of Hecke modifications $\cY(B)$, are called \emph{Schubert cells}. 

Unless $B$ is the highest weight of a minuscule representation of the Langlands dual group (for us, $G^\vee=U(N)$), Schubert cells $\cY(B)$ are not compact, but they admit a natural compactification in terms of their closures 
\be\label{Schubert_cycle}
\overline{\cY(B)} = \bigsqcup_{v\le B} \cY(v)~,
\ee
which in addition to $\cY(B)$ contain lower-dimensional strata $\cY(v)$ for every dominant weight $v$ in the irreducible representation of $G^\vee$ with highest weight $B$. The closure (\ref{Schubert_cycle}) of a Schubert cell is called a \emph{Schubert cycle}. The lower-dimensional Schubert cells $\cY(v)$ are obtained by suitable scaling limits towards the boundaries of the top-dimensional cell $\cY(B)$. One says that the $G$-bundle associated to Hecke modifications of type $B$ is unstable towards becoming a $G$-bundle associated to Hecke modifications of type $v$. The process continues until a stable bundle is reached, corresponding to a $v$ which is the highest weight of a minuscule representation. The physical interpretation of this process is monopole bubbling, whereby dynamical pointlike monopoles screen the 't Hooft monopole operator, thus reducing its charge. In our physical setup of 3d $\cN=4$  SQCD with gauge group $G=U(N)$ on $\bC_\eps \times \bR$, the stratification (\ref{Schubert_cycle}) of the Schubert cycle into several Schubert cells manifests itself in the decomposition (\ref{locresult}) of the VEV of a bare monopole operator of charge $B$. The bubbling contribution $Z_{\rm bub}(B,v)$ is associated to the slice transverse to $\cY(v)$ inside $\overline{\cY(B)}$. 

In the compactification (\ref{Schubert_cycle}), the lower-dimensional strata $\cY(v)$ appear as singular loci of $\overline{\cY(B)}$. The singularities of $\overline{\cY(B)}$ can be resolved or deformed by splitting the single insertion of a monopole operator of charge $B$ into the insertion at separated points of several constituent monopole operators of lower charges $B^{(1)},\dots, B^{(k)}$, such that $\sum_{j=1}^k B^{(j)}$. This configuration then computes a correlation function of $k$ monopole operators. Separating the insertion points along $C$ is a complex structure deformation, which is obstructed in the case of interest to us, where $C$ is replaced by the Omega-background $\bC_\eps$. We can however separate the constituent monopole operators along the transverse $\bR$, where different orderings lead to (in principle) different resolutions. 

The singularities in (\ref{Schubert_cycle}) are what makes it difficult to compute the VEV (\ref{locresult}) and similarly correlation functions involving non-minuscule monopole operators. In terms of the SMM encoding the bubbling contributions, we are on a wall in FI space. However, if the monopole operator of charge $B$ is split into separate constituent monopole operators with charges $B^{(j)}$ which are all highest weights of minuscule representations, then we are at a generic point in the K\"ahler moduli space of $\overline{\cY(B)}$ and the singularity is completely resolved. Correspondingly, the SMM is in the interior of a maximal-dimensional chamber in FI space and we can compute the correlation function (\ref{general_expansion}) using the Jeffrey-Kirwan prescription in the SMM. 

Now that we have introduced all the necessary ingredients in the geometrical description of the affine Grassmannian, we can relate them to their counterparts in the brane construction that we have introduced in sections \ref{sec:Branes_Bubbing_WC} and \ref{sec:SMM_general_U(N)}. To keep the discussion simple, we will only consider positive magnetic charges and therefore NS5$_+$ branes, which also allows us to neglect the D5 branes. This assumption can be relaxed and we leave the general case as an exercise to the reader. We will also omit the $+$ subscript for NS5$_{+}$ branes in the following.

The Schubert cycle $\overline{\cY(B)}$ in the left hand side of (\ref{Schubert_cycle}) is associated to a configuration of NS5-branes with (ordered) linking numbers $\rho$. Viewed as partitions, $B$ and $\rho$ are related by transposition: $\wat{B}=\rho$. We then consider all possible D3-D1-NS5 supersymmetric brane configurations with such NS5 brane linking numbers. The Schubert cells $\cY(v)$ in the right hand side of (\ref{Schubert_cycle}) are associated to the configurations with D3 brane linking numbers $\sigma$ equal to $v$ up to permutations. The requirement that $v\le B$, or equivalently $\sigma\le \wat{\rho}$, is nothing but the s-rule. Interestingly, the dimension (\ref{Schubert cell}) of the Schubert cell $\cY(v)$ can be interpreted in terms of branes as 
\be \label{dim_Schubert_cell_branes}
\dim\cY(v) =\sum_{i<j} |v_i-v_j| = \sum_{i<j} \#(\mathrm{NS5~between~D3_i~and~ D3_j}) \ ,
\ee 
where the number of NS5 branes between a pair of D3 branes in the right hand side is counted once the D3 branes have undergone sufficient Hanany-Witten transitions so that none of them have D1 strings attached. Finally, we can think of the affine Grassmannian $\Gr_N$ in terms of branes as describing all the possible D3-D1-NS5 brane configurations compatible with the s-rule, for a \emph{fixed} number $N$ of D3 branes and an \emph{arbitrary} number of NS5 branes and D1 strings. 

A complete resolution of the singular Schubert cycle $\overline{\cY(B)}$ is achieved by separating the NS5 branes in the $x^0$ direction. Then NS5 branes with linking number $\rho_i$ located at $x^0=x^0_i$ insert minuscule monopole operators $V_{(1^{\rho_i},0^{N-{\rho_i}})}(x^0_i)$. As we mentioned above, the effect of monopole bubbling in the resulting correlation function of monopole operators is controlled by the slice transverse to the Schubert cell $\cY(v)$ in $\overline{\cY(B)}$. In the brane interpretation, the degrees of freedom of the transverse slice are the open strings that start or end on the D1 branes in the resulting intersecting D3-D1-NS5 brane system, which in the IR describes our zero-dimensional bubbling SMM. The Higgs branch of this SMM, which can be visualised by the motions of D1 branes along the D3 branes, is the slice transverse to the Schubert cell $\cY(v)$ in $\overline{\cY(B)}$, and the monopole bubbling contribution $Z_{\rm bub}(B,v)$ is the partition function of this SMM.%
\footnote{If a combination of positively and negatively charged monopole operators are inserted then we also need to include NS5$_-$ branes in the construction and the D5 branes are no longer spectators. The extra D5-D1 strings add fermionic degrees of freedom, which do not change the Higgs branch of the SMM but affect the SMM partition function, making it dependent on the chamber in FI space or equivalently on the specific resolution of $\overline{\cY(B)}$.}

Finally, we note that the well-known fact that the space of Hecke modifications is independent of tensoring the vector bundle $E$ with a power of $\cO(z_0)$ also has a natural brane interpretation: it corresponds to inserting an extra NS5 brane with linking number $N$. After a Hanany-Witten transition across all $N$ D3 branes, this NS5 brane is disconnected from the active part of the brane configuration and plays no role in the bubbling SMM.

%
%\begin{itemize}
%\item Brane construction
%\begin{itemize}
%%	\item Schubert cycle $\overline{\cY(B)}$ $\leftrightarrow$ NS5-brane linking numbers
%%	\item Schubert cells $\cY(v)$ in (\ref{Schubert_cycle}) $\leftrightarrow$ D3-brane linking numbers compatible with s-rule given the NS5-brane linking numbers 
%%	\item Affine Grassmannian $\Gr_N$ $\leftrightarrow$ All NS5-D1 brane configurations compatible with the s-rule in the presence of $N$ D3-branes
%%	\item dimension of Schubert cell 
%%	\be 
%%	\dim\cY(v) \leftrightarrow \sum_{i<j} \#(\mathrm{NS5~between~D3_i~and~ D3_j})
%%	\ee 
%%	once the D3 branes have no D1 attached (or differences of linking numbers of pairs of D3-branes) \SC{Why?}.
%%	\item Minuscule monopole operators $V_{(1^k,0^{N-k})}$ $\leftrightarrow$ NS5 with linking number $k$. 
%	\item Bubbling is independent of tensoring with $\cO(z_0)$ $\leftrightarrow$ Hanany-Witten transition across all $N$ D3-branes.
%%	\item Slice transverse to $\cY(v)$ in $\overline{\cY(B)}$ $\leftrightarrow$ Higgs branch of bubbling SMM on D1-D3-NS5-D5 (D5 can be ignored for Higgs branch).
%%	\item Monopole bubbling factor $\leftrightarrow$ Partition function of bubbling SMM on D1-D3-NS5-D5.
%\end{itemize}  
%\end{itemize}
%

\subsection{Results for $U(N)$ SQCD}

In this section we discuss our results for SQCD theory with gauge group $U(N)$ and $N_f$ flavours of hypermultiplets in the fundamental representation. We first evaluate correlators of bare monopole operators of positive charge and then generalise to correlators of monopole operators of positive and negative charges. 
We then analyse in more detail some examples involving the products of bare monopole operators of minimal positive and negative charge. We evaluate the bubbling factors using the JK prescription, and our results are verified by comparison with the vacuum expectation values found using the star product from section \ref{sec:CBandLoc}. Finally, we discuss wall-crossing phenomena, which are related to the exchange of operators in the VEV.

\subsubsection{A warm-up: powers of the $U^+_1$ bare monopole operator}

Let us start by computing the VEV of the product of $n$ identical operators $V_{(1,0^{N-1})}=U^+_1$, which corresponds to having $\rho=\rho^+=(1^n)$ and empty $\rho^-$. In the brane construction we can neglect the NS5$_-$ branes and the D5 branes, which are spectators. The active branes consist of $n$ D1 strings stretched between $n$ NS5$_+$ branes and $N$ D3 branes, as encoded by an ordered partition $\sigma=\sigma^+$ of $n$ and permutations thereof ($\sigma^-$ is empty).  

A configuration in which the $N$ D3 branes have $v=(v_a)_{a=1}^N$ D1 branes attached realises the abelian monopole $u_v$ with $\sum_{a=1}^N v_a=n$. The SMM whose partition function dresses $u_v$ is the 0d $\cN=2^*$ version of the $T^{p(v)}[SU(n)]$ theory of \cite{Gaiotto:2008ak}, with complex masses $p(\varphi)$  for the flavour symmetry acting on the fundamental hypermultiplets and $\epsilon$ for $F$. There are no Fermi multiplets, so the masses $m$ do not appear. 
We therefore deduce that 
\be\label{V_10...^n}
\left\langle T(\prod_{i=1}^n U^+_1(x_i)) \right\rangle= \sum_{v} Z_{T^{p(v)}[SU(n)]}(p(\varphi),\epsilon;\xi) \cdot u_{v}~
\ee
where the sum is over unordered partitions of $n$ with at most $N$ parts. As we show in appendix \ref{sec:TSUn}, the partition functions in the RHS are the same in the interior of all chambers, in agreement with the fact that the operators in the LHS commute.

We can compute the SMM partition functions in several ways, finding agreement. First, we note that the partition function of the 0d $T^{\sigma}[SU(n)]$ theory is the equivariant volume of the Higgs branch of the theory, with equivariant parameters identified with the complex masses. The Higgs branch of $T^{\sigma}[SU(n)]$ is the intersection of the S\l odowy slice $S_\sigma$ with the nilpotent cone $\cN$ of $SU(n)$ \cite{Gaiotto:2008sa}. Its equivariant volume can be obtained by equivariant localization similarly to the Hilbert series $H_{S_{\sigma}\cap\cN}(t,x)$, which can be found in (3.29) of \cite{Cremonesi2014b} (see \cite{Cremonesi:2014uva} for a derivation), or by extracting the coefficient of the leading pole in the small chemical potential expansion of the Hilbert series:%
\footnote{We rescale $t\to t^2$ in \cite{Cremonesi2014b} to have integer powers of $t$.}
\be\label{HS_to_equiv_volume}
Z_{T^{\sigma}[SU(n)]}(\varphi,\eps) = \lim_{R\to 0} R^{{\mathrm{dim}_\bC}(S_{\sigma}\cap \cN)} H_{S_{\sigma}\cap \cN}(x =e^{-R{\varphi}},t=e^{-R\eps})~,
\ee
where the order of the pole at $R=0$ is the complex dimension of $S_{\sigma} \cap \cN$ 
\be\label{dim_S_sigma}
{\mathrm{dim}_\bC}(S_{\sigma}\cap \cN) = \sum_i \wat\sigma_i^2 -n~.
\ee	
Either way, we find the result 
\be\label{Z_T^sigma}
\begin{split}
	Z_{T^\sigma[SU(n)]}(\varphi,\eps)&=  \frac{\prod\limits_{k=1}^n (2k\eps)}{\prod\limits_{i=1}^{l(\wat\sigma)} \prod\limits_{a,b=1}^{\wat\sigma_i} [\varphi_{ab}+(\sigma_a+\sigma_b-2i+2)\eps]}\\
	&=\frac{n!}{\prod\limits_{a=1}^{l(\sigma)} \sigma_a!}%\binom{n}{\sigma_1,\sigma_2,\dots}
	\cdot\frac{1}{\prod\limits_{i=1}^{l(\wat\sigma)} \prod\limits_{1\le a<b\le \wat\sigma_i} [\pm \varphi_{ab}+(\sigma_a+\sigma_b-2i+2)\eps]}~ ,
\end{split}
\ee
where $\varphi_{ab}:=\varphi_a-\varphi_b$. The multinomial coefficient in the second line of (\ref{Z_T^sigma}) arises from the numerator and the $a=b$ factors in the denominator of the first line.

We can compare the RHS of (\ref{V_10...^n}) with the $n$-th Moyal star product of the VEV (\ref{V+-}) of the bare monopole operator $U^+_1$. First, we insert our result (\ref{Z_T^sigma}) and the definition of the abelian variables (\ref{abelvar}) into the RHS of (\ref{V_10...^n}). This can be massaged into the form
\be\label{star_guess}
\vev{(U^+_1)^n} = \sum_{\substack{v_1,\dots,v_N\ge 0\\ \sum_a v_a=n}} e^{v\cdot \chi} \frac{n!}{\prod_a v_a!} \prod_{a=1}^N \prod_{i_a=0}^{v_a-1} \left[\frac{P(\varphi_a+(v_a-1-2i_a)\eps)}{\prod\limits_{b\neq a}[\pm \varphi_{ab}+(v_a+v_b-2i_a)\eps)]} \right]^{1/2}~.
\ee
One can then prove by induction that (\ref{star_guess}) equals the $n$-th star power $\vev{U^+_1}^{\star n}$. Indeed (\ref{star_guess}) reduces to (\ref{V+-}) for $n=1$. Using (\ref{starprod}), one can compute $\vev{U^+_1}^{\star n}\star \vev{U^+_1}$. This can be shown to equal (\ref{star_guess}) with $n\to n+1$ with the help of the identity 
\be
\sum_{a=1}^N v_a \prod_{b\neq a} \left(1+ \frac{v_b}{y_b-y_a}\right) = \sum_{b=1}^N v_b
\ee 
for all $y=(y_a)$ and $v=(v_a)$. This identity follows from the simpler fact\footnote{We thank Patrick Dorey for providing this proof.}
\be
\sum_{a=1}^n \prod_{b\neq a} \frac{1}{y_b-y_a} = 0    \qquad \forall n>1~,
\ee
which can be proven by showing that the left hand side has no poles when $y_b=y_a$ for all $b\neq a$, as can be easily seen. Since the rational function on the left hand side is homogeneous of degree $1-n<0$ in $y$, it must vanish, which proves the desired identity.

An alternative simple method to derive the result (\ref{Z_T^sigma}),  following the logic of \cite{Gaiotto2012,Cremonesi:2014uva}, is to start from the partition function of the $T[SU(n)]$ SMM
\be\label{Z_TSUn}
\begin{split}
	Z_{T[SU(n)]}(\varphi,\eps)= \sum_{\w\in S_n} \frac{1}{\prod\limits_{1\le a<b\le n} \varphi_{ab}^{\w}(\varphi_{ba}^{\w}+2\eps \sgn(x_{ab}))}=\frac{n!}{ \prod\limits_{1\le a<b\le n} [\pm \varphi_{ab}+2\eps]}~,
\end{split}
\ee
which we compute in appendix \ref{sec:TSUn} by evaluating the JK residues, and then take appropriate residues in the flavour parameters $\varphi$ to obtain the partition function of $T^\sigma[SU(n)]$. Each residue implements the move of a single box in the Young tableaux of $\sigma$, reducing the flavour symmetry. We refer to appendix \ref{sec:residues} for details of this residue calculation.

\subsubsection{Bare monopole operators of positive charge}\label{sec:positive_charge}

The computation of the previous subsection can be generalised to a correlator involving only bare monopole operators of positive charge, by allowing a general unordered partition $\rho=\rho^+$ of magnitude $|\rho^+|=n$ and parts smaller than or equal to $N$, we find that 
\be\label{expansion_positive}
\left\langle T\bigg(\prod_{i=1}^{R}U^+_{\rho_i}(x_i) \bigg)\right\rangle
= \sum_{v} Z_{T^{p(v)}_{ \rho}[SU(n)]}(p(\varphi),\eps; \xi) \cdot u_v~,
\ee
where the relevant SMM is the 0d $\cN=2^*$ version of the  $T^{\sigma}_{\rho}[SU(n)]$ theory of \cite{Gaiotto:2008ak} with $\rho=\rho^+$ and $\sigma=\sigma^+=p(v)$,  and $R=\ell(\rho)$ is the length of $\rho$. The sum is effectively restricted to positive magnetic charges $v$ such that $\wat{p(\rho)}\ge p(v)$. It turns out that the relevant SMM partition functions are chamber independent, hence we omit the FI parameter $\xi$ in the argument of the partition functions in the following.

As in the case of trivial $\rho$, the partition function of the  $T^{\sigma}_{ \rho}[SU(n)]$ SMM is the equivariant volume of its Higgs branch, which is $S_\sigma \cap \overline{O}_{\wat\rho}$, the intersection of the S\l odowy slice of type $\sigma$ with the closure of the nilpotent orbit  of type $\wat\rho$.%
\footnote{To be precise, of type $\wat{p(\rho)}$. We abuse notation in this paragraph for the sake of readability.} See \cite{Maffei2005,Mirkovica,Webster:2020qnz} for a discussion of the isomorphism of $S_\sigma \cap \overline{O}_{\wat\rho}$ with slices to Schubert varieties in the affine Grassmannian of $PGL_n$. The equivariant volume can again be obtained directly by a fixed point formula, analogously to the  Hilbert series (4.2)  in \cite{Cremonesi:2014uva}, or from the coefficient of the leading pole of that same Hilbert series at $R=0$,
\be\label{HS_to_equiv_volume_2}
Z_{T^{\sigma}_\rho[SU(n)]}(\varphi,\eps) = \lim_{R\to 0} R^{{\mathrm{dim}_\bC}(S_{\sigma}\cap \overline{\cO}_{\wat\rho})} H_{S_{\sigma}\cap \overline{\cO}_{\wat\rho}}(x =e^{-R{\varphi}},t=e^{-R\eps})~,
\ee
where the order of the pole is controlled by the complex dimension of $S_{\sigma}\cap \overline{\cO}_{\wat\rho}$ 
\be\label{dim_S_sigma_int_Obar_rhohat}
{\mathrm{dim}_\bC}(S_{\sigma}\cap \overline{\cO}_{\wat\rho}) = \sum_i \wat\sigma_i^2 -\sum_j \rho_j^2~.
\ee	
The result is expressed in terms of the data of the partitions $\sigma$ and $\rho$ as follows:
\be\label{Z_T^sigma_rho}
\begin{split}
	Z_{T^\sigma_\rho[SU(n)]}(\varphi,\epsilon)&=  \frac{\wat q_\rho(\text{a}_\sigma(\varphi,\epsilon),\epsilon)}{\prod\limits_{a=1}^{l(\sigma)}\sigma_a! \prod\limits_{i=1}^{l(\wat\sigma)} \prod\limits_{1\le a<b\le \wat\sigma_i} [\pm\varphi_{ab}+(\sigma_a+\sigma_b-2i+2)\epsilon]}~.
\end{split}
\ee
Here 
\be\label{qhat}
\wat q_\rho (y,\eps) = \frac{1}{\prod\limits_{j=1}^{l(\rho)}
	\rho_j!} \sum_{\w\in S_n} \prod_{\alpha\in\Delta_\rho} (-\alpha\cdot y^\w)(\alpha\cdot y^\w +2\eps) \prod_{\gamma \in\Delta_+} \frac{\gamma\cdot y^\w -2\eps}{\gamma\cdot y^\w}
\ee
where $y=(y_1,\dots,y_n)$, $\Delta_+$ is the set of positive roots of $SU(n)$ and $\Delta_\rho$ is the set of positive roots in the Jordan blocks associated to $\rho$:
\be\label{roots_in_prod}
\begin{split}
	\Delta_+&=\{e_a-e_b ~|~1\le a<b\le n\}\\
	\Delta_\rho&=\{e_a-e_b~|~\sum\limits_{j=1}^{k-1}\rho_j+1\le a<b\le  \sum\limits_{j=1}^{k}\rho_j~~\text{for~some~}k\}~.
\end{split}
\ee
Finally, the first argument $\text{a}_\sigma(\varphi,\eps)$ of $\wat q_\rho$ has components
\be\label{a_sigma}
(\text{a}_\sigma(\varphi,\eps))_{a, h_a} = \varphi_a-(\sigma_a-2h_a+1)\eps~,  \quad h_a=1,\dots,\sigma_a~, \quad a=1,\dots,\ell(\sigma)~.
\ee
Because of the sum over permutations in (\ref{qhat}), the order of the components does not matter, so $\text{a}_\sigma(\varphi,\eps)$ is better thought of as a set rather than a vector. 

As a check, note that if $\rho=(1^n)$, $\Delta_\rho$ is empty and $\wat q_{(1^n)}=n!$, since the $\eps$-dependent terms average out to zero. Formula (\ref{Z_T^sigma_rho}) then reduces correctly to (\ref{Z_T^sigma}). 

As in the previous subsection, the partition function (\ref{Z_T^sigma_rho}) can be computed in two steps, first calculating it for trivial $\sigma=(1^n)$ and then taking residues in the flavour fugacities to obtain the result for a general $\sigma$. We refer the reader to appendices \ref{sec:TrhoSUn} and \ref{sec:residues} for details. There we show explicitly that the relevant SMM partition functions do not depend on the chamber in FI space in which they are computed, in agreement with the (not so obvious) field theory fact that the monopole operators in the LHS of (\ref{expansion_positive}) commute, and neither on the ordering of $\rho$.

\subsubsection{General correlators of non-bubbling bare monopole operators}\label{sec:positive_negative}

Finally, the general topological correlation function (\ref{general_expansion}) of bare monopole operators of positive and negative charges is expressed in terms of partition functions (\ref{Z_general}) of the general $T^\sigma_{\rho,L}[SU(n)]$ SMMs, which now exhibit chamber dependence and wall-crossing due to the extra Fermi multiplets. We compute the SMM partition functions in appendix \ref{sec:TrhoLSUn} for the case of a trivial D3 brane partition $\sigma=(1^n)$ and we explain how to introduce a non-trivial $\sigma$ by computing residues in the flavour fugacities of the SMM in appendix \ref{sec:residues}. Putting together these results, we can write the partition function of the $T^\sigma_{\rho,L}[SU(n)]$ SMM as
\be\label{ZTsigmarhoL}
\begin{split}
	Z_{T^\sigma_{\rho,L}[SU(n)]}(\varphi,m,\eps;x)&=  \frac{q_{\rho,L}(\text{a}_\sigma(\varphi,\epsilon),m,\eps;x)}{\prod\limits_{a=1}^{\ell(\sigma)}\sigma_a! \prod\limits_{i=1}^{\ell(\wat\sigma)} \prod\limits_{1\le a<b\le \wat\sigma_i} [\pm\varphi_{ab}+(\sigma_a+\sigma_b-2i+2)\epsilon]}\\
	&\cdot \frac{1}{\prod\limits_{a=1}^{\ell(\sigma)} \prod\limits_{h_a=L+1}^{\sigma_a} P((\text{a}_\sigma(\varphi,\eps))_{a,h_a}-L\eps)} \ ,
\end{split}
\ee 
where $q_{\rho,L}$ is defined in (\ref{q_general}), $\text{a}_\sigma$ is defined in (\ref{a_sigma}), $P(x)=\prod_{k=1}^{N_f} (x-m_k)$ as in (\ref{P_poly}) and it is understood that products over empty sets are equal to one.  Note that the Fermi factors of $P$  in the second line  all cancel against equal factors in $q_{\rho,L}$ when evaluated at  $\text{a}_\sigma$, as is clear from the original integral formula (\ref{Z_general}).  

The SMM partition function (\ref{ZTsigmarhoL}) depends on the FI parameters $\xi$ or the insertion points $x$ through $q_{\rho,L}$ as in the case of trivial $\sigma$ obtained in appendix \ref{sec:TrhoLSUn}. Every time we cross a wall at which two monopole operators of opposite charge (or equivalently an NS5$_+$ and an NS5$_-$ brane) change order, the partition function jumps. If instead the sign of the charge of the two monopole operators that are exchanged is the same, as in the previous subsection, the partition function does not jump across the wall.

\subsection{Examples: wall-crossing, poles at infinity and star product}

To conclude our analysis of topological correlation functions of bare monopole operators, we study a few correlators with a low number of bare monopole operators of minimal positive and negative charge.  The bubbling terms are determined by following the JK prescription to compute the partition functions for the relevant SMMs. We focus in particular on the relationship between the correlation functions containing a commutator of monopole operators, the non-zero contributions to the partition functions from poles at infinity and wall-crossing phenomena.

\subsubsection{One positive and one negative minimal monopole operator}

Firstly, we consider the VEV of the product of two minimal bare monopole operators of opposite charge,  $U^+_1\equiv V_{(1,0^{N-1})}$ and $U^-_1 \equiv V_{(0^{N-1},-1)}$. Depending on the order of these operators, we obtain two different results. To compute these results we require a setup containing an NS5$_+$ and an NS5$_-$ brane, from each of which emanates a D1 string, that is $\rho^+=\rho^-=(1)$. In total, there are $N(N-1) + 1$ configurations contributing to the VEVs. There are $N(N-1)$ configurations with the NS5$_+$ and the NS5$_-$ connected to different D3 branes and the remaining $N-2$ D3 branes are unconnected ($\sigma^+=\sigma^-=(1)$), corresponding to an abelian magnetic charge $e_a - e_b$, with $a\neq b$. There is one configuration where the NS5 branes are connected by the D1 strings joining and the $N$ D3 branes remain unconnected, with vanishing abelian magnetic charge ($\sigma^+=\sigma^-=()$). This tells us that 
\begin{equation}
\begin{aligned}
\langle U^+_1 U^-_1\rangle = \sum_{a \neq b} u_{e_a - e_b} + Z^{+-}(\varphi,m,\epsilon) \ , \\
\langle  U^-_1 U^+_1 \rangle = \sum_{a \neq b} u_{e_a - e_b} +Z^{-+}(\varphi,m,\epsilon) \ , \\
\end{aligned}
\end{equation}
where $e_a  := (0^{a-1}, 1, 0^{N-a})$ and $a,b = 0, \ldots ,N$. 

As expected, the first term in each VEV has no bubbling factor. The bubbling factors, $Z^{\pm \mp}(\varphi,m,\eps)$, arise from the configuration where the D1s are connected to each other. These are computed as the SMM described by the abelian quiver (in matrix notation)
\begin{equation}
\begin{bmatrix}
N \\
\underline{1}
\end{bmatrix} \ ,
\end{equation}
whose partition function is given by (we omit unnecessary subscripts in $z_1$ and $\xi_1$)
\begin{equation}
\label{Z1N}
Z^{\pm \mp}(\varphi,m,\epsilon)= \oint_{\mathrm{JK}(\pm \xi > 0)} \frac{dz}{2\pi i}\frac{(2\epsilon )\prod\limits_{k=1}^{N_f}\left[z-m_k\right]}{\prod\limits_{a=1}^N \left[\pm(z-\varphi_a)+\epsilon\right]} \ ,
\end{equation}
where $\pm$ in $Z^{\pm \mp}$ corresponds to the FI chamber $\pm\xi >0$ used in evaluating the integral, which is directly linked to the order of the operators in the VEV (but it is unrelated to the product over $\pm$ in the integrand). The FI parameter is given by the difference in the positions of the NS5 branes, $\xi = x_2 - x_1$, where $x_2/x_1$ labels the NS5$_+$/NS5$_-$ position along $x^0$. This JK integral is ill-defined at the codimension-one wall corresponding to the FI parameter $\xi = 0$, which is the situation where the NS5$_+$ and NS5$_-$ are at the same $x^0$ position and the operators $V_{(1,0^{N-1})}$ and $V_{(0^{N-1},-1)}$ collide (such a configuration realises the bare monopole $V_{(1,0^{N-2},-1)}$).

The poles contributing in FI chamber $\pm \xi>0$ are at $z=\varphi_a \mp \epsilon$, where $a = 1,\ldots,N$, and the partition function evaluates to 
\begin{equation}
\label{Z+N}
Z^{\pm \mp}(\varphi,m,\epsilon) = (-1)^{N-1}\sum_{a=1}^N \frac{\prod\limits_{k=1}^{N_f}\left[\varphi_a-m_k\mp\epsilon\right]}{\prod\limits_{b\neq a} \left[\varphi_{ab}\left(\varphi_{ab}\mp 2\epsilon\right)\right]}\ .
\end{equation}
As expected, the two results are related by sending $\epsilon \rightarrow -\epsilon$ (see \eqref{PTrel2}).

Consequently, for the product of one minimal positive operator and one minimal negative operator we find
\begin{equation}
\begin{aligned}
\label{one_positive_one_negative}
\langle U^+_1 U^-_1\rangle = \sum_{a \neq b} u_{e_a - e_b} +(-1)^{N-1}\sum_{a=1}^N \frac{\prod\limits_{k=1}^{N_f}\left[\varphi_a-m_k-\epsilon\right]}{\prod\limits_{b\neq a} \left[\varphi_{ab}\left(\varphi_{ab}-2\epsilon\right)\right]} \ , \\
\langle U^-_1 U^+_1  \rangle = \sum_{a \neq b} u_{e_a - e_b} +(-1)^{N-1} \sum_{a=1}^N \frac{\prod\limits_{k=1}^{N_f}\left[\varphi_a-m_{k}+\epsilon\right]}{\prod\limits_{b \neq a}\left[\varphi_{ab}\left(\varphi_{ab}+2\epsilon\right)\right]} \ , \\
\end{aligned}
\end{equation}
which matches the results found previously using the star product in \eqref{star_product_2_minimal}. These results are our 3d analogue of the results obtained in section 3.2.2 of \cite{Hayashi:2019rpw}. 

The vacuum expectation value of the commutator  $\left[U^+_1, U^-_1\right]$, which is the difference between the results computed in the two chambers, is related to the non-zero contribution $Z^{\infty}(\varphi,m,\eps)$ from evaluating the residue of the integrand in \eqref{Z1N} at $z = \infty$,
\begin{equation}\label{z_inf_result}
\left\langle \left[ U^+_1, U^-_1 \right]\right\rangle = Z^{+-}(\varphi,m,\eps) - Z^{-+}(\varphi,m,\eps)
= - Z^{\infty}(\varphi, m,\eps)  \ ,
\end{equation}
%\begin{equation}
%\begin{aligned}
%\label{z_inf_result}
%\left\langle \left[ V_{(1,0^{N-1})}, V_{(0^{N-1},-1)} \right]\right\rangle &= Z^{+}(\bphi,\bm,\epsilon) - Z^{-}(\bphi,\bm,\epsilon) \\
%&= - Z^{\infty}(\bphi,\bm,\epsilon)  \ ,
%\end{aligned}
%\end{equation}
where
\begin{equation}
\label{z_inf}
Z^{\infty}(\varphi, m,\epsilon) =\underset{z=\infty}{\operatorname{\mathrm{Res}}} \frac{(2\epsilon )\prod\limits_{k=1}^{N_f}\left[z-m_k\right]}{\prod\limits_{a=1}^N \left[\pm(z-\varphi_a)+\epsilon\right]} \ .
\end{equation}
Therefore, to obtain the result for the partition function of the SMM in one chamber from the other chamber, we add or subtract the contribution from evaluating the residue of the pole at infinity. This corresponds to crossing the codimension-one wall where the FI parameter is zero, which is the location where the 0d Coulomb branch opens up. 

For low values of $N_f$, there is no pole at infinity and the two monopole operators commute. The first non-zero contribution from the pole at infinity occurs at $N_f = 2N - 1$. This gives a polynomial of degree 1 in $\epsilon$. In general, the contribution from evaluating the residue of the pole at infinity will be a polynomial in $\varphi$, $m$ and $\epsilon$ of total degree $N_f -2N+2$. This follows from an R-symmetry selection rule and can be seen by Taylor expanding the integrand of $\eqref{Z1N}$ about $z = \infty$. It is possible to express the general form of the contribution from the pole at infinity in terms of a sum of symmetric polynomials. However, we find this to be unilluminating. Instead, we finish this discussion by writing the explicit contributions from the pole at infinity, which computes the monopole commutator in \eqref{z_inf_result}, for small values of $N_f$,
\begin{itemize}
	\item $N_f = 0,1,\ldots,2N-2$ 
	\begin{equation}\label{vanishing_res_infty}
	Z^{\infty}(\varphi, m,\epsilon) = 0 \ .
	\end{equation}
	\item $N_f = 2N-1$
	\begin{equation}
	Z^{\infty}(\varphi, m,\epsilon) = (-1)^{N-1}(2\epsilon) \ .
	\end{equation}
	\item $N_f = 2N$
	\begin{equation}
	Z^{\infty}(\varphi,m,\epsilon) = (-1)^{N-1}(2\epsilon)\left(2\sum_{a=1}^N\varphi_a - \sum_{k=1}^{N_f}m_{k} \right)\ .
	\end{equation}
	\item $N_f = 2N+1$ 
	\begin{equation}
	\hspace{-5pt} Z^{\infty}(\varphi,m,\epsilon) = (-1)^{N-1}\epsilon \left[2N\epsilon^2+\left(2\sum_a \varphi_a^2-\sum_k m_k^2\right)+\left(2\sum_a \varphi_a-\sum_k m_k\right)^2\right] \ .
	\end{equation}
\end{itemize}

\subsubsection{Two positive and one negative minimal monopole operators}

We now expand our analysis by introducing a second minimal positive operator: we compute the VEV of the product of two minimal positive operators and one minimal negative operator. In this scenario we find three different results depending on the order of the operators. We require a setup containing two NS5 pairs and we sum over configurations with three D1 strings, where a single string emanates from the two NS5$_+$, one string emanates from the innermost NS5$_-$ and the remaining NS5$_-$ is a spectator.  \begin{figure}[t]
	\centering
	\includegraphics[scale=0.75]{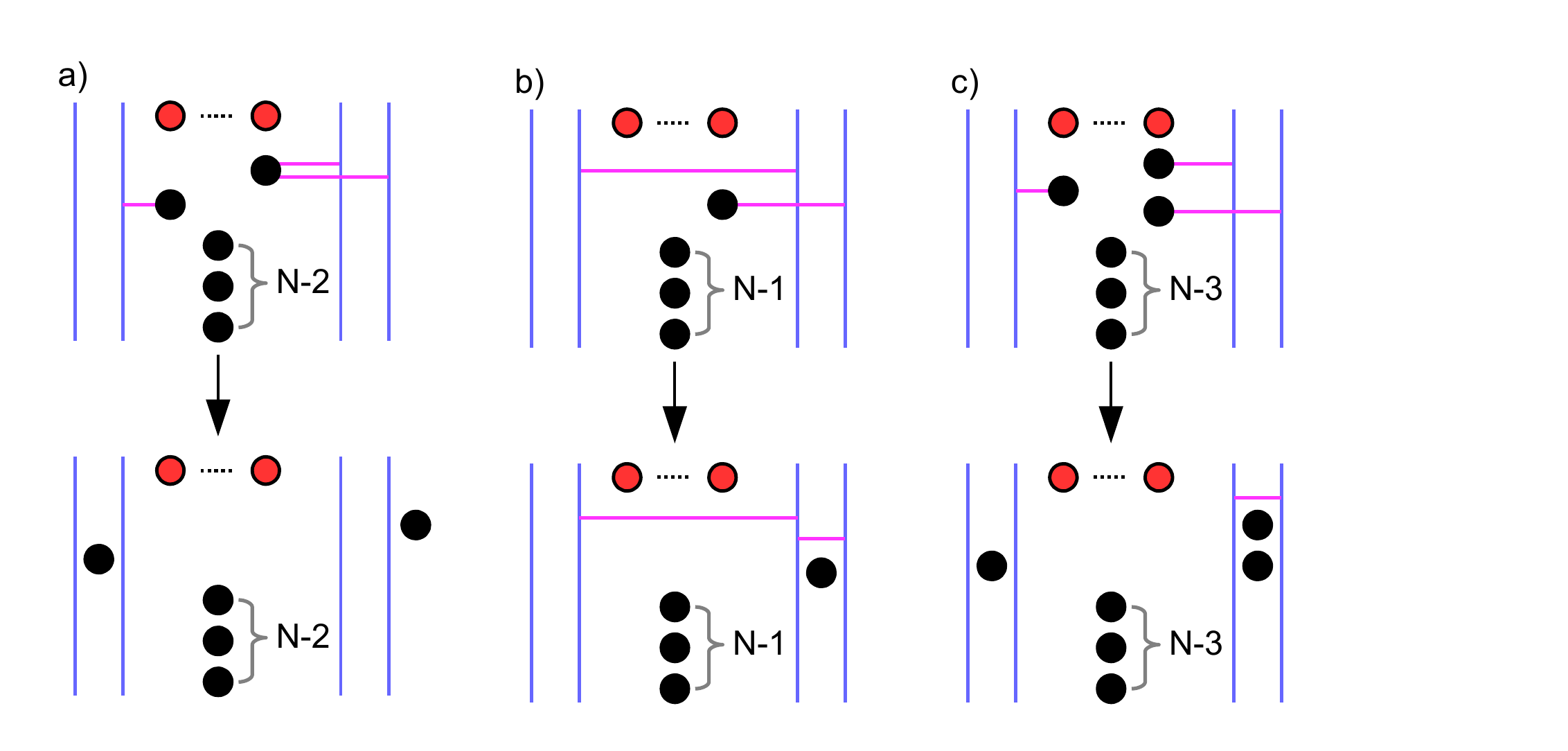}
	\vspace{-1cm}
	\caption{\footnotesize{This figure illustrates the brane setup for the configurations contributing to the vacuum expectation value of two minimal positive operators and one minimal negative operator. % NS5 branes are depicted as blue lines, D5 branes as red dots, D3 branes as black dots and D1 strings as purple lines. 
	The setup is depicted in the $x^{78}$ plane, where the NS5$_-$/NS5$_+$ branes are drawn to the left/right of the D5 branes in each diagram. a), b) and c) are an example of one of the diagrams contributing to each of the sums in the first, second and third term on the RHS of \eqref{VEV_+-+}. The other terms contributing to these sums are given by permutations of the D3 branes.}}
	\label{fig:++-}
\end{figure}
The configurations contributing to the VEV are shown in Figure \ref{fig:++-}. There are $N(N-1)$ configurations with both the NS5$_+$ connected to the same D3 brane, the NS5$_-$ connected to a different D3, and $N-2$ D3s remain unconnected. Additionally, there are $N$ configurations where the NS5$_-$ is connected to one of the NS5$_+$, the other NS5$_+$ is connected to a D3 and the remaining $N-1$ D3 branes are unconnected. Finally, there are $N(N-1)(N-2)/2$ configurations where the three NS5 branes are all connected to three different D3 branes and the remaining $N-3$ D3s are unconnected. This tells us that the vacuum expectation value for the product of these monopole operators is given by
\begin{equation}
\label{VEV_+-+}
\left\langle T\left((U^+_1)^2 U^-_1
\right)\right\rangle = \sum_{a\neq b} u_{2e_a-e_b} +\sum_{a} u_{e_a} Z_a(\varphi,m,\eps; \xi)+\sum_{a \neq b,c \atop b \neq c} u_{e_a + e_b - e_c} Z_{ab}(\varphi,m,\eps) \ .
\end{equation}

The first term on the RHS in the VEV has no monopole bubbling contribution, the partition function of the associated SMM is trivial. The third term in the VEV comes from the cases where the NS5s are all connected to different D3s and the bubbling factor is computed as the SMM described by the 
matrix notation 
\begin{equation}
\begin{bmatrix}
 2 \\
1 
\end{bmatrix} \ ,
\end{equation}
so the SMM is the 0d SQED theory with two hypermultiplets of masses $\varphi_a$ and $\varphi_b$. The partition function of this quiver was computed earlier in (\ref{Z1}) and we simply state the result again, 
\begin{equation}
Z_{ab}(\varphi,m,\eps)=\oint_{\mathrm{JK}} \frac{dz}{2\pi i}\frac{2\epsilon}{\left[\pm(z-\varphi_a)+\epsilon \right]\left[\pm(z-\varphi_b)+\epsilon \right]}=\frac{2}{\left(\pm\varphi_{ab}+2\epsilon \right)} \ .
\end{equation}
where $a$ and $b$ are the D3 branes that are connected to the two NS5$_+$ branes in the construction, see Figure \ref{fig:++-}-c. It is important to highlight that this result is the same regardless of the chamber in which we compute the JK integral. 

The other bubbling factor in the VEV \eqref{VEV_+-+} is computed as the SMM described by the quiver 
\begin{equation}
\begin{bmatrix}
 N-1 & 1   \\
 \underline{1} & 1 \\ 
\end{bmatrix}\ ,
\end{equation}
where we underline the $U(1)$ gauge node attached to the Fermi multiplets. The partition function of this theory is given by 
\begin{equation}
\label{Z21N}
\hspace{-3pt} Z_{a}(\varphi,m,\eps; \xi) = \oint_{\mathrm{JK}(\xi)} \frac{dz_0dz_1}{(2\pi i)^2} \frac{(2\epsilon)^2\prod\limits_{k=1}^{N_f}(z_0-m_{k})}{\prod\limits_{b\neq a}\left[\pm(z_0-\varphi_b)+\epsilon \right] \left[\pm(z_0-z_1)+\epsilon\right]\left[\pm(z_1-\varphi_a)+\epsilon\right]} \ ,
\end{equation} 
where $a$ labels the single D3 brane that is located in the interval between the two NS5$_+$ branes and the remaining $N-1$ D3s are between the innermost NS5$_-$ and the innermost NS5$_+$, see Figure \ref{fig:++-}-b. The bubbling factors for the different orderings of the monopole operators are obtained by evaluating this integral in the different FI chambers. The ordering of the operators is linked to the order of the NS5 branes, which affects the sign of the FI parameters and leads to the different chambers. In this case, there are two FI parameters, which are given by
\begin{equation}
\xi_0 = x_1 - x_{-1} \ , \ \xi_1 = x_2 - x_1 \ ,
\end{equation}
where $x_1, x_2$ are the $x^0$ coordinates of the two NS5$_+$ branes and $x_{-1}$ of the inner NS5$_-$. The outer NS5$_-$ is a spectator and plays no role here.

\begin{figure}[t]
	\center{\includegraphics[scale=0.4]{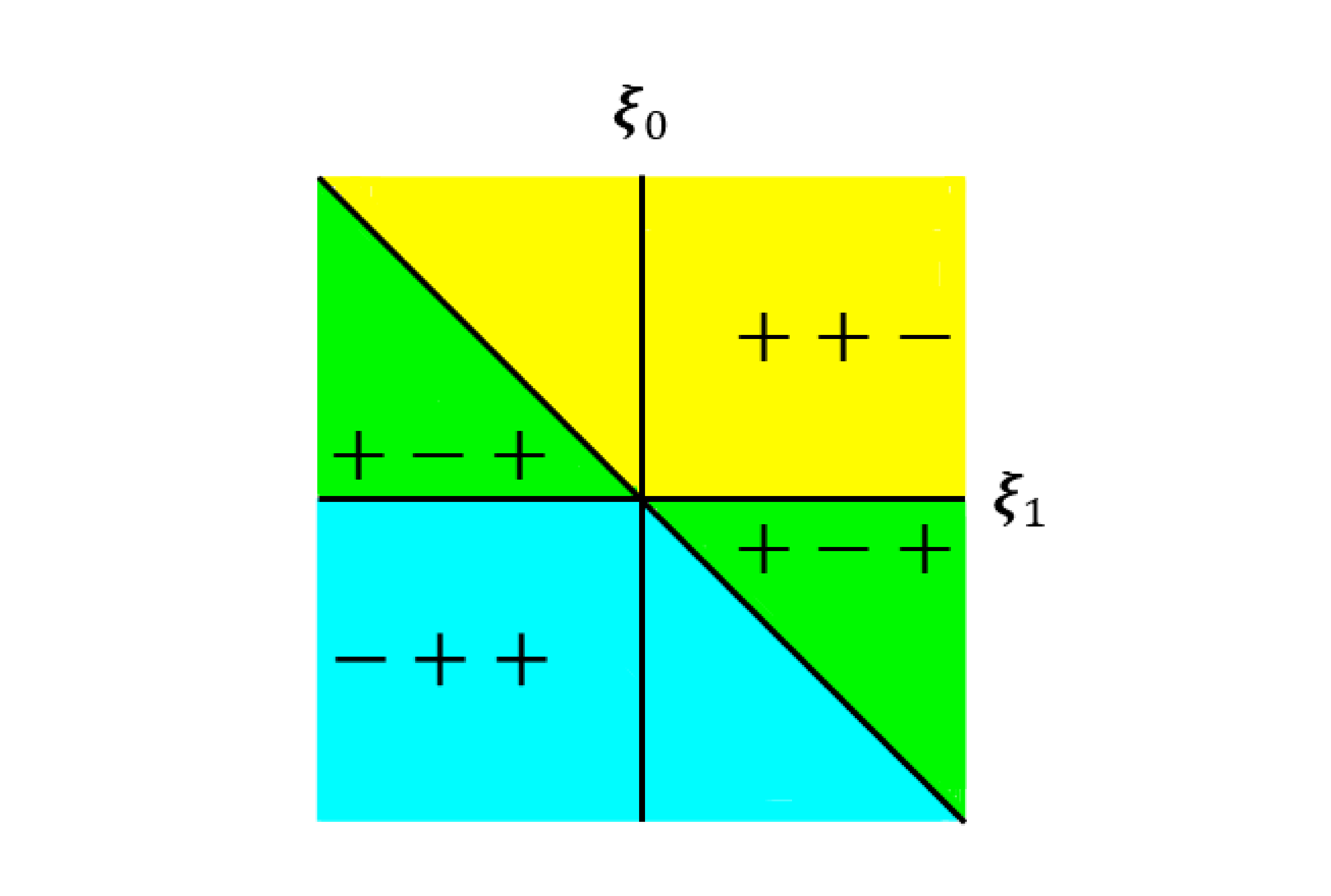}}
	\vspace{-.5cm}
	\caption{\footnotesize{This figure illustrates how the 6 regions of the FI space (separated by the solid black lines) are grouped into 3 distinct chambers, due to the symmetry under the exchange of the two NS5$_+$ branes. Each chamber corresponds to a different ordering of the operators in the correlator \eqref{VEV_+-+}.}}
	\label{chambers}
\end{figure}

By naively considering the order of these NS5 branes, one expects to find 6 chambers from the permutations of $x_1, x_2, x_{-1}$. However, there is a symmetry under the exchange of $x_1$ and $x_2$, which tells us that $(\xi_1, \xi_0)$ is equivalent to $(-\xi_1, \xi_0 + \xi_1)$. Consequently, there are only 3 distinct chambers, which are illustrated in Figure \ref{chambers}, where: 
\begin{itemize}
	\item The $+ + -$ chamber satisfies the region $\xi_0 > 0, \xi_0 + \xi_1 > 0$.
	\item The $- - +$ chamber satisfies the region $\xi_0 < 0, \xi_0 + \xi_1 < 0$.
	\item The final chamber, $+ - +$, contains the remaining regions described by $\xi_1 > 0 , \xi_0 < 0 , \xi_0 + \xi_1 > 0$ and $\xi_1 < 0 , \xi_0 > 0 , \xi_0 + \xi_1 < 0$. 
\end{itemize}
In general, for a VEV $\langle T \left((U^+_1)^A (U^-_1)^B \right) \rangle$ there will be $\frac{(A+B)!}{A!B!}$ inequivalent chambers, in correspondence to all the orderings of $A$ $U^+_1$ operators and $B$ $U^-_1$ operators. 

The JK prescription tells us that the multi-dimensional poles contributing to the JK integral in \eqref{Z21N} are different in the different regions of the FI space. In Table \ref{table_of_poles} we list the poles contributing to the integral in each chamber. We apply the constructive definition of the JK residue \citep{Szenes2004}, where each term in the sum of residues comes with an appropriate $\pm$ sign, which is called $\nu(F)$ in section 2.4.3 of \citep{Benini:2013xpa}, and depends on the orientation of the ordered basis used to determine the order in which to perform the iterated residue for each pole. 
\begin{table}[h!]
	\begin{center}
		\begin{tabular}{ |c|c|c|c| } 
			\hline
			Region of FI space & Chamber & Multi-dimensional Pole $(z_0 , z_1)$ \\
			\hline
			$\xi_0 > 0$& &$(\varphi_a - 2\epsilon, \varphi_a - \epsilon)$ \\ 
			$\xi_1 > 0$  & $+ + -$ & $(\varphi_b - \epsilon , \varphi_b -2 \epsilon)$\\ 
			$\xi_0 + \xi_1 > 0$&  & $(\varphi_b - \epsilon , \varphi_a -\epsilon)$ \\ 
			\hline
			$\xi_0 > 0$ & &$(\varphi_a - 2\epsilon, \varphi_a - \epsilon)$ \\ 
			$\xi_1 < 0$ & $+ + -$ & $(\varphi_b -\epsilon  , \varphi_b)$\\ 
			$\xi_0 + \xi_1 > 0$&  & $(\varphi_b - \epsilon , \varphi_a +\epsilon)$ \\ 
			\hline
			$\xi_0 < 0$& &$(\varphi_a , \varphi_a - \epsilon )$ \\ 
			$\xi_1 > 0$  & $+ - +$ & $(\varphi_b - \epsilon , \varphi_b - 2\epsilon)$\\ 
			$\xi_0 + \xi_1 > 0$&  & $(\varphi_b + \epsilon , \varphi_a -\epsilon)$ \\ 
			\hline
			$\xi_0 > 0$ & &$(\varphi_a , \varphi_a + \epsilon)$ \\ 
			$\xi_1 < 0$ & $+ - +$ & $(\varphi_b + \epsilon , \varphi_b +2 \epsilon)$\\ 
			$\xi_0 + \xi_1 < 0$&  & $(\varphi_b - \epsilon , \varphi_a +\epsilon)$ \\ 
			\hline
			$\xi_0 < 0$ & &$(\varphi_a +2 \epsilon, \varphi_a + \epsilon)$ \\ 
			$\xi_1 < 0$ & $- + +$ & $(\varphi_b + \epsilon , \varphi_b + 2\epsilon)$\\ 
			$\xi_0 + \xi_1 < 0$&  & $(\varphi_b + \epsilon , \varphi_a +\epsilon)$ \\ 
			\hline
			$\xi_0 < 0$ & &$(\varphi_a + 2\epsilon, \varphi_a + \epsilon)$ \\ 
			$\xi_1 > 0$& $- + +$ & $(\varphi_b  +\epsilon, \varphi_b)$\\ 
			$\xi_0 + \xi_1 < 0$&  & $(\varphi_b + \epsilon , \varphi_a -\epsilon)$ \\ 
			\hline
		\end{tabular}
		\caption{\footnotesize{Multi-dimensional poles contributing to the integral \eqref{Z21N} in each chamber, where $a$ has a fixed value and $b = 1, \ldots, a-1, a+1, \ldots, N$.}}
		\label{table_of_poles}
	\end{center}
\end{table}

Evaluating \eqref{Z21N} in each of the three chambers we find
\begin{align}
\label{Z++-}
Z^{++-}_a &= \frac{(-1)^{N-1}P(\varphi_a-2\eps)}{\prod\limits_{b\neq a} \left[\left(\varphi_{ab}-\epsilon\right)\left( \varphi_{ab}-3\epsilon\right) \right]} +\sum_{b\neq a} \frac{2(-1)^{N-1}P(\varphi_b -\eps)}{\left(\varphi_{ab}-\epsilon\right)\left( \varphi_{ab}+3\epsilon\right)\prod\limits_{c\neq b,a \atop b \neq a} \left[\varphi_{bc}\left(\varphi_{bc}-2\epsilon \right) \right]  }\\
\label{Z+-+}
Z^{+-+}_a &= \frac{P(\varphi_a)}{\prod\limits_{b \neq a} \left[\pm\varphi_{ab}+\epsilon \right]} +\sum_{b\neq a}\Bigg[ \frac{(-1)^{N-1}P(\varphi_b +\eps)}{\left(\varphi_{ab}-\epsilon\right)\left( \varphi_{ab}-3\epsilon\right)\prod\limits_{c \neq b,a \atop b \neq a} \left[\varphi_{bc}\left(\varphi_{bc}+2\epsilon \right) \right]} +(\eps\to -\eps)\Bigg]\\
\label{Z-++}
Z^{--+}_a &= \frac{(-1)^{N-1}P(\varphi_a+2\eps)}{\prod\limits_{b\neq a} \left[\left(\varphi_{ab}+\eps\right)\left( \varphi_{ab}+3\eps\right) \right]} +\sum_{b\neq a} \frac{2(-1)^{N-1}P(\varphi_b +\eps)}{\left(\varphi_{ab}+\eps\right)\left( \varphi_{ab}-3\eps\right)\prod\limits_{c\neq b,a \atop b \neq a} \left[\varphi_{bc}\left(\varphi_{bc}+2\eps \right) \right]  }
\end{align}
As expected, $Z^{++-}_a$ and $Z^{-++}_a$ are related by $\epsilon \rightarrow -\epsilon$ while $Z^{+-+}_a$ is invariant.

Consequently, we find the following results for the correlator (\ref{VEV_+-+}):
\begin{equation}
\begin{aligned}
\label{two_positive_one_negative}
\left\langle U^+_1 U^+_1 U^-_1\right\rangle &= \sum_{a\neq b} u_{2e_a-e_b}+\sum_{a} u_{e_a}  Z^{++-}_a+\sum_{a\neq b,c \atop b \neq c} u_{e_a + e_b - e_c} \frac{2}{\left(\pm\varphi_{ab}+2\epsilon \right)} \ ,\\
\left\langle U^+_1 U^-_1 U^+_1\right\rangle &= \sum_{a\neq b} u_{2e_a-e_b} +\sum_{a} u_{e_a}  Z^{+-+}_a +\sum_{a\neq b,c \atop b \neq c} u_{e_a + e_b - e_c} \frac{2}{\left(\pm\varphi_{ab}+2\epsilon \right)}\ ,\\
\left\langle U^-_1 U^+_1 U^+_1\right\rangle &=\sum_{a\neq b} u_{2e_a-e_b} +\sum_{a} u_{e_a} Z^{-++}_a+\sum_{a\neq b,c \atop  b \neq c} u_{e_a + e_b - e_c} \frac{2}{\left(\pm\varphi_{ab}+2\epsilon \right)} \ ,
\end{aligned}
\end{equation}
where $ Z^{++-}_a$, $Z^{+-+}_a$ and $Z^{-++}_a$ are given in \eqref{Z++-}, \eqref{Z+-+} and \eqref{Z-++}. We have checked that these results agree with those obtained using (\ref{V+-}) and the star product. These correlation functions are our 3d analogue of the results obtained in section 4.2.2 of \citep{Hayashi:2019rpw}. The correlators of two minimal negative operators and one minimal positive operator can also be obtained from these results by sending $U^+_1 \leftrightarrow U^-_1$ and $e_a \leftrightarrow e_{-a}$.

We can now study the relationship between the results computed in the different chambers and wall-crossing.  % We focus on codimension-one wall-crossing, since any codimension-$N$ crossing can be split into $N$ codimension-one wall jumps.
The jump between chamber $++-$ and $+-+$ is given by 
\begin{equation}
\begin{aligned}
\left\langle U^+_1 \left[U^+_1 , U^-_1\right] \right\rangle &= \sum_{a} u_{e_a} \left(Z^{+ + -}_a(\varphi,m,\eps) - Z^{+ - +}_a(\varphi,m,\eps)\right)\ .
\end{aligned}
\end{equation}
The difference between the SMM partition functions $Z_a$ of (\ref{Z21N}) computed in the $++-$ and $+-+$ chambers is captured by the residue of a pole at infinity. To obtain $Z^{+ - +}_a$ from $Z^{+ + -}_a$ there are two options. The first option involves setting $\xi_0 = 0$ and crossing the $\xi_1$ axis, see Figure \ref{chambers}. This corresponds to adding the contribution from the pole where $z_1$ is finite and $z_0 \rightarrow \infty$,
\begin{equation}
Z^{+ + -}_a(\varphi,m,\eps)- Z^{+ - +}_a(\varphi,m,\eps) = -\underset{z_0=\infty}{\operatorname{\mathrm{Res}}}\ \underset{z_1=\varphi_a-\epsilon}{\operatorname{\mathrm{Res}}} I_a(z,\varphi,m,\eps) \ ,
\end{equation}
where $ I_a(z,\varphi,m,\eps)$ is the integrand in \eqref{Z21N}. The second option involves crossing the line $\xi_0 + \xi_1 = 0$, which corresponds to adding the contribution from the pole where $z_1 \rightarrow \infty , z_0 \rightarrow \infty$ with finite $z_1 - z_0$,
\begin{equation}
Z^{+ + -}_a(\varphi,m,\eps)-Z^{+ - +}_a(\varphi,m,\eps)= - \underset{z_1=\infty}{\operatorname{\mathrm{Res}}}\ \underset{z_0=z_1-\epsilon}{\operatorname{\mathrm{Res}}} I_a(z,\varphi,m,\eps)  \ .
\end{equation}
These two options are identical, since the contributions from the poles at infinity obey the following relations,
\begin{equation}
\underset{z_0=\infty}{\operatorname{\mathrm{Res}}}\ \underset{z_1=\varphi_a-\epsilon}{\operatorname{\mathrm{Res}}} I_a =  \underset{z_1=\varphi_a-\epsilon}{\operatorname{\mathrm{Res}}}\ \underset{z_0=\infty}{\operatorname{\mathrm{Res}}} I_a=  \underset{z_1=\infty}{\operatorname{\mathrm{Res}}}\ \underset{z_0=z_1-\epsilon}{\operatorname{\mathrm{Res}}} I_a = - \underset{z_0=\infty}{\operatorname{\mathrm{Res}}}\ \underset{z_1 = z_0 + \epsilon}{\operatorname{\mathrm{Res}}} I_a \ .
\end{equation}
Consequently, we can write
\begin{equation}
\begin{aligned}
\left\langle U^+_1 \left[U^+_1, U^-_1\right] \right\rangle &=-\sum_{a} u_{e_a}\underset{z_0=\infty}{\operatorname{\mathrm{Res}}}\ \underset{z_1=\varphi_a-\epsilon}{\operatorname{\mathrm{Res}}} I_a(z,\varphi,m,\eps) = -\sum_a u_{e_a} Z^{\infty}(\varphi - e_a \eps,m,\epsilon)\ ,
\end{aligned}
\end{equation}
which is obtained by evaluating the residue for the pole $z_1 = \varphi_a -\epsilon$ and then using the definition of $Z^{\infty}(\varphi, m, \eps)$ in \eqref{z_inf}. We can now link this with our discussion of the star product by using the quantized abelian relations from section \ref{sec:Star product as a Moyal product and abelian relations}. In particular, using $u_{e_a} \cdot f(\varphi - e_a \epsilon) = u_{e_a} \star f(\varphi)$, we find 
\begin{equation}
\begin{aligned}
\left\langle U^+_1 \left[U^+_1 , U^-_1\right] \right\rangle &= \sum_a u_{e_a} \star \left[-Z^{\infty}(\varphi,m,\eps)\right] = \left\langle U^+_1 \right\rangle \star \left\langle\left[U^+_1,U^-_1\right]\right\rangle \ ,
\end{aligned}
\end{equation}
where the last equality is obtained using the VEV of the commutator in \eqref{z_inf_result}. 

Similarly, we find that the jump between chambers $+-+$ and $-++$ is 
\begin{equation}
\hspace{-3pt} \left\langle  \left[U^+_1 , U^-_1\right]U^+_1 \right\rangle = \sum_{a} u_{e_a} \left(Z^{+ - +}_a - Z^{- + +}_a\right)= -Z^{\infty}(\varphi,m,\eps) \star \sum_a u_{e_a}  = \left\langle\left[U^+_1,U^-_1\right]\right\rangle\star \left\langle U^+_1 \right\rangle 
\end{equation}
after applying the relation $u_{e_a} \cdot f(\varphi + e_{a}\epsilon) = f(\varphi) \star u_{e_a}$ in \eqref{q_abelian_relations}. Lastly, 
\begin{equation}
\langle  [(U^+_1)^2,U^-_1]\rangle = 
\langle  U^+_1 \left[U^+_1 , U^-_1\right]\rangle + \langle \left[U^+_1 , U^-_1\right]U^+_1 \rangle 
= \sum_{a} u_{e_a} \left(Z^{+ + -}_a- Z^{- + +}_a\right)\ ,
\end{equation}
which involves a sum of two commutators since to obtain $Z^{- + +}_a$ from $Z^{+ + -}_a$ we must cross two codimension-one walls. 

In this discussion we have not mentioned what happens when we set $\xi_1 = 0$ and cross the $\xi_0$ axis. This situation corresponds to reversing the order of two NS5$_+$ branes. Hence, crossing the codimension-one wall given by $\xi_1 = 0$ is related to exchanging the two minimal positive operators in the VEV, which does not change the result since they obviously commute. We can also explain this by looking at the contribution from the appropriate pole at infinity. Crossing the $\xi_0$ axis corresponds to adding the contribution from the pole where $z_0$ is finite and $z_1 \rightarrow \infty$. Evaluating the residue of the integrand in \eqref{Z21N} at this pole, we find zero. 

Consequently, this analysis is consistent with the ordering of the NS5 branes described earlier, which explained how the 6 regions of FI space are actually grouped into only 3 inequivalent chambers, due to the symmetry under the exchange of the two NS5$_+$ branes.  The difference in the ordering of the operators in the vacuum expectation value is related to the signs and relevant magnitudes of the FI parameters. The codimension-one wall-crossing phenomenon is manifest when we exchange two operators of a different type, and the contribution from evaluating the residue of the appropriate pole at infinity is non-zero. The pole at infinity is specified entirely by crossing the codimension-one wall between the relevant chambers. When the contribution from a particular pole at infinity is zero, the VEV is unchanged which corresponds to the exchange of two commuting (and in this case identical) operators. 

\subsubsection{Two positive and two negative minimal monopole operator}

We finish our analysis by studying the VEV of the product of two minimal positive operators and two minimal negative operators, which is an example of a computation where a bubbling term with a non-abelian unitary gauge node emerges. Depending on the order of the operators, we obtain six different results. We require a setup containing two NS5 pairs and we sum over configurations with four D1 strings, where a single string emanates from each of the NS5 branes. 

\begin{figure}[t]
	\centering
	
	\hspace*{-2.5cm}
	\includegraphics[scale=0.8]{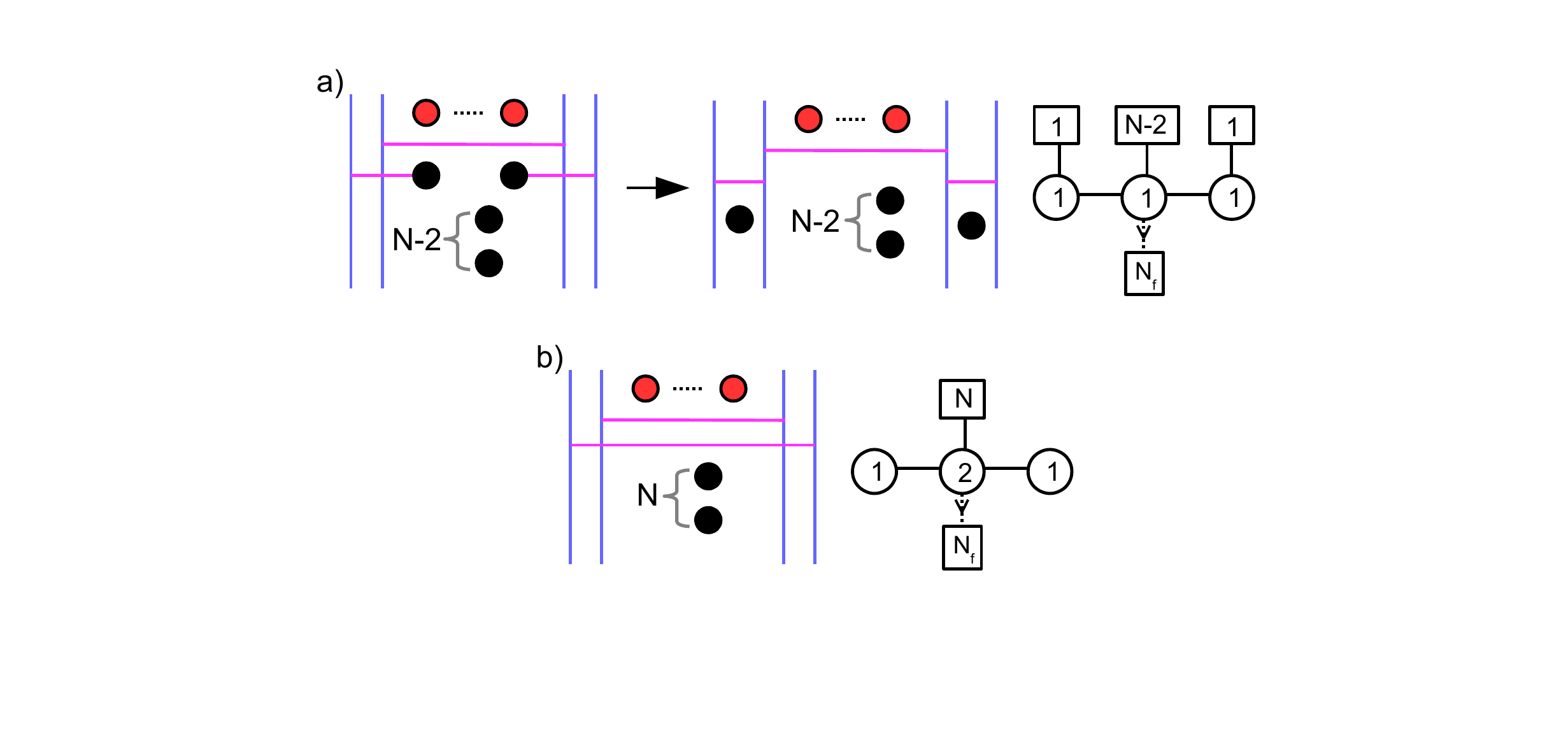}
	\vspace{-3cm}
	\caption{\footnotesize{This figures illustrates the brane configurations and their associated SMMs for the chamber dependent terms that arise in the correlator $\left\langle T \left( (U^+_1)^2 (U^-_1)^2  \right) \right\rangle$ \eqref{VEV_+-+-}.}}
	\label{fig:++--}
\end{figure}

The VEV of the product of these monopole operators is given by
\begin{equation}
\begin{aligned}
\label{VEV_+-+-}
\left\langle T \left( (U^+_1)^2 (U^-_1)^2  \right) \right\rangle &=  \sum_{a,b} u_{2e_a - 2e_b} + \sum_{a,b,c}  u_{2e_a - e_b - e_c} Z^{-}_{bc}+ \sum_{a,b,c} u_{-2e_a + e_b +e_c} Z^{+}_{bc} \\
&+ \sum_{a,b,c,d} u_{e_a + e_b - e_c -e_d} Z_{ab,cd}+ \sum_{a,b} u_{e_a - e_b} Z_{ab}(\xi)+ Z(\xi) \ ,
\end{aligned}
\end{equation}
where  $a,b,c,d = 1, \ldots, N$ and it is understood that all indices that are summed over are different. We have indicated explicitly the SMM partition functions that exhibit a non-trivial dependence on the FI parameters $\xi$ and the brane configurations for these two terms are illustrated in Figure \ref{fig:++--}.

The first term on the RHS in the VEV, which has no monopole bubbling factor, arises from the configurations where one of the D3 branes is connected to both NS5$_+$, another D3 is connected to both NS5$_-$ and $N-2$ D3s remain unconnected.  The second term on the RHS in \eqref{VEV_+-+-} comes from the cases where both of the NS5$_+$ are connected to the same D3, the two NS5$_-$ are connected to different D3s and $N-3$ branes are unconnected. The bubbling factor is computed as the SMM described by the quiver
\begin{equation}\label{TSU2}
\begin{bmatrix}
2 \\
1  
\end{bmatrix} \ .
\end{equation}
The partition function of this theory has already been computed previously and is
\begin{equation}
Z^{-}_{bc} = \oint_{\mathrm{JK}} \frac{d z}{2\pi i}\frac{2\epsilon}{\left[\pm(z-\varphi_b)+\epsilon \right]\left[\pm(z-\varphi_c)+\epsilon \right]} =\frac{2}{\left(\pm \varphi_{bc}+2\epsilon \right)}\ ,
\end{equation}
where $b$ and $c$ are the D3 branes that are connected to the two NS5$_-$ in the construction. The third term in the VEV is given by a left-right mirror configuration, where both of the NS5$_-$ are connected to the same D3, the two NS5$_+$ are connected to different D3s and $N-3$ branes remain unconnected. The bubbling factor is computed as the partition function of SMM described by the same quiver as (\ref{TSU2}). Its partition function is $Z^{+}_{bc}=Z^{-}_{bc}$, 
where now $b$ and $c$ represent the D3 branes that are connected to the two NS5$_+$ branes in the construction. The next term in \eqref{VEV_+-+-} arises from the cases where all of the NS5 branes are connected to different D3s and $N-4$ D3s are unconnected. The resulting bubbling term factorises into a product of the two previous bubbling factors,
\begin{equation}
Z_{ab,cd} = Z^{+}_{ab} Z^{-}_{cd} = \frac{4}{\left(\pm\varphi_{ab}+2\epsilon \right)\left(\pm\varphi_{cd}+2\epsilon\right)} \ ,
\end{equation}
where $a,b \ (c,d)$ denote the D3 branes that are connected to the NS5$_+$ (NS5$_-$).

All of the terms discussed so far are chamber independent. The different orderings of the operators in the VEV are encoded in the last two terms of \eqref{VEV_+-+-}, where the bubbling factors carry FI chamber dependence. The first term in this final line arises from the configurations where a single NS5$_+$ and a single NS5$_-$ are connected, and the other NS5$_+$ and NS5$_-$ are connected to different D3s, with $N-2$ D3s remaining unconnected, see Figure \ref{fig:++--}-a. The bubbling factor is computed as the SMM described by the quiver
\begin{equation}
\begin{bmatrix}
1 & N-2 & 1 \\
1 & \underline{1} & 1 
\end{bmatrix} \ ,
\end{equation}
where again we underline the gauge node that is attached to the Fermi multiplets. The partition function of this theory is given by 
\begin{equation}
\label{Z111}
Z_{ab}(\xi) = \oint_{\mathrm{JK}(\xi)} \frac{dz_{-1}dz_0dz_1}{(2\pi i)^3}\frac{(2\epsilon)^3 P(z_0)}{\Delta(z)} \ ,
\end{equation}
where
\begin{equation}
\hspace{-2pt} \Delta(z) = \left[\pm(z_{-1}-\varphi_b)+\epsilon \right]\left[\pm(z_1-\varphi_a)+\epsilon \right] \prod_{s=\pm 1}\left[\pm(z_0-z_s)+\eps\right]\prod\limits_{c\neq a,b}\left[\pm(z_0-\varphi_c)+\epsilon\right] ~,
\end{equation}
where $a (b)$ denotes the single D3 brane that is connected to the NS5$_+$ (NS5$_-$), namely the D3 brane that is located in the interval between the two NS5$_+$ (NS5$_-$), after applying a Hanany-Witten transition. Finally, there is one configuration where all of the NS5s are connected by the D1 strings and $N$ D3 branes remain unconnected, as shown in Figure \ref{fig:++--}-b. This gives the remaining bubbling factor, which is computed as the SMM described by the quiver
\begin{equation}
\begin{bmatrix}
0 & N & 0 \\
1 & \underline{2} & 1 
\end{bmatrix} \ ,
\end{equation}
whose partition function is given by
\begin{equation}
\label{Z4}
Z(\xi) = \oint_{\mathrm{JK}(\xi)}\frac{dz_{-1}dz_1 \prod\limits_{i=1}^2 dz_{0,i} }{2(2\pi i)^4}\frac{(2\epsilon)^4\left[\pm z_{0,12}\right]\left[\pm z_{0,12}+2\epsilon\right]\prod\limits_{i=1}^2 P(z_{0,i}) }{\prod\limits_{i=1}^2 \left[ \prod\limits_{s=\pm 1} \left[\pm(z_{0,i}-z_s)+\epsilon \right] \prod\limits_{a=1}^N\left[\pm(z_{0,i}-\varphi_a)+\epsilon \right]\right]  } \ .
\end{equation}
All of the $N$ D3 branes are located in the interval between the innermost NS5$_+$ and the innermost NS5$_-$. The explicit details of the computation of the monopole bubbling factors \eqref{Z111} and \eqref{Z4} has been relegated to appendix \ref{sec:bub_factors}.  

The ordering of the NS5 branes affects the signs of the FI parameters, which leads to the different chambers. In this case, there are three FI parameters given by
\begin{equation}
\xi_{-1} = x_{-1} - x_{-2} \ , \  \xi_0 = x_1 - x_{-1} \ , \  \xi_1 = x_2 - x_1 \ ,
\end{equation}
where $x_1 ,x_2$ are the $x^0$ coordinates of the two NS5$_+$ branes and $x_{-1}, x_{-2}$ of the two NS5$_-$ branes. These FI parameters tell us that
\begin{equation}
\xi_{-1} + \xi_{0} = x_1 - x_{-2} \ , \ \xi_{0} + \xi_1 = x_2 - x_{-1} \ , \ \xi_{-1} + \xi_0 + \xi_1 = x_2 - x_{-2} \ .
\end{equation} 

By naively considering the order of these four NS5 branes, one expects to find 24 chambers from the permutations of $x_{-2}, x_{-1}, x_1, x_2$. However, there is a symmetry under the exchange of $x_1$ and $x_2$ and another symmetry under the exchange of $x_{-1}$ and $x_{-2}$. Consequently, there are only 6 distinct chambers, which are given up to the aforementioned permutations by:
\begin{itemize}
	\item $+ + - -$ chamber: $x_{-2} < x_{-1} < x_1 < x_2$.  \item $+ - + -$ chamber: $x_{-2} < x_{1} < x_{-1} < x_2$.
	\item $+ - - +$ chamber: $x_{1} < x_{-2} < x_{-1} < x_2$.
	\item $- - + +$ chamber: $x_{1} < x_{2} < x_{-2} < x_{-1}$.
	\item $- + - +$ chamber: $x_{1} < x_{-2} < x_{2} < x_{-1}$.
	\item $- + + -$ chamber:  $x_{-2} < x_{1} < x_{2} < x_{-1}$.
\end{itemize}

For the chamber dependent SMM partition functions $Z_{ab}(\xi)$ and $Z(\xi)$ in (\ref{VEV_+-+-}), we will denote the chamber in which the partition function is computed by a superscript, \emph{e.g.} $Z^{++--}_{ab}$ is $Z_{ab}$ evaluated in the $++--$ chamber.  

We conclude our analysis by studying one example of wall-crossing between the results for (\ref{VEV_+-+-}) computed in two adjacent chambers, $++--$ and $+-+-$. Then 
\begin{equation}
\left\langle U^+_1 \left[U^+_1,U^-_1\right]U^-_1\right\rangle = \sum_{a \neq b} u_{e_a - e_b} \left(Z^{+ + - -}_{ab} - Z^{+ - + -}_{ab}\right) + \left(Z^{+ + - -} - Z^{+ - + -} \right) \ .
\end{equation}
Firstly, to obtain $Z^{+ - + -}_{ab}$ from $Z^{+ + - -}_{ab}$ we cross the plane $\xi_0=0$, which corresponds to adding the contribution from the pole where $z_1$ and $z_{-1}$ are finite and $z_0 \rightarrow \infty$, 
\begin{equation}
Z^{+ + - -}_{ab} - Z^{+ - + -}_{ab} =-\Res_{z_0=\infty}~ \Res_{z_1=\varphi_a - \epsilon}~ \Res_{z_{-1}=\varphi_b-\epsilon} I_{ab} \ ,
\end{equation}
where $I_{ab}$ is the integrand in \eqref{Z111}. Analogously, to obtain $Z^{+ - + -}$ from $Z^{+ + - -}$ we must also cross $\xi_0=0$. Taking into account both the gauge and flavour symmetry, we find
\begin{equation}
\begin{aligned}
Z^{+ + - -} - Z^{+ - + -}  &= -\sum_{a=1}^N \left(\Res_{z_{0,1}=\infty}~ \Res_{z_{-1} = \varphi_a - 2\epsilon}~ \Res_{z_1=\varphi_a-2\epsilon}~ \Res_{z_{0,2}=\varphi_a-\epsilon} I + \left(z_{0,1}\leftrightarrow z_{0,2} \right) \right) \ ,
\end{aligned}
\end{equation}
where $I$ is the integrand in \eqref{Z4}. Due to the gauge symmetry, we find
\begin{equation}
\Res_{z_{0,1}=\infty}~ \Res_{z_{-1} = \varphi_a - 2\epsilon}~ \Res_{z_1=\varphi_a-2\epsilon}~ \Res_{z_{0,2}=\varphi_a-\epsilon} I = \Res_{z_{0,2}=\infty}~ \Res_{z_{-1} = \varphi_a - 2\epsilon}~ \Res_{z_1=\varphi_a-2\epsilon}~ \Res_{z_{0,1}=\varphi_a-\epsilon} I \ .
\end{equation}

By evaluating the residues of the poles not at infinity and using our definition of $Z^{\infty}$ in \eqref{z_inf}, it is possible to write 
\begin{equation}
\begin{aligned}
\left\langle U^+_1 \left[U^+_1,U^-_1\right] U^-_1\right\rangle &=-\sum_{a \neq b}^N u_{e_a - e_b} Z^{\infty}(\varphi-(e_a+e_b)\eps,m,\eps)\\  
& -\sum_{a=1}^N Z^{\infty}(\varphi-2e_a\eps,m,\eps) \frac{(-1)^{N-1}\prod\limits_{k=1}^{N_f}\left[\varphi_a-m_k-\epsilon\right]}{\prod\limits_{b\neq a}\varphi_{ab}\left(\varphi_{ab}-2\epsilon\right)}  \ .
\end{aligned}
\end{equation}
Comparing with \eqref{starproducts}, we see that the final term in bracket is just $u_{e_a} \star u_{-e_a}$ and using our quantized abelian relations from section \ref{sec:Star product as a Moyal product and abelian relations} along with (\ref{z_inf_result}), it is easy to see that
\begin{equation}
\left\langle U^+_1 \left[U^+_1,U^-_1\right] U^-_1\right\rangle = \left\langle U^+_1\right\rangle \star \left\langle \left[ U^+_1,U^-_1\right] \right\rangle \star \left\langle U^-_1\right\rangle \ ,
\end{equation}
showing agreement with the result expected from the star product.

%%%%%%%%%%%%%%%%%%%%%%%

\section{Casimir and Dressed Monopole Operators}
\label{sec:CasimirAndDressedMonop}

In this section we discuss the brane realisation of Casimir  and dressed monopole operators in the $U(N)$ SQCD theory and the evaluation of correlators involving them.

\subsection{Brane realisations} \label{sec:brane_realisations}

The realisation of Casimir operators in the type IIB setup is achieved by adding D3' branes oriented as descibed in Table \ref{tab:orientations}, along $x^{4567}$. 
To be more precise, there will be D3' branes stretched between pairs of NS5 branes with a fixed number of F1 strings (fundamental strings) attached. This corresponds to an electric charge in the D3' worldvolume theory that can be measured at spatial infinity in $x^{456}$. In the following, we may omit mentioning the pair of NS5 branes that (always) accompany a given D3' brane. Unless otherwise stated explicitly, we always insert the D3' at $x^8=x^9=0$.

For the $U(N)$ SQCD theory, we propose that a single D3' brane  with $n$ units of electric flux realises the insertion of the Casimir operator  
\be
\label{casimir}
\Phi_n := \sum_{a_1 < a_2 < \ldots < a_n} \varphi_{a_1}\varphi_{a_2}\ldots\varphi_{a_n}  \quad \longleftrightarrow \quad 
\text{D3' brane with $n$ units of flux,}
\ee
where we assume $n \le N$. 

\begin{figure}[t]
	\centering
	\vspace{-.5cm}
	\includegraphics[scale=1]{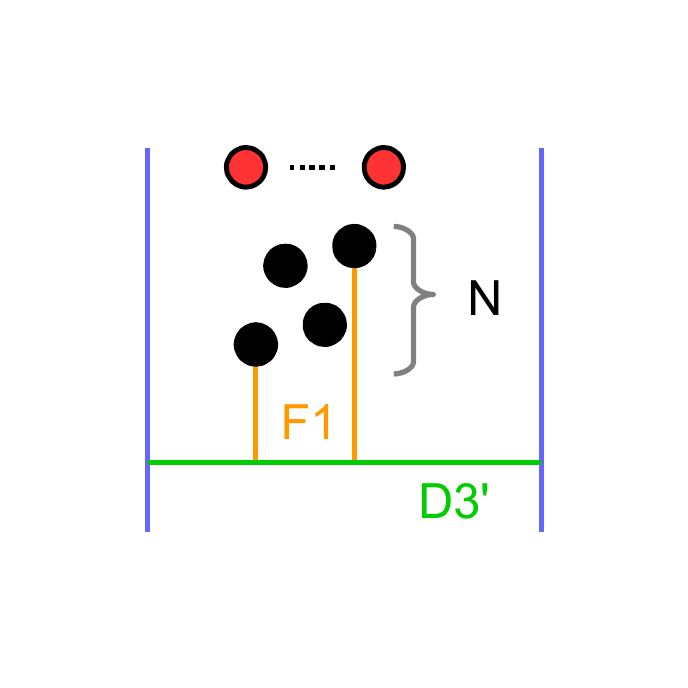}
	\vspace{-1cm}
	\caption{\footnotesize{Brane setup realising the insertion of the Casimir operator $\Phi_2 = \sum\limits_{a < b} \varphi_a \varphi_b$, where the D3' brane has two F1 strings attached. Only one pattern of strings is shown here.}}
	\label{CasimirOp}
\end{figure}

This is understood as follows. The $n$ units of flux correspond to $n$ F1 strings emanating from the D3' brane. These strings must end on the $N$ D3 branes realising the SQCD theory, however, because of the s-rule (the D3'-D3-F1 system is of Hanany-Witten type), the $n$ F1 strings must end on different D3 branes. The lowest excitation on such a string is a complex fermion of mass $\varphi_{a_i}$, where $a_i$ labels the D3 brane with the $i$-th string attached. Integrating out this fermion leads to the insertion of $\varphi_{a_i}$. Summing over all the possible patterns of F1s ending on D3s leads to the expression for $\Phi_n$ given above. The antisymmetric index structure in (\ref{casimir}) is a consequence of the s-rule of \cite{Hanany:1996ie} (or Pauli exclusion). The situation when $n=2$ is illustrated in Figure \ref{CasimirOp}. 

Note that the Casimir operators $\Phi_n$ with $1 \le n \le N$ generate the whole ring of Casimir operators. If $n>N$, the operator $\Phi_n$ vanishes by antisymmetry. Correspondingly, inserting a D3' brane with $n>N$ units of flux leads to a violation of the s-rule.

\medskip

For dressed monopole operators, it is natural to propose that they are realised in the brane picture with both D1 strings emanating from NS5 branes and F1 strings emanating from D3' branes.

We will only provide the brane realisation of ``minimal" dressed monopoles in the $U(N)$ SQCD theory, which are monopoles with magnetic charges generating the chambers in the magnetic charge lattice and dressed with polynomials in $\varphi_a$ of degree less than $N$. They form a natural extension of the bare monopole and Casimir operators discussed so far.
Concretely we define, for $0 \le m + n  \le N$,
\bea\label{dressed_mon}
U^+_{m,n} &:= V_{(1^m,0^{N-m}),P_n(\varphi)} = \sum_{a_1 < \ldots < a_m} u_{e_{a_1} + \ldots + e_{a_m}} \sum\limits_{b_1 < \ldots < b_n \atop b_i \neq a_j } \varphi_{b_1} \ldots \varphi_{b_n} \,, \cr
U^-_{m,n} &:= V_{(0^{N-m},(-1)^m),P_n(\varphi)} = \sum_{a_1 < \ldots < a_m} u_{-e_{a_1} - \ldots - e_{a_m}} \sum\limits_{b_1 < \ldots < b_n \atop b_i \neq a_j} \varphi_{b_1} \ldots \varphi_{b_n} \,.
\eea
Note that $U^\pm_{m,0} = U^\pm_{m}$, with $m>0$, are the bare monopoles defined in \eqref{monop_generators} and $U^\pm_{0,n} = \Phi_n$, with $n >0$, are the Casimir operators discussed above (by convention $U^\pm_{0,0} = 1$).

\begin{figure}[t]
	\centering
	\vspace{-.5cm}
	\includegraphics[scale=1]{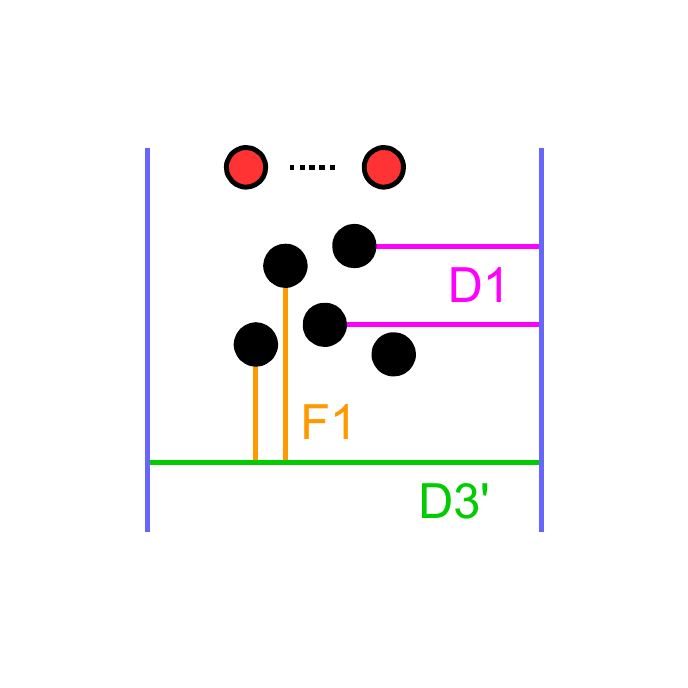}
	\vspace{-1cm}
	\caption{\footnotesize{Brane setup realising the insertion of the dressed monopole $U^+_{2,2}$, with two D1s emanating from the NS5$_+$ and two F1s emanating from the D3'. Only one pattern of strings is shown here, with strings (D1 and F1) ending on different D3s.}}
	\label{DressedMonop}
\end{figure}

We propose that the insertion of the dressed monopole $U^\pm_{m,n}$ is realised by the configuration containing a pair of NS5 branes with $m$ D1 strings emanating from the NS5$_\pm$ and with a D3' brane stretched between the same two NS5s, with $n$ units of flux. Schematically, we have
\be\label{dressed_brane}
U^\pm_{m,n} \quad \longleftrightarrow \quad 
\text{NS5 pair with a stretched D3' + $m$ D1s from NS5$_\pm$ + $n$ F1s from  D3'.}
\ee
We provide an example in Figure \ref{DressedMonop}.

The argument for this proposal is the following. Because of the s-rule, the D1 strings emanating from the NS5$_\pm$ must end on different D3s, so each configuration is associated to an abelian magnetic charge $\pm(e_{a_1} + \ldots + e_{a_m})$, with $a_i$ all different. In such a setup, all of the D3s with a D1 string attached may be pulled out of the central region, crossing the NS5$_\pm$ and annihilating the D1s. There are $N-m$ D3s remaining in the central region, on which the $n$ F1 strings emanating from the D3' can terminate. Again each F1 must end on a different D3 (so that we need $n \le N-m$). The resulting setup is associated to the dressed abelian monopole
\begin{equation}
u_{\pm(e_{a_1} + \ldots + e_{a_m})} \sum\limits_{b_1 < \cdots < b_n \atop b_i \neq a_j} \varphi_{b_1} \ldots \varphi_{b_n} \ .
\end{equation} 
Summing over all the possible patterns of strings one obtains the expression for $U^{\pm}_{m,n}$.
Importantly, from this discussion, we understand that the F1 and D1 strings are not allowed to end on the same D3 brane (although this was not a priori obvious).

\medskip 

	The discussion so far has involved a single D3' brane located at the origin in $x^{89}$. If we instead consider the D3' inserted at a generic position $x^8+ix^9=w$, then the LHS of \eqref{casimir} becomes
	\be\label{Casimir_gen}
	\Phi_n(w) := \sum_{a_1 < a_2 < \ldots < a_n}( \varphi_{a_1}-w)(\varphi_{a_2}-w)\ldots(\varphi_{a_n}-w)  \ .
	\ee
	This defines polynomials in $w$ valued in $\cC_\epsilon[\cC]$. In particular, the coefficients of the generating polynomial $\Phi_N(w)$ are the Casimir operators $\Phi_n$. 
	
	Similarly, if the D3' in (\ref{dressed_brane}) is placed at $x^8+ix^9=w$, then  (\ref{dressed_mon}) becomes
	\bea\label{monopole_gen}
	U^+_{m,n}(w) &:=  \sum_{a_1 < \ldots < a_m} u_{e_{a_1} + \ldots + e_{a_m}} \sum\limits_{b_1 < \ldots < b_n \atop b_i \neq a_j } (\varphi_{b_1}-w) \ldots (\varphi_{b_n}-w) \,, \cr
	U^-_{m,n}(w) &:=  \sum_{a_1 < \ldots < a_m} u_{-e_{a_1} - \ldots - e_{a_m}} \sum\limits_{b_1 < \ldots < b_n \atop b_i \neq a_j} (\varphi_{b_1}-w) \ldots (\varphi_{b_n}-w) ~.
	\eea
	We can then package all the bare and dressed monopole operators $U^\pm_{m,n}$ as coefficients of two generating polynomials $U^\pm_{m,N}(w)$. 
	
	We will not use these generating polynomials any further, but computing correlators of such generating polynomial operators may prove very useful in finding Coulomb branch relations efficiently, as shown in \cite{Bullimore:2015lsa,Assel:2017jgo,Assel:2018exy}.

\subsection{Correlators in $U(N)$ SQCD}

We extend here the computations of the previous sections to correlators involving Casimir operators and dressed monopoles, in simple examples. We will show that our computations are compatible with the Moyal product representation of the star product on the quantized Coulomb branch.

\medskip

\begin{figure}[t]
	\centering
	\vspace{-.5cm} \hspace{5pt}
	\includegraphics[scale=1]{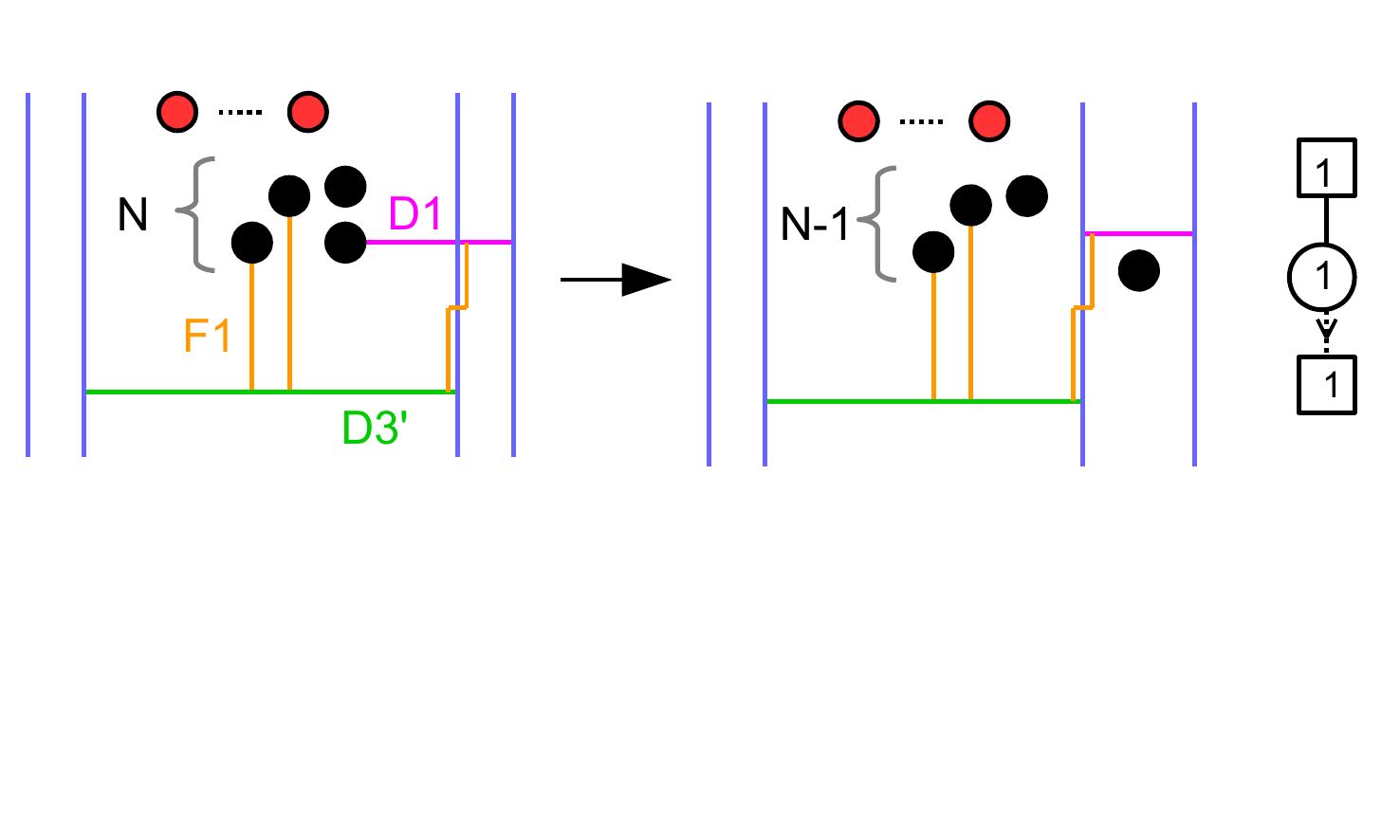}
	\vspace{-3.8cm}
	\caption{\footnotesize{A string pattern contributing to the correlator $\vev{U^+_1\Phi_n}$ (here $n=3$). The figure on the right (after Hanany-Witten move) allows us to read the SMM on the D1 segment.}}
	\label{CasMonopCorrel}
\end{figure}

\noindent\underline{$\vev{U^+_1\Phi_n }$}:

First we consider the correlator $\vev{U^+_1\Phi_n}$ between a Casimir operator $\Phi_n$ and the bare monopole operator $U^+_1 := U^+_{1,0}$. The brane configuration realising this correlator has two pairs of NS5s, one for each operator insertion. Here we have several choices of ordering along $x^7$ for the NS5 branes. We will consider the inner pair as associated with the $\Phi_n$ insertion and the outer pair to be associated with the $U^+_1$ insertion. In this case there is a D3' stretched between the inner pair of NS5s, with $n$ F1 strings attached, and one D1 string emanating from the outer NS5$_+$. This is the setup of Figure \ref{CasMonopCorrel}.

The D1 string can end on a D3 brane -- this gives $N$ possibilities -- which can be moved from the central region to the region between the two NS5$_+$ (see Figure \ref{CasMonopCorrel}). This configuration allows for patterns of F1 strings ending on either the remaining $N-1$ D3s in the central region or ending on the remaining D1 segment that is stretched between the two NS5$_+$. Because of the s-rule we only have two sectors: either all of the $n$ F1 strings end on the D3s or $n-1$ of them end on the D3s and the remaining string ends on the D1 segment, so that it is stretched across the inner NS5$_+$. For the case $n=N$, only the latter sector is possible, since there are only $N-1$ D3s in the central region.

The abelian magnetic flux is $v = e_a$, where $a$ labels the D3 brane that has been moved in-between the NS5$_+$ branes. In the sector where all of the F1s end on D3s, the dressing factor is $\varphi_{b_1}\varphi_{b_2}\ldots \varphi_{b_n}$ with $b_1 < b_2 < \ldots < b_n$ and $b_i \neq a$, where the $\varphi_{b_i}$ factors arise from the D3'-D3 fermion modes. In the sector with a F1 string stretched across the inner NS5$_+$, the dressing factor is $\varphi_{b_1}\varphi_{b_2}\cdots \varphi_{b_{n-1}} Z_{\rm SMM}(\varphi_a)$ with $b_1 < b_2 < \ldots < b_{n-1}$ and $ b_i \neq a$, where $Z_{\rm SMM}(\varphi_a)$ is the SMM living on the D1 segment. 
In this case the SMM has a $U(1)$ gauge group, one fundamental hypermultiplet of mass $\varphi_a$ from the D3-D1 lowest excitations and one fundamental Fermi multiplet of zero mass, which is the D3'-D1 lowest excitation (from strings stretched across the left NS5$_+$).%
\footnote{We deduce that this excitation is a Fermi multiplet by thinking of the same configuration but with the D1 segment ending on a (new) D3 instead of the right NS5$_+$. Moving the D3 past the left NS5$_+$, the D1 is annihilated and the lowest string mode corresponds to the D3'-D3 excitation, which is a Fermi multiplet.}
The SMM is given by the quiver in Figure \ref{CasMonopCorrel}, which we encode in matrix notation by
\be
\begin{bmatrix}
	1 \\
	1 \\
	1
\end{bmatrix} \ ,
\ee
where now the top line denotes the number of fundamental hypers, the middle line denotes the rank of the gauge group and the bottom line denotes the number of fundamental Fermi multiplets. We introduced the bottom line for future convenience, departing from the matrix notation (\ref{matrix_notation_2}). This is nothing but the $Z_3$ matrix model, with $N_f=1$ and vanishing Fermi mass, computed in \eqref{Z3} for $\xi >0$.

We thus obtain
\bea
\vev{U^+_1\Phi_n} &= \sum_{a=1}^N u_{e_a} \Big[\sum\limits_{b_1 < \ldots < b_n \atop b_i \neq a} \varphi_{b_1}\ldots \varphi_{b_n} + \sum\limits_{b_1 <\ldots < b_{n-1} \atop b_i \neq a} \varphi_{b_1}\ldots \varphi_{b_{n-1}} Z_3(\varphi_a, N_f=1, \xi > 0) \Big] \cr
&= \sum_{a=1}^N u_{e_a} \Big[\sum\limits_{b_1 < \ldots < b_n \atop b_i \neq a} b_1\ldots b_n + \sum\limits_{b_1 < \ldots < b_{n-1} \atop b_i \neq a} \varphi_{b_1}\ldots \varphi_{b_{n-1}} (\varphi_a -\epsilon)\Big]  \cr 
& = \sum_{a=1}^N u_{e_a} \sum\limits_{b_1 < \ldots <  b_n} \varphi_{b_1} \ldots \varphi_{b_n} \left.\right|_{\varphi_a \rightarrow \varphi_a - \epsilon}                   \ .
\eea
Reversing the order of the insertions along $x^0$ affects the sign of the FI parameter in $Z_3$ and the computation leads to
\bea
\vev{\Phi_n U^+_1} &=   \sum_{a=1}^N u_{e_a} \sum\limits_{ b_1 < \ldots < b_n} \varphi_{b_1} \ldots \varphi_{b_n} \left.\right|_{\varphi_a \rightarrow \varphi_a + \epsilon}                \ . 
\eea
This agrees very nicely with the Moyal product formula \eqref{starprod} applied to these two Coulomb branch operators.

The right hand side of these correlators can be recognised as a linear combination of the dressed monopoles $U^+_{1,n}$, $U^+_{1,n-1}$ and another dressed monopole  that is not in the basis discussed so far (it is generated in these products).
The commutator is simply expressed as
\be
[U^+_1,\Phi_n] = -2\epsilon U^+_{1,n-1} \,.
\ee

\bigskip

\noindent\underline{$\vev{U^+_m\Phi_n }$}:

The $\left\langle U^+_1 \Phi_n \right\rangle$ computation generalises to the case $\left\langle U^+_m \Phi_n\right\rangle$, where we recall that $U_m^+ := U^+_{m,0}$. The brane configuration realising this correlator again has two pairs of NS5s and we choose the inner pair to be associated with the $\Phi_n$ insertion and the outer pair as associated with the $U^+_m$ insertion. This setup is identical to the one in Figure \ref{CasMonopCorrel}, except now we have $m$ D1 strings emanating from the outermost NS5$_+$. The correlator  $\left\langle U^+_m \Phi_n\right\rangle$ is computed in an identical fashion to  $\left\langle U^+_1 \Phi_n\right\rangle$ and, skipping the explicit details of the brane configurations, we find
\begin{equation}
\left\langle U^+_m \Phi_n\right\rangle = \sum\limits_{a_1 < \ldots < a_m} u_{e_{a_1} + \ldots e_{a_m}} \left[\sum\limits_{j=0}^m \sum\limits_{b_1 < \ldots < b_{n-j} \atop b_i \neq a_k} \varphi_{b_1}\ldots \varphi_{b_{n-j}} Z_{m,j} \right] \ .
\end{equation}
The bubbling term $Z_{m,j}$ is given by
\begin{equation}
\begin{bmatrix}
m \\
m \\
\bigwedge^j  
\end{bmatrix} \ ,
\end{equation}
where the Fermi multiplets transform in a single $j$-th antisymmetric tensor power of the fundamental representation, $\bigwedge^j$, with $j \in [0,m]$. The partition function of the SMM for this bubbling term is
\begin{equation}
Z_{m,j} = \oint_{\mathrm{JK}(\xi)} \frac{\prod\limits_{i=1}^m dz_i}{(2\pi i)^m} \frac{(2\epsilon)^m}{m!}\frac{\prod\limits_{1 \leq i < k \leq m}\left[\pm z_{ik}\left(\pm z_{ik}+2\epsilon\right) \right]}{\prod\limits_{i,k=1}^m \left[\pm(z_i-\varphi_{a_k}) + \epsilon \right]} \sum\limits_{c_1 < \ldots < c_j}^m z_{c_1}\ldots z_{c_j} \ .
\end{equation}
The poles for this integral, in the positive FI chamber, are given by $z_1 = \varphi_{a_1} - \epsilon, z_2 = \varphi_{a_2} - \epsilon , \ldots , z_m = \varphi_{a_m}-\epsilon$ plus permutations of the mass parameters for the $U(m)$ gauge node. Performing the JK integral we find
\begin{equation}
Z_{m,j} = \sum\limits_{c_1 < \ldots < c_j}^m \left(\varphi_{c_1}-\epsilon \right)\ldots \left(\varphi_{c_j}-\epsilon \right)  \ ,
\end{equation}
and combining everything we obtain the general result
\begin{equation}
\left\langle U^+_m \Phi_n\right\rangle = \sum\limits_{a_1 < \ldots < a_m} u_{e_{a_1} + \ldots e_{a_m}}\sum\limits_{b_1 < \ldots < b_n} \varphi_{b_1}\ldots \varphi_{b_n} \left.\right|_{\varphi_{a_i}\rightarrow \varphi_{a_i}-\epsilon} \ ,
\end{equation}
which agrees with the star product computation. Sending $\epsilon \to -\epsilon$, we obtain the result for the correlator $\left\langle  \Phi_n U^+_m \right\rangle$, where the order of the operators has been reversed.

\bigskip

\noindent\underline{$\vev{U^+_{1,1} U^+_{1,0}}$}:

Finally, we study the correlator $\vev{U^+_{1,1} U^+_{1,0}}$ between the dressed minimal positive operator $U^+_{1,1}$ and the bare monopole operator $U^+_{1,0}$ for $N=2$. Explicit computation using the star product tells us that
\begin{equation}
\begin{aligned}
\vev{U^+_{1,1} U^+_{1,0}} = U^+_{1,1}\star U^+_{1,0} = u_{2e_1}\varphi_2 + u_{2e_2}\varphi_1 - u_{e_1+e_2}\left(\frac{(\varphi_2+\epsilon)}{\varphi_{12}\left(\varphi_{12}-2\epsilon \right)}+\frac{(\varphi_1+\epsilon)}{\varphi_{12}\left(\varphi_{12}+2\epsilon\right)} \right) \ , \\
\vev{U^+_{1,0} U^+_{1,1}} = U^+_{1,0}\star U^+_{1,1} = u_{2e_1}\varphi_2 + u_{2e_2}\varphi_1 - u_{e_1+e_2}\left(\frac{(\varphi_2-\epsilon)}{\varphi_{12}\left(\varphi_{12}+2\epsilon \right)}+\frac{(\varphi_1-\epsilon)}{\varphi_{12}\left(\varphi_{12}-2\epsilon\right)} \right) \ .
\end{aligned}
\end{equation}

To compute these correlators using the brane construction, we require two pairs of NS5 branes. There are four different options for the NS5$_+$ and NS5$_-$ pair that the D3' (with one F1 string attached) stretches between: inner NS5$_-$-inner NS5$_+$, inner NS5$_-$-outer NS5$_+$, outer NS5$_-$-inner NS5$_+$, and outer NS5$_-$-outer NS5$_+$. The final result is the same regardless of the option that we choose so we will pick the inner-inner case and leave it to the enthusiastic reader to check the validity of the other options. 

\begin{figure}[t]
	\centering
	\vspace{-.5cm}
	\includegraphics[scale=0.8]{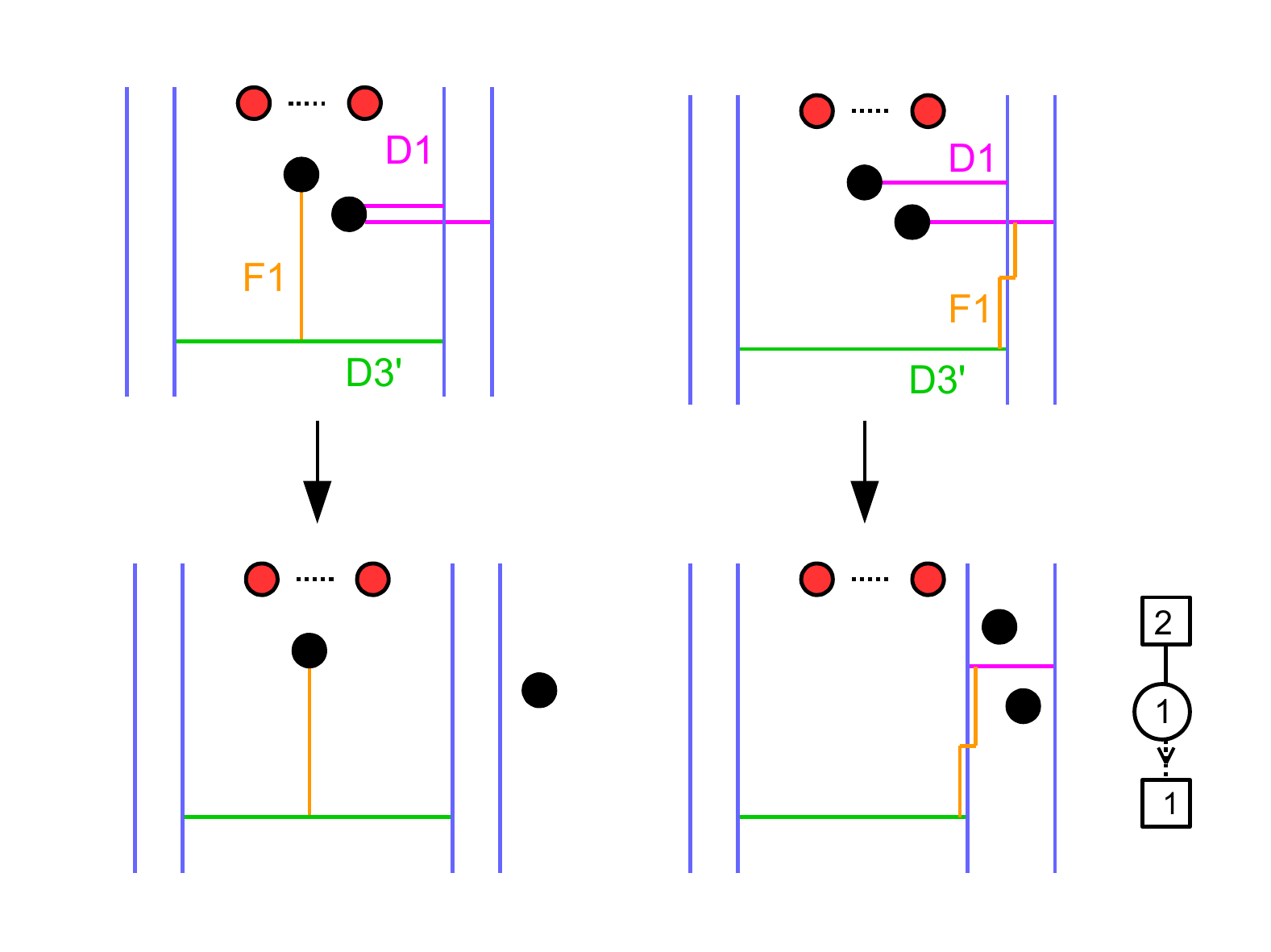}
	\vspace{-1cm}
	\caption{\footnotesize{Two brane setups contributing to $\vev{U^+_{1,1} U^+_{1,0}}$.}}
	\label{DressedCorrel}
\end{figure}

For our case, we have a D1 string emanating from each of the NS5$_+$, the D3' is stretched between the inner NS5$_-$ and the inner NS5$_+$ with one F1 string attached and two D3 branes are located in this interval. There are three configurations that contribute to the VEV: two configurations with both the D1s connected to the same D3 and one configuration where the D1s connect to different D3s, as shown in Figure \ref{DressedCorrel}. This tells us that
\begin{equation}
\left\langle T\left(U^+_{1,1} U^+_{1,0} \right) \right\rangle = u_{2e_1}\varphi_2 + u_{2e_2} \varphi_1 + u_{e_1+e_2} Z_2(\varphi,N_f=1;\xi)
\end{equation}
where $Z_2$ is the bubbling term computed in \eqref{Z2}, with $N_f=1$ and vanishing Fermi mass, whose SMM is given by
\begin{equation}
\begin{bmatrix}
2 \\
1 \\
1 
\end{bmatrix} \ .
\end{equation}
The FI parameter is given by the difference in the $x^0$ positions of the $U^+_{1,0}$ NS5 pair and the $U^+_{1,1}$ pair, $\xi=x^0_{U^+_{1,0}} - x^0_{U^+_{1,1}}$. For $x^0_{U^+_{1,0}} > x^0_{U^+_{1,1}}$ we compute $Z_2$ in the positive FI chamber and match the result for $\vev{U^+_{1,0} U^+_{1,1}}$. On the other hand, computing the bubbling term in the negative FI chamber corresponds to the scenario $x^0_{U^+_{1,0}} < x^0_{U^+_{1,1}}$ and we match the star product result for $\vev{U^+_{1,1} U^+_{1,0}}$. 

\bigskip

So far we have found perfect agreement between star product computations and the brane construction. However, we begin to run into issues when we consider combinations of positive and negative dressed and bare monopole operators. 

\begin{figure}[t]
	\centering
	\vspace{-.5cm}
	\includegraphics[scale=1]{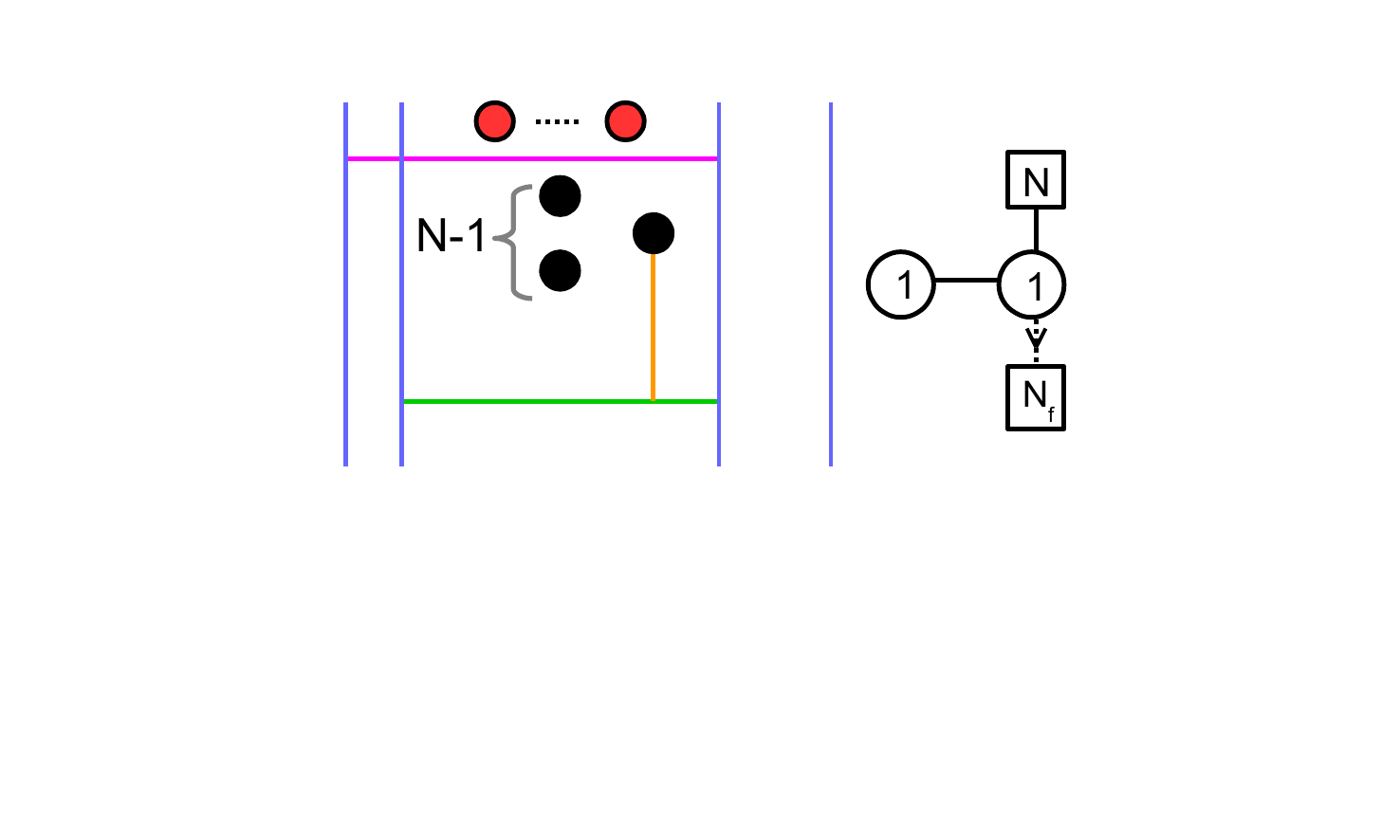}
	\vspace{-3.8cm}
	\caption{\footnotesize{One of the brane setups contributing to the correlator $\left\langle U^+_{1,1} U^-_{1,0}\right\rangle$. The partition function of the SMM given in the quiver diagram is an example of an on-the-wall partition function, which cannot be computed using the JK prescription, since the inner NS5$_-$ and inner NS5$_+$ are connected by the D3' and thus the FI parameter for the central $U(1)$ gauge node is zero.}}
	\label{dressed}
\end{figure}

For example, consider the correlator $\left\langle U^+_{1,1} U^-_{1,0}\right\rangle$, containing a dressed minimal positive operator $U^+_{1,1}$ and a negatively charged bare monopole operator $U^-_{1,0}$, with a D3' (with one F1 attached) stretched between the inner NS5$_-$ and the inner NS5$_+$, a D1 string emanating from the inner NS5$_+$ and the outer NS5$_-$, and $N$ D3 branes. One of the configurations contributing to this VEV is given by the case where the D1s are reconnected and the F1 string is attached to one of the D3s, see Figure \ref{dressed}.
This configuration is dressed by a bubbling term, whose SMM is 
\begin{equation}
\begin{bmatrix}
0 & N  \\
1 & \underline{1} 
\end{bmatrix} \ .
\end{equation}
The inner NS5$_-$ and NS5$_+$ are connected by the D3', which means that they share the same $x^0$ position and the corresponding FI parameter for the central $U(1)$ gauge node is zero. Consequently, the partition function for this SMM cannot be computed using the JK residue prescription, since it is an example of an on-the-wall partition function. This situation arises generically, pointing again to the missing ingredient in our construction: the computation of $Z_{\rm SMM}$ on the FI walls.

The computation of this correlator using the star product  hints that something is missing in the previous analysis. Perhaps the missing ingredients could come from the 3d theory living on the D3' branes, which we have neglected so far according to the (naive?) logic that higher-dimensional modes are frozen in the eyes of a lower-dimensional observer. This deserves further investigation.

%%%%%%%%%%%%%%%%%%%%%%%

\section{Discussion}
\label{sec:Discussion}

In this paper we have proposed a method to compute correlators of monopole (and Casimir) operators, based on localization results and the brane realisation of these operators in type IIB string theory. This method has various limitations. In particular, it requires a brane realisation of the 3d gauge theory, which is only available for quiver-like theories with classical gauge group factors. Even with a brane realisation, the method may become tedious for long quivers. One may argue that for complicated theories, any method will necessarily be tedious anyway.
On the other hand, for simple theories, like the SQCD theory that we have focused on, our method has several advantages. First, it is a direct approach, computing correlators from supersymmetric localization, so it does not rely on assumptions, unlike other methods. It can also be applied for arbitrary gauge group rank. In addition, it provides evidence for the Moyal product realisation of the non-commutative product. Finally, most importantly for us, it makes the physics of the results transparent.

 Nevertheless there is still quite some work left to be done to complete this work. The main point is the computation of the matrix models $Z_{\rm SMM}$ at zero FI parameters, which is the key to the final expressions for generic monopole VEVs and correlators. This is a technical task that we believe can be accomplished in the near future. 
Another issue, to complete nicely the picture that we have developed, would be to resolve the tension that we observe in the evaluation of dressed monopole correlators using different brane setups, as discussed at the end of section \ref{sec:CasimirAndDressedMonop}.
 
This work can be extended to study the quantized Coulomb branch of various 3d gauge theories (with brane realisations). Furthermore, these exact results for correlators of Coulomb branch operators should be used to address various physics questions, in particular how to determine the quantized monopole relations and the precise map of operators under mirror symmetry. In section \ref{sec:brane_realisations} of this paper we have identified the brane realisation of the generating polynomials of monopole operators and Casimir operators, which feature prominently in the algebraic definition of the Coulomb branch of $U(N)$ SQCD \cite{Bullimore:2015lsa}. It would be desirable to determine the relations that the generating polynomials satisfy and their map under mirror symmetry directly from the brane construction, as was achieved in the context of abelian theories in \cite{Assel:2017hck}. This could then be generalised to a large class of quiver gauge theories which can be engineered using Hanany-Witten brane setups. 

Finally, while much is known on the Coulomb branch of 3d $\cN=4$ theories thanks to the recent mathematical progress, a lot remains to be understood about moduli spaces of vacua, chiral rings and their quantization in the context of non-abelian gauge theories with $\cN=3$ or $\cN=2$ supersymmetry (see \cite{Cremonesi2015,Hwang:2017kmk,Cremonesi:2016nbo} for some partial results using Hilbert series). One of the main motivations for our work was that Hanany-Witten brane constructions can be easily generalised to configurations with lower supersymmetry \cite{Bergman:1999na}, so it is natural to expect that our approach to the computation of correlation functions of chiral operators can lead to new exact results for three-dimensional theories with $\cN=3$ or $\cN=2$ supersymmetry. We hope to address some of these questions in the future.

\section*{Acknowledgements}

We thank Patrick Dorey, Daniele Dorigoni, I\~naki Garc\'ia Etxebarria, Philip Glass and Antonio Sciarappa for useful discussions. 
SC thanks the Galileo Galilei Institute in Florence and the organisers of the workshop `Supersymmetric Quantum Field Theories in the Non-perturbative Regime' for hospitality during the preliminary stages of this work. MR is funded by an EPSRC PhD scholarship.

%%%%%%%%%%%%%%%%%

\appendix

\section{SMM partition functions for trivial $\sigma$}\label{sec:trivial_sigma}

In this appendix we compute explicitly the partition functions of the zero-dimensional $T[SU(n)]$, $T_\rho [SU(n)]$ and $T_{\rho,L}[SU(n)]$  matrix models. We show that the former two are chamber independent, whereas the latter is not.

\subsection{$T[SU(n)]$} \label{sec:TSUn}

Consider first the $\cN=2^*$ $T[SU(n)]$ matrix model, the linear balanced quiver with $SU(n)$ flavour symmetry, which is described in matrix notation by 
\begin{equation}\label{matrix_TSUn}
\begin{bmatrix}
n & 0 & 0 & \dots& 0 & 0 & 0  \\
n-1 & n-2 & n-3 & \dots & 3 & 2 & 1  \\ 
\end{bmatrix}\ ,
\end{equation}
starting from node 1 on the left. The partition function of this theory is given by
\bea\label{TSUN_integ}
&Z_{T[SU(n)]}(\varphi,\eps;\xi) = \oint_{\mathrm{JK}(\xi)} \prod_{k=1}^{n-1} \Bigg[ \frac{d^{n-k} z_k}{(2\pi i)^{n-k} (n-k)!} \frac{(2\eps)^{n-k} \prod\limits_{i\neq j}^{n-k} z_{k,ij}(z_{k,ij}+ 2\eps)}{\prod\limits_{i=1}^{n-k}\prod\limits_{j=1}^{n-k+1}(\pm (z_{k,i}-z_{k-1,j})+\eps)}\Bigg]\\
&\qquad= \oint_{\mathrm{JK}(\xi)} \prod_{k=1}^{n-1} \Bigg[ \frac{d^{n-k} (2\eps z_k)}{(2\pi i)^{n-k} (n-k)!} \prod\limits_{1\le i< j\le n-k} \hspace{-5pt} W(z_{k,ij})\cdot \prod\limits_{i=1}^{n-k}\prod\limits_{j=1}^{n-k+1} H(z_{k,i}-z_{k-1,j}) \Bigg] \ ,
\eea
where $z_{0,a}:=\varphi_a$ are the complex hypermultiplet masses. In the second line we introduce the shorthand notation 
\be\label{H}
H(x):=\frac{1}{\pm x+\eps}~, \qquad W(x):=(\pm x)(\pm x+2\eps)=H(x\pm \eps)^{-1} \ ,
\ee
for the one-loop determinants of the hypermultiplets and the W-boson pairs, respectively. 

In this case it is easy to see that the integrand has no poles at infinity, which implies that the partition function is the same in all chambers. We will however show this by calculating the integral explicitly in all chambers, a method which is best suited for generalization. 

We first identify the multi-dimensional poles that contribute to the JK residue, which are in one-to-one correspondence with the Higgs vacua in each chamber, as collections of one-dimensional poles. We start by discussing these poles in a chamber independent way and we explain the chamber-dependence only afterwards. One-dimensional poles in the integral (\ref{TSUN_integ}) only arise due to the hypermultiplet factors. A one-dimensional pole associated to a hypermultiplet factor $H(z_{k+1,i_{k+1}}-z_{k,i_{k}})$ determines the integration variable $z_{k,i_k}$ in terms of $z_{k+1,i_{k+1}}$ .  Iterating this procedure starting from the rightmost (abelian) gauge node and moving towards the left of the quiver, we span a linear abelian subquiver of length $n$, with $n-1$ gauge nodes and one flavour node on the left: the edges of the subquiver are associated to the hypermultiplets responsible for the poles (or the Higgs VEV); the nodes of the subquiver are associated to the $z$ integration variables which are iteratively determined in terms of the flavour mass $\varphi$. We then repeat the procedure starting from the leftover integration variable of the $U(2)$ gauge node, spanning an abelian linear subquiver of length $n-1$. The two abelian linear quivers cannot overlap because the W-boson factors in (\ref{TSUN_integ}) cancel the would-be poles. Iterating this procedure, we can associate to each multi-dimensional pole contributing to the JK residue (\ref{TSUN_integ}) a collection of linear abelian subquivers of decreasing lengths. We can therefore denote a multi-dimensional pole diagrammatically as follows:
\be\label{pole_1}
\hspace{-5pt}
\begin{array}{ccccccccc}
	\varphi_{n} & \stackrel{x_n-x_1}{\text{--------}}  & z_{1,n-1}\equiv \varphi_{n,1} & \stackrel{x_n-x_2}{\text{--------}} &  \dots & \stackrel{x_{n}-x_{n-2}}{\text{--------}} & z_{n-2,2}\equiv \varphi_{n,n-2} & \stackrel{x_{n}-x_{n-1}}{\text{--------}} & z_{n-1,1}\equiv \varphi_{n,n-1}  \vspace{2mm} \\
	\varphi_{n-1} & \stackrel{x_{n-1}-x_1}{\text{--------}}  & z_{1,n-2}\equiv \varphi_{n-1,1} & \stackrel{x_{n-1}-x_2}{\text{--------}} & \dots & \stackrel{x_{n-1}-x_{n-2}}{\text{--------}} & z_{n-2,1}\equiv \varphi_{n-1,n-2} &  &   \vspace{-1mm} \\
	\vdots & & \vdots & & &  \iddots & & &\vspace{4mm} \\
	\varphi_{2} & \stackrel{x_2-x_1}{\text{--------}} & z_{1,1}\equiv \varphi_{2,1} & & & & & & \vspace{2mm} \\
	\varphi_{1} & & & & & & & &   
\end{array} 
\ee
where the left column denotes the flavour nodes and the other columns denote the gauge nodes (the expressions above the edges are for future reference).  We use the invariance under the Weyl subgroup of the gauge group to permute the $z$ variables within each column and we chose a particular ordering of the $\varphi_a$. Other multi-dimensional poles are obtained by permuting the $\varphi_a$, for a total of $n!$  multi-dimensional poles in each chamber. In terms of branes, a linear abelian subquiver corresponds to a D1 string suspended between an NS5 brane and a D3 brane, intersecting a subset of the other NS5 branes. The $n!$ multi-dimensional poles correspond to all the possible pairings of NS5 branes with D3 branes.

We have not yet specified which of the two chiral multiplets in the hypermultiplet associated to an edge is responsible for the pole. This depends on the chamber in FI space and determines the precise multi-dimensional pole. It is easier to see how this comes about if we observe that a $[1]-(1)-\dots-(1)$ linear abelian subquiver with $n$ gauge nodes decouples into $n$ copies of SQED with a single charged hypermultiplet: the hypermultiplet is an edge in the linear abelian subquiver, and the gauge $U(1)$ is the diagonal combination of the gauge $U(1)$ factors associated to the nodes to the right of the edge in the subquiver. We indicate the FI parameter for the relevant diagonal $U(1)$ gauge group above each edge of a linear abelian subquiver in (\ref{pole_1}), and use the third line of (\ref{ranks_SMM_alt}) to express it as a difference $x_a-x_i$ between the insertion points of two monopole operators (or the position of two NS5 branes) along $x^0$. It is then clear that the chambers in FI space correspond to the orderings of the NS5 branes or monopole operators along $x^0$, in agreement with field theory expectations.

Denoting a bifundamental chiral multiplet by an arrow using standard quiver notation, we can therefore be more precise and replace each edge in (\ref{pole_1}), which represents a one-dimensional pole due to a hypermultiplet, by an arrow representing a pole due to one of the two chiral multiplets in the hypermultiplet as follows: 
\be\label{edge}
(~\varphi_{a,i-1} \stackrel{x_a-x_i}{\text{--------}}  \varphi_{a,i} ~)~~~:= ~~~\begin{cases}
	~~\varphi_{a,i-1}  %\stackrel{x_i<x_a}{\longleftarrow}
	\longleftarrow  \varphi_{a,i}=\varphi_{a,i-1}-\eps~, & x_a>x_i \\
	~~\varphi_{a,i-1}  %\stackrel{x_i>x_a}{\longrightarrow} 
	\longrightarrow \varphi_{a,i}=\varphi_{a,i-1}+\eps ~, & x_a<x_i 
\end{cases} ~.
\ee
In the first case, it is the chiral multiplet in the fundamental representation of the $i$-th gauge node and antifundamental of the $(i-1)$-th gauge node that is responsible for the 1d pole (or equivalently, which takes VEV). This occurs when $x_a-x_i=\sum_{i\le j <a} \xi_j >0$.  If instead  $x_a-x_i<0 $, it is the oppositely charged chiral that is responsible for the pole. The multi-dimensional pole associated to the trivial (identity) permutation of the $\varphi_a$ is therefore encoded by a collection of abelian subquivers of increasing lengths as in (\ref{pole_1}), together with the assignment of an arrow to each edge, the orientation of which is determined by the chamber in FI space as in (\ref{edge}). We can therefore summarise the chamber dependence of the value of the gauge parameters at the multi-dimensional pole encoded by the diagram (\ref{pole_1}) using
\be\label{general_pole_TSU}
z_{i,a-i}\equiv \varphi_{a,i} = \varphi_a + \eps \sum_{j=1}^i \sgn(x_j-x_a)~.
\ee

Next, we compute the residue at the pole (\ref{general_pole_TSU}). Let us consider a pair of rows (or abelian subquivers) in (\ref{pole_1}), starting from $\varphi_a$ and $\varphi_b$, respectively. The pair contributes various factors to the multi-dimensional residue due to the massive W-bosons (which we can picture as vertical links between the two rows) and massive bifundamental hypermultiplets (diagonal links between two entries in different rows and adjacent columns). 
Due to the fundamental relation $W(x)=H(x\pm\eps)^{-1}$ in (\ref{H}), the one-loop determinants of massive W-bosons and hypermultiplets charged under the two abelian subquivers give telescopic products that largely cancel out, leaving an overall factor of $H(\varphi_{ab}-\eps\sgn(x_{ab}))$ for each pair $(a,b)$ of rows in (\ref{pole_1}). Taking into account all pairs and summing over the $n!$ multi-dimensional poles labelled by the permutations of the $\varphi_a$, we obtain the result 
\be
\begin{split}\label{ZTSUn-final}
	Z_{T[SU(n)]}(\varphi,\eps;\xi)&= \sum_{\w\in S_n}\prod\limits_{1\le a<b\le n} H(\varphi_{ab}^\w-\eps\sgn(x_{ab}))\\
	&= \sum_{\w\in S_n} \frac{1}{\prod\limits_{1\le a<b\le n} \varphi_{ab}^{\w}(\varphi_{ba}^{\w}+2\eps \sgn(x_{ab}))} \\
	&=  \frac{n!}{\Delta(\varphi)\prod\limits_{1\le a<b\le n} (\pm\varphi_{ab}+2\epsilon)} \cdot\frac{1}{n!}\sum_{\w\in S_n}(-1)^{\sigma(\w)}\prod\limits_{1\le a<b\le n}(\varphi_{ab}^{\w}+2\eps \sgn(x_{ab}))  \\
	&=\frac{n!}{ \prod\limits_{1\le a<b\le n} (\pm \varphi_{ab}+2\epsilon)}~,
\end{split}
\ee
where $\sigma(\w)$ is the signature of the permutation $\w$. 
The average over the permutations in the third line reproduces the Vandermonde determinant $\Delta(\varphi)=\prod_{a<b} \varphi_{ab}$: the $\eps$- (and $x$-) dependent terms average out to zero for symmetry reasons. The final result is therefore independent of $x$ or $\xi$, showing explicitly that the partition function of $T[SU(n)]$ is chamber independent, as expected from the commutativity of the monopole operators inserted in the correlation function.

\subsection{$T_\rho[SU(n)]$} \label{sec:TrhoSUn}

We now generalise to the $T_\rho[SU(n)]$ matrix models, which are described in matrix notation by 
\begin{equation}\label{matrix_T_rhoSUn}
\begin{bmatrix}
n & 0 & 0 & \dots& 0 & 0   \\
N_1 & N_2 & N_3 & \dots & N_{l(\rho)-2} & N_{l(\rho)-1}  \\ 
\end{bmatrix}\ ,
\end{equation}
with $N_k=\sum_{i>k} \rho_i$ and $n\equiv N_0=|\rho|$. Their partition functions are given by 
\be\label{Z_Trho_integ}
\begin{split}
	&Z_{T_{\rho}[SU(n)]}(\varphi,\epsilon;\xi) = \oint_{\mathrm{JK}(\xi)} \prod_{k=1}^{\ell(\rho)-1} \Bigg[ \frac{d^{N_k} z_k}{(2\pi i)^{N_k} N_k!} \frac{(2\epsilon)^{N_k} \prod\limits_{i\neq j}^{N_k} z_{k,ij}(z_{k,ij}+ 2\eps)}{\prod\limits_{i=1}^{N_k}\prod\limits_{j=1}^{N_{k-1}}(\pm (z_{k,i}-z_{k-1,j})+\eps)}\Bigg]~\\
	&\qquad= \oint_{\mathrm{JK}(\xi)} \prod_{k=1}^{\ell(\rho)-1} \Bigg[ \frac{d^{N_k} (2\eps z_k)}{(2\pi i)^{N_k} N_k!} \prod\limits_{1\le i< j\le N_k} \hspace{-5pt} W(z_{k,ij})\cdot \prod\limits_{i=1}^{N_k}\prod\limits_{j=1}^{N_{k-1}} H(z_{k,i}-z_{k-1,j}) \Bigg]~.
\end{split}
\ee

It turns out again that these partition functions are the same in all FI chambers, in agreement with the field theory fact that bare monopole operators of positive charge $U^+_a$ mutually commute. We will show this by explicit computation of the partition function (\ref{Z_Trho_integ}), since some intermediate results will be useful later when we consider the more general $T^\sigma_{\rho,L}[SU(n)]$ SMM. We will also derive the related result that the partition function (\ref{Z_Trho_integ}) does not depend on how the parts of $\rho$ are ordered. 

The multi-dimensional poles can again be encoded by a collection of linear subquivers, which are now generically non-abelian. Let $\rho=(\rho_1,\rho_2,\dots,\rho_R)$, with $R=\ell(\rho)$. We then associate to each part $\rho_A$ of $\rho$ a linear subquiver $[\rho_A]-(\rho_A)-\dots-(\rho_A)$ with $A-1$ unitary gauge nodes and a single unitary flavour node, all of equal rank $\rho_A$. This linear subquiver is realised on the worldvolume of the $\rho_A$ D1 strings suspended between the $A$-th NS5 brane and $\rho_A$ different D3 branes, intersecting $A-1$ other NS5 branes along the way. Analogously to the abelian case, once the mass parameters for the $U(\rho_A)$ flavour node are specified, the partition function of the linear subquiver is computed by a single multi-dimensional residue in each chamber, corresponding to the fact that the subquiver has a single Higgs vacuum. The pole can be determined by starting from the rightmost gauge node, which has effectively as many flavours as colours, and iterating the procedure as one moves towards the left in the quiver. 
Putting together the linear subquivers associated to the different NS5 branes, we can therefore specify a multi-dimensional pole diagrammatically as follows:
\be\label{pole_2}
\hspace{-5pt}
\begin{array}{ccccccccccc}
	\varphi_{\rho_1+\dots+\rho_R} & \stackrel{x_R-x_1}{\text{--------}}  & \bullet & \stackrel{x_R-x_2}{\text{--------}} & \bullet & ~~~\dots~~~ & \bullet & \stackrel{x_{R}-x_{R-2}}{\text{--------}}& \bullet & \stackrel{x_{R}-x_{R-1}}{\text{--------}} & \bullet   \\
	\vdots &  & \vdots &  & \vdots &  & \vdots & & \vdots & & \vdots \\
	\varphi_{\rho_1+\dots+\rho_{R-1}+1} & \stackrel{x_R-x_1}{\text{--------}}  & \bullet & \stackrel{x_R-x_2}{\text{--------}} & \bullet & \dots &  \bullet & \stackrel{x_{R}-x_{R-2}}{\text{--------}}& \bullet & \stackrel{x_{R}-x_{R-1}}{\text{--------}} & \bullet  \vspace{2mm} \\
	\varphi_{\rho_1+\dots+\rho_{R-1}} & \stackrel{x_{R-1}-x_1}{\text{--------}}  & \bullet & \stackrel{x_{R-1}-x_2}{\text{--------}} & \bullet & ~~~\dots~~~ & \bullet & \stackrel{x_{R-1}-x_{R-2}}{\text{--------}}& \bullet &  &  \\
	\vdots &  & \vdots &  & \vdots &  & \vdots & & \vdots & &  \\
	\varphi_{\rho_1+\dots+\rho_{R-2}+1} & \stackrel{x_{R-1}-x_1}{\text{--------}}  & \bullet & \stackrel{x_{R-1}-x_2}{\text{--------}} & \bullet & \dots &  \bullet & \stackrel{x_{R-1}-x_{R-2}}{\text{--------}}& \bullet & &   \\
	\vdots & & \vdots & & \vdots &  & \vdots & & & &  \\
	\vdots & & \vdots & & \vdots &  & & & & &  \\	
	\varphi_{\rho_1+\rho_2+\rho_3} & \stackrel{x_3-x_1}{\text{--------}}  & \bullet & \stackrel{x_{3}-x_2}{\text{--------}} & \bullet & & & & & & \\
	\vdots &  & \vdots &  & \vdots & & & & & &  \\
	\varphi_{\rho_1+\rho_2+1} & \stackrel{x_3-x_1}{\text{--------}}  & \bullet & \stackrel{x_{3}-x_2}{\text{--------}} & \bullet & & & & & & \vspace{2mm} \\
	\varphi_{\rho_1+\rho_2} & \stackrel{x_2-x_1}{\text{--------}}  & \bullet & &  & & & & & & \\
	\vdots &  & \vdots &  &  &  &  & &  & &  \\
	\varphi_{\rho_1+1} & \stackrel{x_2-x_1}{\text{--------}}  & \bullet & & & & & & & &  \vspace{2mm}\\
	\varphi_{\rho_1} & & & & & & & & & & \\
	\vdots & & & & & & & & & &  \\
	\varphi_{1} & & & & & & & & & & 
\end{array} 
\ee
We replace the gauge variables $z$ by bullets for brevity: they are determined in each chamber as in (\ref{edge}). We also pick a particular ordering of the flavour parameters $\varphi_a$ for definiteness. Other multi-dimensional poles are obtained by permuting the $\varphi_a$, except that permutations within each block of size $\rho_A$ can be undone by a  Weyl transformation in the gauge group. Multi-dimensional poles (or Higgs vacua) are therefore in one-to-one correspondence with elements of $S_n/\times_A S_{\rho_A}$. This is manifest in the brane construction, where we are partitioning the $n$ D3 branes in parts of $\rho_A$, each of which is paired with a different NS5 brane.    

In view of the block (or strip) structure associated to the linear subquivers, which is visible in (\ref{pole_2}), it is useful to introduce a shorthand block notation. Let 
\be\label{S_A}
S_A = \{ K \in \bN ~|~ \sum_{J<A} \rho_J < K \le \sum_{J\le A} \rho_J  \}
\ee 
be the set of integer labels in each block, and
\be\label{varphi_A}
\underline{\varphi}_A = \{ \varphi_a ~|~ a \in S_A \}~
\ee 
be the set of associated $\varphi_a$ mass parameters, for the ordering chosen in (\ref{pole_2}). We can then rewrite (\ref{pole_2}) more simply as 
\be\label{pole_2_short}
\hspace{-5pt}
\begin{array}{ccccccccccc}
	\underline{\varphi}_R & \stackrel{x_R-x_1}{\text{--------}}  & \bullet & \stackrel{x_R-x_2}{\text{--------}} & \bullet & ~~~\dots~~~ & \bullet & \stackrel{x_{R}-x_{R-2}}{\text{--------}}& \bullet & \stackrel{x_{R}-x_{R-1}}{\text{--------}} & \bullet  \vspace{2mm} \\
	\underline{\varphi}_{R-1} & \stackrel{x_{R-1}-x_1}{\text{--------}}  & \bullet & \stackrel{x_{R-1}-x_2}{\text{--------}} & \bullet & ~~~\dots~~~ & \bullet & \stackrel{x_{R-1}-x_{R-2}}{\text{--------}}& \bullet &  &   \\
	\vdots & & \vdots & & \vdots &  & \vdots & & & &  \\
	\vdots & & \vdots & & \vdots &  & & & & &  \\	
	\underline{\varphi}_3 & \stackrel{x_3-x_1}{\text{--------}}  & \bullet & \stackrel{x_{3}-x_2}{\text{--------}} & \bullet & & & & & &  \vspace{2mm} \\
	\underline{\varphi}_2 & \stackrel{x_2-x_1}{\text{--------}}  & \bullet & &  & & & & & & \vspace{2mm}\\
	\underline{\varphi}_1 & & & & & & & & & & 
\end{array} 
\ee

It is straightforward to compute the residue at the multi-dimensional pole (\ref{pole_2}) or equivalently (\ref{pole_2_short}). The linear subquivers have unit partition functions, so the only contribution comes from massive hypermultiplets and W-bosons connecting different linear subquivers. After cancellations, we find that each pair $(A,B)$ of subquivers in (\ref{pole_2_short}) contributes the factor
\be\label{hyper_block_notation}
H(\underline{\varphi}_{AB}-\eps\sgn(x_{AB})):=\prod_{\substack{a \in S_A\\ b\in S_B}}H(\varphi_{ab}-\eps \sgn(x_{AB}))
\ee 
to the SMM partition function. Summing over all such pairs and over all multi-dimensional poles, labelled by elements of $S_n/\times_A S_{\rho_A}$, we obtain the partition function 
\be\label{Z_Trho}
Z_{T_\rho[SU(n)]}(\varphi,\eps;x)= \frac{1}{\prod_A \rho_A!} \sum_{\w\in S_n} \prod_{1\le A<B\le n} H(\underline{\varphi}_{AB}^\w - \eps \sgn(x_{AB}))~,
\ee
where the permutation $\w$ acts on the $n$-tuple $(\varphi_1, \dots,\varphi_n)$. 

Note that the SMM partition function (\ref{Z_Trho}) is manifestly invariant under a common permutation of $(\rho_A)$, $(x_A)$ and $(\underline{\varphi}_A)$, which corresponds to permuting the NS5 branes. So the $T_\rho[SU(n)]$ SMM is invariant under \emph{naive} Seiberg duality (\emph{naive} because the Higgs branches agree only for generic FI parameters, as in \cite{Assel:2017jgo}).  
We can also see that the partition function is chamber independent, due to the average over permutations $\w$. This is perhaps more manifest if we rewrite the partition function as  
\be\label{Z_Trho_2}
Z_{T_\rho[SU(n)]}(\varphi,\eps;x)= \frac{1}{\prod_A \rho_A!} \sum_{\w\in S_n} \prod_{1\le A<B\le n} H(\underline{\varphi}_{P(A)P(B)}^\w + \eps)~,
\ee
with the permutation $P\in S_R$ defined by $x_{P(A)}<x_{P(A+1)}$ for all $A=1,\dots,R$. 

Having proven chamber independence, we can work in the chamber where all the FI parameters are positive for definiteness and rewrite more explicitly the SMM partition function  (\ref{Z_Trho}) as 
\be\label{Z_TrhoSUn_2}
\begin{split}
	Z_{T_\rho[SU(n)]}(\varphi,\epsilon)&=  \frac{1}{\prod_A\rho_A!}\sum_{\w\in S_n} \prod_{\alpha\in \Delta_+\setminus \Delta_\rho} \frac{1}{(-\alpha\cdot \varphi^\w)(\alpha\cdot \varphi^\w+2\eps)}\\
	&=\frac{1}{\prod\limits_{1\le a<b\le n}(\pm \varphi_{ab}+2\eps)} \wat q_\rho(\varphi,\epsilon)~,
\end{split}
\ee
where $\wat q_\rho(\varphi,\eps)$ and $\Delta_+$, $\Delta_\rho$ are defined as in (\ref{qhat}) and (\ref{roots_in_prod}), respectively.

\subsection{$T_{\rho,L}[SU(n)]$} \label{sec:TrhoLSUn}

Finally, we extend the result to the general situation with trivial $\sigma$ and compute the partition function of the $T_{\rho,L}[SU(n)]$ SMM. The difference compared to the case of the previous subsection is that there are $N_f$ fundamental Fermi multiplets charged under the $L$-th gauge node in the quiver, arising from D5-D1 strings in the brane construction, and there are now $L+R-1$ gauge nodes instead of $R-1$. Recall that $L=\ell(\rho^-)$, $R=\ell(\rho^+)$ and $\rho=(N-\rho^-,\rho^+)$. 

The quiver for the $T_{\rho,L}[SU(n)]$ matrix models is described in matrix notation by 
\begin{equation}\label{matrix_T_rhoL}
\begin{bmatrix}
n & 0 & \dots& 0& 0 & 0 & \dots & 0   \\
N_1 & N_2 & \dots & N_{L-1}& \underline{N_L} & N_{L+1}& \dots & N_{L+R-1}  \\ 
\end{bmatrix}\ 
\end{equation}
with $N_k=\sum_{i>k}\rho_i$ and $n\equiv N_0=|\rho|$. Its partition function is computed by the integral
\be\label{Z_TrhoL}
\begin{split}
	&Z_{T_{\rho,L}[SU(n)]}(\varphi,\epsilon;\xi) = \\ 
	&~~ \oint_{\mathrm{JK}(\xi)} \prod_{k=1}^{L+R-1} \Bigg[ \frac{d^{N_k} (2\eps z_k)}{(2\pi i)^{N_k} N_k!} \prod\limits_{1\le i< j\le N_k} \hspace{-5pt} W(z_{k,ij})\cdot \prod\limits_{i=1}^{N_k}\prod\limits_{j=1}^{N_{k-1}} H(z_{k,i}-z_{k-1,j}) \Bigg] \prod_{h=1}^{N_L} P(z_{L,h})~,
\end{split}
\ee
where the $P$ factors of (\ref{P_poly}) account for the $N_f$ fundamental Fermi multiplets. We will see shortly that the Fermi multiplets generically make the partition function (\ref{Z_TrhoL}) dependent on the FI parameters, with jumps when certain walls in FI space are crossed.  

To compute the partition function (\ref{Z_TrhoL}) we can recycle most of the calculations in the previous subsection. The poles are unchanged since they are only due to hypermultiplets, so we just need to evaluate the Fermi factors $P(z_{L,h})$ at each multi-dimensional pole. Using the block notation 
\be\label{block_not_1}
f(\underline{\varphi}_A):=\prod_{a\in S_A} f(\varphi_a)~,
\ee
the result is that the partition function becomes (compare with (\ref{Z_Trho}))
\be\label{Z_TrhoL_2}
\begin{split}
	Z_{T_{\rho,L}[SU(n)]}(\varphi,\eps;x)= \frac{1}{\prod_A \rho_A!} \sum_{\w\in S_n} &\prod_{A<B} H(\underline{\varphi}_{AB}^\w - \eps \sgn(x_{AB})) \\
	\cdot& \prod_{C>L} P(\underline{\varphi}_C^\w + \eps \sum_{D\le L} \sgn(x_{DC}))~.
\end{split}
\ee

The partition function (\ref{Z_TrhoL_2}) is invariant under common permutations of the first $L$ elements or last $R$ elements of $(\rho_A)$, $(\underline{\varphi}_A)$ and $x_A$. This corresponds to the separate permutations of NS5$_-$ branes (or bare monopole operators of negative charge) and of NS5$_+$ branes (or bare monopole operators of positive charge). If elements of the positive and negative sets are instead permuted amongst themselves, the partition function (\ref{Z_TrhoL_2}) changes because of the $P$ factors. This reflects the fact that two bare monopole operators commute if their charges are both positive or both negative, but not if they are oppositely charged. 

Finally, we can write the partition function (\ref{Z_TrhoL_2}) similarly to (\ref{Z_TrhoSUn_2}) as follows
\be\label{Z_TrhoL_3}
Z_{T_{\rho,L}[SU(n)]}(\varphi,\eps;x)= \frac{1}{\prod\limits_{1\le a<b\le n}(\pm \varphi_{ab}+2\eps)} q_{\rho,L}(\varphi,m,\eps;x)~,
\ee
where now 
\be\label{q_general}
\begin{split}
	q_{\rho,L}(\varphi,m,\eps;x)= \frac{1}{\prod_A \rho_A!} \sum_{\w\in S_n} &\left[\prod_{A<B} \frac{\underline{\varphi}_{AB}^\w +2 \eps \sgn(x_{AB})}{\underline{\varphi}_{AB}^\w} \right] \left[\prod_A (\underline{\varphi}_{AA}+2\eps)' \right] \\
	\cdot &\left[\prod_{C>L} P(\underline{\varphi}_C^\w + \eps \sum_{D\le L} \sgn(x_{DC}))\right]~,
\end{split}
\ee
depends on the mass parameters $m$ and the FI parameters $\xi$ (or equivalently the monopole insertion points $x$), in addition to $\varphi$ and $\eps$. In (\ref{q_general}) we use the block notation (\ref{block_not_1}) and 
\be\label{block_not_2}
g(\underline{\varphi}_{AB}):=\prod_{\substack{a\in S_A\\ b\in S_B}} g(\varphi_{ab})~, \qquad h(\underline{\varphi}_{AA})':=\prod_{\substack{a,a'\in S_A\\ a\neq a'}} h(\varphi_{aa'})~.
\ee
Stripping out the prefactor in front of $q$ in (\ref{Z_TrhoL_3}) will prove useful in the next appendix when we introduce a non-trivial D3 brane partition $\sigma$.

\section{Residues in flavour fugacities and changing $\sigma$}\label{sec:residues}

In this appendix we show how the partition function of the $T^\sigma_{\rho, L}[SU(n)]$ matrix model (\ref{Z_general})  can be obtained from the partition function for the trivial D3 brane partition $\sigma=(1^n)$ by taking appropriate residues in the flavour fugacities, which have the effect of moving boxes in the Young tableaux associated to $\sigma$. We follow a related computation performed for Hilbert series in appendix C of \cite{Cremonesi:2014uva}.

Any partition $\sigma$ of $n$ can be obtained from $\sigma = (1^n)$ by repeatedly moving the last box to a previous row which is  followed by rows of a single box only, so it is enough to consider the move%
\footnote{In the language of Kraft-Procesi transitions \cite{Cabrera:2016vvv}, in which moving a box up by $k$ rows and to the right by one column is an $A_{k}$ transition, whereas moving a box up by one row and to the right by $l$ columns is an $a_l$ transition, the move (\ref{sigmasigma}) realises an $A_{h-1}$ transition followed by an $a_{H-1}$ transition. }
\begin{equation}
\label{sigmasigma}
\sigma = (\sigma_1,\ldots,\sigma_{d-h},H,1^h) \ \rightarrow \  \sigma'=(\sigma_1,\ldots,\sigma_{d-h},H+1,1^{h-1}) \ ,
\end{equation}
where the lengths of the partitions $\sigma$ and $\sigma'$ are $d+1$ and $d$, respectively. For brevity, we will denote  $X=d-h+1$, so that $\sigma_X=H$ and $\sigma'_X=H+1$. We assume $\sigma_{X-1}>H$ so that the move is allowed.
%\begin{equation}
%\label{H}
%H = \sigma_{d-h+1} \ .
%\end{equation}
We claim that for such $\sigma$ and $\sigma'$, 
\be\label{flavour_res_prescription}
\Res_{z=0}~ Z_{T^\sigma_{\rho, L}[SU(n)]}\big| =  \frac{P(\varphi_X +(H-L)\eps)}{ (H+1)(2\epsilon)\prod\limits_{a=X+1}^d [\pm(\varphi_{Xa}+(H-1)\epsilon)+\epsilon]} Z_{T^{\sigma'}_{\rho,L}[SU(n)]}~,
\ee
where $\big|$ denotes the substitution $\varphi_X\to\varphi_X-(\eps-z), \varphi_{d+1}\to\varphi_X+H(\eps-z)$ and $N_f$ fundamental Fermi multiplets are attached to gauge node $L$. For $L \leq H$, the term containing the Fermi multiplets is present since the rank of the gauge node attached to these multiplets reduces during the transition \eqref{sigmasigma}. In contrast, when $L > H$, the Fermi multiplets are unaffected by the transition and it is understood that $P\left(\varphi_X+(H-L)\epsilon\right) \to 1$ in \eqref{flavour_res_prescription}. 

To see this, recall that the quiver diagram of $T^{\sigma}_\rho[SU(n)]$ is represented by the matrix 
\begin{equation}\label{matrix_Tsigma}
\begin{bmatrix}
0 & h & 0 & \dots & 1 & M_{H+1} & \dots \\
0 & N_1 & N_2 & \dots & N_H & N_{H+1} & \dots \\ 
\end{bmatrix}\ ,
\end{equation}
where the first/second row in the matrix denotes the rank of a unitary flavour/gauge group, starting from nodes labelled by  $i=0$.  In the case $L = H$, for example, the Fermi multiplets are attached to the $U(N_H)$ gauge group with a single flavour of fundamental hypermultiplets with mass parameter $\varphi_X$, which can be illustrated by underlining $N_H$ in the matrix above. The matrix notation for $T^{\sigma'}_\rho[SU(n)]$ is given by
\begin{equation}\label{matrix_Tsigma'}
\begin{bmatrix}
1& h-1 & 0 & \dots & 0 & M_{H+1}+1 & \dots \\
0&N_1 - 1 & N_2 - 1 & \dots & N_H - 1 & N_{H+1} & \dots \\ 
\end{bmatrix} \ .
\end{equation}
The ellipses on the right remain unchanged in the transition. In terms of these 0d quivers and their partition functions (\ref{Z_general}), we identify $\varphi_{d+1}=\tilde\varphi_{1,h}$ and $\varphi_X=\tilde\varphi_{H,1}$, so $\varphi_{d+1}$ is the mass parameter for one of the $h$ fundamental hypermultiplets of gauge group $1$ and $\varphi_{X}$ is the mass parameter for the fundamental hypermultiplet of gauge group $H$. 

The partition function in the LHS of (\ref{flavour_res_prescription}) has a simple pole at $z=0$ if the term 
\begin{equation}\label{desired_pole}
\frac{1}{\varphi_{X,d+1}+(H+1)\epsilon} \ ,
\end{equation}
is present before the substitution. This term appears if the abelian subquiver
\begin{equation}
\begin{bmatrix}
1 & 0 & \ldots & 0 & 1 \\
1 & 1 & \ldots & 1 & 1 
\end{bmatrix} \ ,
\end{equation}
where the left/right flavour node has mass parameter $\varphi_{d+1}$/$\varphi_{X}$ and the gauge nodes have parameters $z_{i,N_i}$, leads to a multi-dimensional pole in the SMM integral given by
\begin{equation}\label{multi-pole}
z_{i,N_i}=\begin{cases}
\varphi_{d+1}-i\epsilon~, & 1 \leq i \leq a\\
\varphi_X+(H+1-i)\epsilon ~, & a+1 \leq i \leq H  
\end{cases}
\end{equation}
for some integer $a \in \left[0,H\right]$. This multi-dimensional pole of the partition function of the abelian subquiver can be illustrated diagrammatically by
\begin{equation}\label{desired_pole_2}
\begin{array}{cccccccccccc}
\varphi_{d+1} & & & & & & & & & & & \varphi_X \\
\uparrow & & & & & & & & & & & \downarrow \\
\varphi_{d+1}-\epsilon & \leftarrow & \varphi_{d+1}-2\epsilon & \ldots & \leftarrow \varphi_{d+1}-a\epsilon & & & \varphi_X+(H-a)\epsilon & \leftarrow & \ldots & \varphi_X+2\epsilon & \leftarrow \varphi_X+\epsilon \vspace{2mm} \\
1 & &  2 &  \ldots & a & & & a+1 & & \ldots & H+1 &  H 
\end{array} \ ,
\end{equation}
where arrows indicate the chiral multiplets that contribute to the pole and the nodes have been replaced by their corresponding parameters. Due to the massive hypermultiplet associated to the missing link in the quiver (\ref{desired_pole_2}), the residue of the abelian subquiver partition function is $[\pm(\varphi_{X,d+1}+H\eps)+\eps]^{-1}$,  which indeed contains the factor (\ref{desired_pole}). Note that a pole of the type (\ref{desired_pole_2}) for a certain value of $a$ appears in any chamber: the relevant value of $a$ is determined by $x_{a+1}=%\underset{b\in [1,H+1] }
{\mathrm{max}}\{x_b\}_{b=1}^{H+1}$.%
\footnote{Recall that the FI parameter for the $\mathrm{a}$-th gauge node is given by $x_{a+1}-x_{a}$.}

If we now embed the abelian subquiver in the full non-abelian quiver and evaluate the integrals over $z_{i,N_i}$  for $i=1,\dots,H$ at the poles (\ref{multi-pole}) in the full SMM partition functions, the gauge group (\ref{matrix_Tsigma}) of the original SMM reduces to the gauge group  (\ref{matrix_Tsigma'}) of the new matrix model.  Keeping track of the masses of the fields which enter the one-loop determinants and following several cancellations, it is then tedious but straightforward to obtain the RHS of (\ref{flavour_res_prescription}), where the new $\varphi_X$ is now identified with an extra mass parameter for the flavour symmetry at node $H+1$, as expected.

The general formula (\ref{ZTsigmarhoL}) for the partition function of the 0d $\cN=2^*$ version of $T^\sigma_{\rho,L}[SU(n)]$ can then be obtained by induction.  We start from (\ref{Z_TrhoL_3}) and show that the residue formula (\ref{flavour_res_prescription}) holds for any $\sigma$ and $\sigma'$ related by the move (\ref{sigmasigma}). The pole at $z=0$ is due to the denominator of the first line of (\ref{ZTsigmarhoL}), which is independent of $\rho$ and behaves as discussed above. As for the numerator, the only effect of the residue is to change the argument $\text{a}_\sigma(\varphi,\eps)$ of $q_\rho$, which is defined in (\ref{a_sigma}), according to the substitution  $\varphi_X\to\varphi_X-\eps$,  $ \varphi_{d+1}\to\varphi_X+H\eps$. The only changes are in the $a=X$ and $a=d+1$ entries in (\ref{a_sigma}). Upon the previous substitution, they combine to become 
\be
\varphi_X+(H+2-2j)\eps~, \qquad j=1,\dots,H+1~,
\ee
reproducing the $a=X$ entries in $\text{a}_{\sigma'}(\varphi,\eps)$. Finally, the denominator in the second line of (\ref{ZTsigmarhoL}) accounts for the factors of $P$ in the residues (\ref{flavour_res_prescription}). 
Therefore, the formula (\ref{ZTsigmarhoL}) for the partition function of $T^\sigma_{\rho,L}[SU(n)]$ satisfies the residue relation (\ref{flavour_res_prescription}), which proves (\ref{ZTsigmarhoL}) by induction starting from (\ref{Z_TrhoL_3}).

\section{More on the computation of monopole bubbling factors}
\label{sec:bub_factors}

In this appendix we state the results for the partition functions \eqref{Z111} and \eqref{Z4} in each of the 6 distinct chambers. These results have been computed by applying the JK prescription and have been verified by following the residues in the flavour fugacities procedure outlined in appendix \ref{sec:residues}.

Evaluating \eqref{Z111} in each of the six chambers we find

\begin{equation}
\begin{aligned}
&Z^{++--}_{ab}(\varphi,m,\eps) = \left(\frac{(-1)^{N-1}\prod\limits_{k=1}^{N_f}\left[\varphi_a - m_k - 2\epsilon\right]}{\varphi_{ab}\left(\varphi_{ab}-2\epsilon \right)\prod\limits_{c \neq b,a}\left[\left(\varphi_{ac}-\epsilon \right) \left( \varphi_{ac}-3\epsilon\right)\right]}\right.\\
&\left.+\frac{(-1)^{N-1}\prod\limits_{k=1}^{N_f}\left[\varphi_a - m_k - 2\epsilon\right]}{\left(\varphi_{ab}-2\epsilon\right)\left(\varphi_{ab}-4\epsilon \right)\prod\limits_{c \neq b,a}\left[\left(\varphi_{ac}-\epsilon \right) \left( \varphi_{ac}-3\epsilon\right)\right]}+(a\leftrightarrow b)\right)\\
&+\sum_{c \neq b,a} \left[ \frac{(-1)^{N-1}\prod\limits_{k=1}^{N_f}\left[ \varphi_c - m_k-\epsilon\right]}{\prod\limits_{r=a,b}\left(\varphi_{rc}+\epsilon \right)\left(\varphi_{rc}+3\epsilon \right) 
\prod\limits_{d \neq c, b, a} \varphi_{cd}\left(\varphi_{cd}-2\epsilon\right)       }   +\frac{(-1)^{N-1}\prod\limits_{k=1}^{N_f}\left[ \varphi_c - m_k-\epsilon\right]}{\prod\limits_{r=a,b}\left(\pm\varphi_{rc}+\epsilon \right)
\prod\limits_{d \neq c, b, a} \varphi_{cd}\left(\varphi_{cd}-2\epsilon\right)                   }                     \right. \\
&\left.+\left(\frac{(-1)^{N}\prod\limits_{k=1}^{N_f}\left[ \varphi_c - m_k-\epsilon\right]}{\left(\pm\varphi_{ac}+\epsilon \right)\left(\varphi_{bc}+\epsilon \right)\left( \varphi_{bc}+3\epsilon\right)\prod\limits_{d \neq c, b, a} \varphi_{cd}\left(\varphi_{cd}-2\epsilon\right)       }                                          + (a\leftrightarrow b) \right) \right] \ . 
\end{aligned}
\end{equation}

\begin{equation}
\begin{aligned}
&Z^{--++}_{ab}(\varphi,m,\eps) = \left(\frac{(-1)^{N-1}\prod\limits_{k=1}^{N_f}\left[\varphi_a - m_k+2\epsilon\right]}{\varphi_{ab}\left(\varphi_{ab}+2\epsilon \right)\prod\limits_{c \neq b,a}\left[\left(\varphi_{ac}+\epsilon \right) \left( \varphi_{ac}+3\epsilon\right)\right]}\right.\\
&\left.+\frac{(-1)^{N-1}\prod\limits_{k=1}^{N_f}\left[\varphi_a - m_k+2\epsilon \right]}{\left(\varphi_{ab}+2\epsilon\right)\left(\varphi_{ab}+4\epsilon \right)\prod\limits_{c \neq b,a}\left[\left(\varphi_{ac}+\epsilon \right) \left( \varphi_{ac}+3\epsilon\right)\right]}+(a\leftrightarrow b) \right)\\
&+\sum_{c \neq b,a} \left[\frac{(-1)^{N-1}\prod\limits_{k=1}^{N_f}\left[ \varphi_c - m_k+\epsilon\right]}{  \prod\limits_{r=a,b}\left(\varphi_{rc}-\epsilon \right)\left(\varphi_{rc}-3\epsilon \right)
\prod\limits_{d \neq c, b, a} \varphi_{cd}\left(\varphi_{cd}+2\epsilon\right)                   } +  \frac{(-1)^{N-1}\prod\limits_{k=1}^{N_f}\left[ \varphi_c - m_k+\epsilon\right]}{\prod\limits_{r=a,b}\left(\varphi_{rc}\pm\epsilon \right)
	\prod\limits_{d \neq c, b, a} \varphi_{cd}\left(\varphi_{cd}+2\epsilon\right)       }  \right. \\
 &\left.+\left(\frac{(-1)^{N}\prod\limits_{k=1}^{N_f}\left[ \varphi_c - m_k+\epsilon\right]}{\left(\varphi_{ac}-\epsilon \right)\left(\varphi_{ac}-3\epsilon \right)\left( \pm\varphi_{bc}+\epsilon\right)\prod\limits_{d \neq c, b, a} \varphi_{cd}\left(\varphi_{cd}+2\epsilon\right)       }                                         +(a\leftrightarrow b) \right) \right]\ .
\end{aligned}
\end{equation}

\begin{equation}
\begin{aligned}
Z^{+-+-}_{ab}&(\varphi,m,\eps) = \left(\frac{(-1)^{N-1}\prod\limits_{k=1}^{N_f}\left[\varphi_a - m_k\right]}{\varphi_{ab}\left(\varphi_{ab}+2\epsilon \right)\prod\limits_{c \neq b,a}\left[\left(\varphi_{ac}-\epsilon \right) \left( \varphi_{ac}+\epsilon\right)\right]}\right.\\
&\left.+\frac{(-1)^{N-1}\prod\limits_{k=1}^{N_f}\left[\varphi_a - m_k - 2\epsilon\right]}{\left(\varphi_{ab}-2\epsilon\right)\left(\varphi_{ab}-4\epsilon \right)\prod\limits_{c \neq b,a}\left[\left(\varphi_{ac}-\epsilon \right) \left( \varphi_{ac}-3\epsilon\right)\right]}+(a\leftrightarrow b)\right)\\
&+\sum_{c \neq b,a} \left[\left( \frac{(-1)^{N-1}\prod\limits_{k=1}^{N_f}\left[ \varphi_c - m_k+\epsilon\right]}{\prod\limits_{r=a,b}\left(\varphi_{rc}-\epsilon \right)\left(\varphi_{rc}-3\epsilon \right) 
\prod\limits_{d \neq c, b, a} \varphi_{cd}\left(\varphi_{cd}+2\epsilon\right)       }     +(\epsilon\rightarrow -\epsilon) \right)                  \right. \\
&+ \left.\left(\frac{(-1)^{N}\prod\limits_{k=1}^{N_f}\left[ \varphi_c - m_k-\epsilon\right]}{\left(\pm\varphi_{ac}+\epsilon \right)\left(\varphi_{bc}+\epsilon \right)\left( \varphi_{bc}+3\epsilon\right)\prod\limits_{d \neq c, b, a} \varphi_{cd}\left(\varphi_{cd}-2\epsilon\right)                       }+(a \leftrightarrow b)\right) \right] \ .
\end{aligned}
\end{equation}

\begin{equation}
\begin{aligned}
Z^{-+-+}_{ab}&(\varphi,m,\eps) =\left( \frac{(-1)^{N-1}\prod\limits_{k=1}^{N_f}\left[\varphi_a - m_k\right]}{\varphi_{ab}\left(\varphi_{ab}-2\epsilon \right)\prod\limits_{c \neq b,a}\left[\left(\varphi_{ac}-\epsilon \right) \left( \varphi_{ac}+\epsilon\right)\right]}\right.\\
&\left.+\frac{(-1)^{N-1}\prod\limits_{k=1}^{N_f}\left[\varphi_a - m_k+2\epsilon \right]}{\left(\varphi_{ab}+2\epsilon\right)\left(\varphi_{ab}+4\epsilon \right)\prod\limits_{c \neq b,a}\left[\left(\varphi_{ac}+\epsilon \right) \left( \varphi_{ac}+3\epsilon\right)\right]}+(a\leftrightarrow b) \right)\\
&+\sum_{c \neq b,a} \left[ \left(\frac{(-1)^{N-1}\prod\limits_{k=1}^{N_f}\left[ \varphi_c - m_k+\epsilon\right]}{\prod\limits_{r=a,b}\left(\varphi_{rc}-\epsilon \right)\left(\varphi_{rc}-3\epsilon \right)
\prod\limits_{d \neq c, b, a} \varphi_{cd}\left(\varphi_{cd}+2\epsilon\right)                       }+(\epsilon \rightarrow -\epsilon)\right) \right. \\
&\left. +\left( \frac{(-1)^{N}\prod\limits_{k=1}^{N_f}\left[ \varphi_c - m_k+\epsilon\right]}{\left(\varphi_{ac}-\epsilon \right)\left(\varphi_{ac}-3\epsilon\right)\left(\pm\varphi_{bc}+\epsilon \right)\prod\limits_{d \neq c, b, a} \varphi_{cd}\left(\varphi_{cd}+2\epsilon\right)  }   +(a \leftrightarrow b) \right) \right] \ .
\end{aligned}
\end{equation}

\begin{equation}
\begin{aligned}
Z^{+--+}_{ab}(\varphi,m,\eps) &= \frac{(-1)^{N-1}\prod\limits_{k=1}^{N_f}\left[\varphi_a - m_k\right]}{\varphi_{ab}\left(\varphi_{ab}-2\epsilon \right)\prod\limits_{c \neq b,a}\left[\left(\varphi_{ac}-\epsilon \right) \left( \varphi_{ac}+\epsilon\right)\right]}\\
&+\frac{(-1)^{N-1}\prod\limits_{k=1}^{N_f}\left[\varphi_b - m_k +2\epsilon\right]}{\left(\varphi_{ab}-2\epsilon\right)\left(\varphi_{ab}-4\epsilon \right)\prod\limits_{c \neq b,a}\left[\left(\varphi_{bc}+\epsilon \right) \left( \varphi_{bc}+3\epsilon\right)\right]}\\
 &+\sum_{c \neq b,a}  \frac{(-1)^{N-1}2\prod\limits_{k=1}^{N_f}\left[ \varphi_c - m_k+\epsilon\right]}{\left(\varphi_{ac}-\epsilon \right)\left(\varphi_{ac}-3\epsilon \right)\left(\varphi_{bc}+\epsilon \right)\left( \varphi_{bc}-3\epsilon\right)\prod\limits_{d \neq c, b, a} \varphi_{cd}\left(\varphi_{cd}+2\epsilon\right)       }          \\
& +(\epsilon \rightarrow -\epsilon )\ .
\end{aligned}
\end{equation}

\begin{equation}
\begin{aligned}
Z^{-++-}_{ab}(\varphi,m,\eps) &=\frac{(-1)^{N-1}\prod\limits_{k=1}^{N_f}\left[\varphi_b - m_k \right]}{\varphi_{ab}\left(\varphi_{ab}-2\epsilon \right)\prod\limits_{c \neq b,a}\left[\left(\varphi_{bc}-\epsilon \right) \left( \varphi_{bc}+\epsilon\right)\right]}\\
&+ \frac{(-1)^{N-1}\prod\limits_{k=1}^{N_f}\left[\varphi_a - m_k-2\epsilon\right]}{\left(\varphi_{ab}-2\epsilon\right)\left(\varphi_{ab}-4\epsilon \right)\prod\limits_{c \neq b,a}\left[\left(\varphi_{ac}-\epsilon \right) \left( \varphi_{ac}-3\epsilon\right)\right]}\\
&+\sum_{c \neq b,a} %\Bigg[
\frac{(-1)^{N-1}2\prod\limits_{k=1}^{N_f}\left[ \varphi_c - m_k-\epsilon\right]}{\left(\varphi_{ac}-\epsilon \right)\left(\varphi_{ac}+3\epsilon \right)\left(\varphi_{bc}+\epsilon \right)\left(\varphi_{bc}+3\epsilon\right)\prod\limits_{d \neq c, b, a} \varphi_{cd}\left(\varphi_{cd}-2\epsilon\right)       }         \\
& +(\epsilon \rightarrow - \epsilon)  ~.
\end{aligned}
\end{equation}

As expected from the action of PT, these results satisfy 
\be\label{PT_inv}
\begin{split}
Z^{++--}_{ab}(\varphi,m,\eps)&=Z^{--++}_{ab}(\varphi,m,-\eps)~,\\
Z^{+-+-}_{ab}(\varphi,m,\eps)&=Z^{-+-+}_{ab}(\varphi,m,-\eps)~,\\
Z^{+--+}_{ab}(\varphi,m,\eps)&=Z^{+--+}_{ab}(\varphi,m,-\eps)~,\\
Z^{-++-}_{ab}(\varphi,m,\eps)&=Z^{-++-}_{ab}(\varphi,m,-\eps)~.
\end{split}
\ee
In addition, the correlator (\ref{VEV_+-+-}) is invariant under a ``charge conjugation'' C which reverses the sign of the charge of monopole operators (and of abelian monopole variables) and sends $\eps \to -\eps$. This symmetry manifests itself in the identities
\be\label{C_inv}
\begin{split}
	Z^{++--}_{ab}(\varphi,m,\eps)&=Z^{--++}_{ba}(\varphi,m,-\eps)~,\\
	Z^{+-+-}_{ab}(\varphi,m,\eps)&=Z^{-+-+}_{ba}(\varphi,m,-\eps)~,\\
	Z^{+--+}_{ab}(\varphi,m,\eps)&=Z^{-++-}_{ba}(\varphi,m,-\eps)~.
\end{split}
\ee

All of these 6 results agree when $N_f=0,1,\dots,2N-2$, in which case the bubbling factor is no longer chamber dependent.

Evaluating \eqref{Z4} in each of the six chambers we find
\begin{equation}
\begin{aligned}
Z^{++--}(\varphi,m,\eps) &= \sum_{a=1}^N \frac{\prod\limits_{k=1}^{N_f}\left[\left(\varphi_a-m_k-\epsilon\right)\left(\varphi_a-m_k-3\epsilon\right) \right]}{\prod\limits_{b \neq a}\left[ \varphi_{ab}\left(\varphi_{ab}-2\epsilon\right)^2\left(\varphi_{ab}-4\epsilon\right)\right]}\\
&+\sum_{a \neq b} \frac{2\prod\limits_{k=1}^{N_f}\left[\left(\varphi_a-m_k-\epsilon\right)\left(\varphi_b-m_k-\epsilon\right) \right]}{\varphi_{ab} \left(\varphi_{ab}+ 2\epsilon\right)^2(\varphi_{ab}-2\eps)\prod\limits_{c\neq a,b} \prod\limits_{r=a,b} \varphi_{rc}(\varphi_{rc}-2\eps)} \ .
\end{aligned}
\end{equation}

\begin{equation}
\begin{aligned}
Z^{--++}(\varphi,m,\eps) &= \sum_{a=1}^N \frac{\prod\limits_{k=1}^{N_f}\left[\left(\varphi_a-m_k+\epsilon\right)\left(\varphi_a-m_k+3\epsilon\right) \right]}{\prod\limits_{b \neq a}\left[ \varphi_{ab}\left(\varphi_{ab}+2\epsilon\right)^2\left(\varphi_{ab}+4\epsilon\right)\right]}\\
&+\sum_{a \neq b} \frac{2\prod\limits_{k=1}^{N_f}\left[\left(\varphi_a-m_k+\epsilon\right)\left(\varphi_b-m_k+\epsilon\right) \right]}{\varphi_{ab} \left(\varphi_{ab}- 2\epsilon\right)^2(\varphi_{ab}+2\eps)\prod\limits_{c\neq a,b} \prod\limits_{r=a,b} \varphi_{rc}(\varphi_{rc}+2\eps)} \ .
\end{aligned}
\end{equation}

\begin{equation}
\begin{aligned}
Z^{+-+-}(\varphi,m,\eps) &= \sum_{a=1}^N \frac{\prod\limits_{k=1}^{N_f} \left(\varphi_a-m_k-\epsilon\right)^2 }{\prod\limits_{b \neq a}\left[ \varphi_{ab}^2\left(\varphi_{ab}-2\epsilon\right)^2\right]}\\
&+\sum_{a \neq b} \frac{\prod\limits_{k=1}^{N_f}\left[\left(\varphi_a-m_k-\epsilon\right)\left(\varphi_b-m_k+\epsilon\right) \right]}{\varphi_{ab}\left(\varphi_{ab}-2\epsilon\right)^2\left(\varphi_{ab}-4\epsilon\right)\prod\limits_{c\neq a,b}\left[\varphi_{ac}\varphi_{bc}\left(\varphi_{ac}-2\epsilon\right)\left(\varphi_{bc}+2\epsilon\right)\right]      }\\
&+\sum_{a \neq b} \frac{\prod\limits_{k=1}^{N_f}\left[\left(\varphi_a-m_k-\epsilon\right)\left(\varphi_b-m_k-\epsilon\right) \right]}{\varphi_{ab}^2\left(\varphi_{ab}\pm 2\epsilon\right)\prod\limits_{c\neq a,b}\left[\varphi_{ac}\varphi_{bc}\left(\varphi_{ac}-2\epsilon\right)\left(\varphi_{bc}-2\epsilon\right)\right]      }\ .
\end{aligned}
\end{equation}

\begin{equation}
\begin{aligned}
Z^{-+-+}(\varphi,m,\eps) &= \sum_{a=1}^N \frac{\prod\limits_{k=1}^{N_f}\left(\varphi_a-m_k+\epsilon\right)^2 }{\prod\limits_{b \neq a}\left[ \varphi_{ab}^2\left(\varphi_{ab}+2\epsilon\right)^2\right]}\\
&+\sum_{a \neq b} \frac{\prod\limits_{k=1}^{N_f}\left[\left(\varphi_a-m_k+\epsilon\right)\left(\varphi_b-m_k-\epsilon\right) \right]}{\varphi_{ab}\left(\varphi_{ab}+2\epsilon\right)^2\left(\varphi_{ab}+4\epsilon\right)\prod\limits_{c\neq a,b}\left[\varphi_{ac}\varphi_{bc}\left(\varphi_{ac}+2\epsilon\right)\left(\varphi_{bc}-2\epsilon\right)\right]      } \\
&+\sum_{a \neq b} \frac{\prod\limits_{k=1}^{N_f}\left[\left(\varphi_a-m_k+\epsilon\right)\left(\varphi_b-m_k+\epsilon\right) \right]}{\varphi_{ab}^2\left(\varphi_{ab}\pm 2\epsilon\right)\prod\limits_{c\neq a,b}\left[\varphi_{ac}\varphi_{bc}\left(\varphi_{ac}+2\epsilon\right)\left(\varphi_{bc}+2\epsilon\right)\right]      }\ .
\end{aligned}
\end{equation}

\begin{equation}
\begin{aligned}
Z^{+--+}&(\varphi,m,\eps) = \sum_{a=1}^N \frac{\prod\limits_{k=1}^{N_f}\left(\varphi_a-m_k\pm \epsilon\right)}{\prod\limits_{b \neq a}\left[ \varphi_{ab}^2\left(\varphi_{ab}\pm 2\epsilon\right)\right]}\\
&+\sum_{a \neq b} \frac{2\prod\limits_{k=1}^{N_f}\left[\left(\varphi_a-m_k-\epsilon\right)\left(\varphi_b-m_k+\epsilon\right) \right]}{\varphi_{ab}^2\left(\varphi_{ab}-2\epsilon\right)(\varphi_{ab}-4\eps)\prod\limits_{c\neq a,b}\left[\varphi_{ac}\varphi_{bc}\left(\varphi_{ac}-2\epsilon\right)\left(\varphi_{bc}+2\epsilon\right) \right]      } \ .
\end{aligned}
\end{equation}

\begin{equation}
\begin{aligned}
Z^{-++-}&(\varphi,m,\eps) = \sum_{a=1}^N \frac{\prod\limits_{k=1}^{N_f}\left(\varphi_a-m_k\pm \epsilon\right)}{\prod\limits_{b \neq a}\left[ \varphi_{ab}^2\left(\varphi_{ab}\pm 2\epsilon\right)\right]} \\
&+\sum_{a \neq b} \frac{2\prod\limits_{k=1}^{N_f}\left[\left(\varphi_a-m_k+\epsilon\right)\left(\varphi_b-m_k-\epsilon\right) \right]}{\varphi_{ab}^2\left(\varphi_{ab}+2\epsilon\right)(\varphi_{ab}+4\eps)\prod\limits_{c\neq a,b}\left[\varphi_{ac}\varphi_{bc}\left(\varphi_{ac}+2\epsilon\right)\left(\varphi_{bc}-2\epsilon\right) \right]      } \ .
\end{aligned}
\end{equation}
These results also transform appropriately under the discrete symmetries PT and C, which manifest themselves in identities identical to (\ref{PT_inv}) and (\ref{C_inv}) with $a$ and $b$ removed. We have also checked that the chamber dependence disappears when $N_f=0,1,\dots,2N-2$, as expected from \eqref{vanishing_res_infty} and the Moyal product.

%%%%%%%%%%%%%%%%%%%%%%%
%%%%%%%%%%%%%%%%%%%%%%%

\bibliography{Bad}
\bibliographystyle{JHEP}

\end{document}